\documentclass[12pt]{article}
\usepackage{amsmath}

\usepackage{setspace}

\usepackage{amssymb}
\usepackage{mathrsfs}
\usepackage{graphicx}
\usepackage[polutonikogreek,english]{babel}

\usepackage[linktocpage]{hyperref}
\usepackage{comment}

\DeclareFontFamily{U}{mathx}{}
\DeclareFontShape{U}{mathx}{m}{n}{<-> mathx10}{}
\DeclareSymbolFont{mathx}{U}{mathx}{m}{n}
\DeclareMathAccent{\widehat}{0}{mathx}{"70}
\DeclareMathAccent{\widecheck}{0}{mathx}{"71}

\newcommand{\rl}{\ell}
\newcommand{\bv}{\mbox{\boldmath$v$}}
\newcommand{\pR}{\,^{+}\!R}
\newcommand{\bl}{\mbox{\boldmath$l$}}
\newcommand{\pOmega}{\,^+\!\Omega}
\newcommand{\mOmega}{\,^-\!\Omega}
\newcommand{\pmOmega}{\,^\pm\!\Omega}
\newcommand{\pmR}{\,^{\pm}\!R}
\newcommand{\tilS}{\tilde{S}}
\newcommand{\overOm}{\overline{\Omega}}
\newcommand{\overT}{\overline{T}}
\newcommand{\trv}{{\widetilde{\mbox{\rm v}}}}
\newcommand{\rv}{\mbox{\rm v}}
\newcommand{\tf}{{\widetilde{f}}}
\newcommand{\tw}{{\widetilde w}}
\newcommand{\tq}{{\widetilde q}}

\newcommand{\oDelta}{{\overline{\Delta}}}
\newcommand{\cS}{{\widecheck{S}}}
\newcommand{\on}{{\overline{n}}}
\newcommand{\cG}{{\widecheck{G}}}
\newcommand{\bp}{{\mbox{\boldmath$p$}}}
\renewcommand{\d}{\mathrm{d}}
\renewcommand{\Im}{\mathop{\rm Im}\nolimits}
\renewcommand{\Re}{\mathop{\rm Re}\nolimits}
\newcommand{\cA}{{\cal A}}
\newcommand{\rlambda}{{\rm \textgreek{l}}}
\newcommand{\rnu}{{\rm \textgreek{n}}}
\newcommand{\cD}{{\widecheck{D}}}
\newcommand{\oT}{{\overline{T}}}
\newcommand{\ralpha}{{\rm \textgreek{a}}}
\newcommand{\rbeta}{{\rm \textgreek{b}}}
\newcommand{\tg}{{\widetilde g}}
\newcommand{\tA}{{\widetilde{A}}}
\newcommand{\tDelta}{{\widetilde{\Delta}}}
\newcommand{\trlambda}{{\widetilde \rlambda}}
\newcommand{\tvareps}{{\widetilde \varepsilon}}
\newcommand{\tn}{{\widetilde n}}
\newcommand{\trnu}{{\widetilde \rnu}}
\newcommand{\tl}{\widetilde{l}}
\newcommand{\otDelta}{{\overline{\widetilde{\Delta}}}}
\newcommand{\ccM}{{\widecheck{\cal M}}}
\newcommand{\cmM}{{\widecheck {\mathfrak M}}}
\newcommand{\cLambda}{{\widecheck \Lambda}}
\newcommand{\cmf}{{\widecheck {\mathfrak f}}}
\newcommand{\ccF}{{\widecheck{\cal F}}}
\newcommand{\crO}{{\widecheck{\mathrm O}}}
\newcommand{\mm}{{\mathfrak m}}
\newcommand{\cM}{{\cal M}}
\newcommand{\mM}{{\mathfrak M}}
\newcommand{\imu}{\mbox{\it\textgreek{m}}}
\newcommand{\bimu}{{\breve {\imu}}}
\newcommand{\ocrO}{{\overline{\widecheck{\mathrm O}}}}
\newcommand{\cimu}{{\widecheck {\imu}}}
\newcommand{\bcimu}{{\breve {\cimu}}}
\newcommand{\chM}{{\widecheck{M}}}
\newcommand{\rO}{{\mathrm O}}
\newcommand{\cPhi}{{\widecheck \Phi}}
\newcommand{\ccO}{{\widecheck{\cal O}}}
\newcommand{\Det}{\mathop{\mathrm{Det}}}
\newcommand{\occO}{{\overline{\widecheck{\cal O}}}}
\newcommand{\rkappa}{{\rm \textgreek{k}}}
\newcommand{\cO}{{\cal O}}
\newcommand{\mf}{{\mathfrak f}}
\newcommand{\Tr}{\mathop{\mathrm{Tr}}}
\newcommand{\co}{{\widecheck {o}}}
\newcommand{\bco}{{\breve {\co}}}
\newcommand{\mD}{{\mathfrak D}}
\newcommand{\gGamma}{\Gamma \! \! \! \Gamma}
\newcommand{\rgamma}{{\rm \textgreek{g}}}
\newcommand{\rV}{\mathrm{V}}
\newcommand{\orO}{{\overline{\mathrm O}}}
\newcommand{\orV}{\overline{\mathrm{V}}}

\allowdisplaybreaks[4]
\begin{document}

\title{Towards the consistent perturbative expansion in discrete gravity}

\author{V.M. Khatsymovsky \\
 {\em Budker Institute of Nuclear Physics} \\ {\em of Siberian Branch Russian Academy of Sciences} \\ {\em
 Novosibirsk,
 630090,
 Russia}
\\ {\em E-mail address: khatsym@gmail.com}}
\date{}
\maketitle

\begin{abstract}

We consider correctly defining the perturbative expansion in a discrete gravity (simplicial or Regge calculus) needed to study physical effects like graviton loop corrections to Newton's potential.

For the symmetric derivative $\Delta^{(s ) }_\lambda = i \sin p_\lambda$ in the finite-difference action, the propagator has a graviton pole at $\sin^2 p_0 = \sum^3_{\alpha = 1} \sin^2 p_\alpha$, or, at small $p_\alpha$, at $p_0$ close to 0 or $\pm \pi$. This pole doubling means doubling the result of integration over $\d p_0$ compared to the continuum.

The usual derivative $\Delta_\lambda = \exp ( i p_\lambda ) - 1$ leads to a tricky analytical structure of the propagator, since $\Delta_\lambda \neq - \oDelta_\lambda$, and again to a discrepancy with the continuum.

The way out is to use an action $\cS_{\rm g}$ with both $\Delta^{(s ) }_\lambda$ and $\Delta_\lambda$ and the synchronous gauge $g_{0 \lambda} = g_{0 \lambda }^{(0 ) }$ (implemented as "soft" by adding to $\cS_{\rm g}$ a term bilinear in $n^\lambda ( g_{0 \lambda} = g_{0 \lambda }^{(0 ) } )$, $n^\lambda = [ 1, - \varepsilon ( \Delta^{(s ) \alpha } \Delta^{(s ) }_\alpha )^{- 1} \Delta^{(s ) \beta } ]$, $\varepsilon \to 0$, thus removing singularities at $p_0 = 0$).

Given the propagator $\cG ( n, \on )$, we form a principal value propagator $\frac{1}{2} [ \cG ( n, n ) + \cG ( \on, \on ) ]$ by analytically continuing from real $n = \on$. Singularities are resolved like $p_0^{-j} \Rightarrow [ (p_0 + i \varepsilon )^{-j} + (p_0 - i \varepsilon )^{-j} ] / 2$ leading to separate diagram finiteness at $\varepsilon\to0$.

We analyze a 1-parameter $k$ family of actions differing in using $\Delta_\lambda$ vs $\Delta^{(s ) }_\lambda$, find the only one reproducing convergent continuum diagrams for small external momenta (which is natural to demand from discretization), consider finiteness of the principal value gauge-fixing term and vanishing the ghost contribution at $\varepsilon \to 0$. The analysis also applies in a broader context and is illustrated by the electromagnetic/Yang-Mills case.

\end{abstract}


MSC classes: 83C27; 81S40

keywords: general relativity; discrete gravity; Regge calculus; Feynman diagrams; synchronous gauge; functional integral

\tableofcontents

\section{Introduction}

The main task of the given paper is to construct a consistent discrete analogue of the continuum Feynman diagrams. Regge calculus is used as the working mechanism for the emergence of discreteness (meanwhile, the analysis also applies in a broader context). Accordingly, as an introductory part, Subsections \ref{Regge}, devoted to the approach using the Regge calculus, and \ref{Feynman}, devoted to the problem of the correct definition of the discrete Feynman diagrams themselves, and which also contains the further plan of the paper, are presented.

Having a discrete system, we can develop a discrete diagram technique. A natural requirement for discretization is that the discrete and continuum diagrams in the region of small external and loop momenta, where lattice effects should be negligible, should be close to each other; in general, there should be consistency between their analytical structure. This imposes severe restrictions on the true zero-approximation finite-difference form of the action (at the level of non-leading orders over metric variations from site to site) and on the gauge-fixing term. (Otherwise, a discretized diagram, even in the low momenta region, may differ from its continuum prototype not by a small value, as in (\ref{int-sin-p=64int-sin-p/2}) below.) The analysis of such a gauge (synchronous) is complicated by the problem of eliminating nonphysical singularities at $p_0 = 0$ (as in the continuum theory), which is solved somehow by bypassing them at $\varepsilon \neq 0$, where $\varepsilon \to 0$ is a gauge "softening" parameter. Remarkably is that the synchronous gauge is also required from the viewpoint of the feasibility of some functional integration (over connection) in closed form in Subsection \ref{Regge}, needed in constructing the perturbative expansion.

\subsection{Regge calculus approach}\label{Regge}

The Regge calculus (RC) approach to general relativity (GR), which restricts the set of Riemannian spacetimes under consideration to the set of piecewise flat spacetimes consisting of flat 4-simplices (4-dimensional tetrahedra) \cite{Regge}, works with a set of invariant variables—edge lengths. This set is countable, which is important in numerical general relativity and, due to the formal non-renormalizability of general relativity, in quantum theory. A continuum manifold can be obtained as a limit in a certain topology of discrete piecewise flat (simplicial) manifolds with edge length scales tending to zero \cite{Fein,CMS}. Quantum theory of a simplicial manifold can make predictions for physical quantities such as the Newtonian potential \cite{HamWil1}. Here the path integral approach appears to be the most fruitful \cite{HamWil2,Ham1}. By restricting ourselves to a few specific types of 4-simplices, we simplify the use of the functional integral to analyze physical effects in the Causal Dynamic Triangulation (CDT) approach \cite{cdt,cdt1}. Here, simplicial spacetime serves as a regularization tool that allows us to describe GR using a countable set of variables (edge lengths). In some other approaches, spacetime is considered to be truly piecewise flat, as in \cite{Mik}.

Some physical effects can also be analyzed within the framework of perturbation theory. Perturbative gravity at the one-loop level, including divergences in it, was considered in \cite{hooft}. One-loop graviton corrections to the above Newtonian potential were studied in \cite{don,don1,don2,don3,muz,akh,hamliu1,kk,kk1}. (Here we have mentioned important works for us on the direct evaluation of graviton loop diagrams describing the scattering of matter fields, while the Newton potential can also be found by other methods, for example, functional ones, as, for example, in \cite{Frob,Tiber}.) At higher orders, diagram divergences may arise, and the simplicial structure of spacetime may matter. The usual (continuum) perturbation theory needs to be modified to the discrete case. Feynman rules for simplicial gravity were discussed in \cite{HamLiu}.

Now we are interested in the diagrammatic technique that arises within the framework of a specific mechanism for fixing an elementary length scale (a specific functional integral measure). Here, a certain quantum non-perturbativity lies in the difference of the typical elementary edge length scale (which depends on $\hbar $ as $\sqrt{\hbar }$) from zero, that is, in the discreteness itself.

Since we assume actual discreteness and piecewise-flat nature and do not pass to the limit of the elementary edge length scale tending to zero, we understand the passage to the continuum limit as considering the system at scales significantly larger than the typical elementary edge length scale. The RC strategy implies summing/averaging over all possible simplicial structures, and this should restore those symmetries that are spontaneously broken by the choice of a given simplicial structure. A simplified example of such summation is averaging over orientations of a given simplicial structure or, equivalently, averaging over the orientations of the studied system, two gravitating masses in the case of the Newtonian potential, with respect to the simplicial structure.

The functional measure in terms of edge lengths turns out to be bell-shaped, thus being a natural source of an ultraviolet (UV) cutoff \cite{our1}. It turns out that this cutoff is at the edge length scale that maximizes the measure. Such a scale is the Planck length multiplied by a large dimensionless value which is (a function of) a large parameter of the theory. Another consequence of the non-Lebesgue form of the measure is the emergence of new graviton interaction vertices.

In the approach that includes the mechanism of a loose fixation of the edge lengths, the Regge action $S_{\rm g}(\rl )$ as a phase in the functional integral can be obtained if we issue from the functional integral in terms of both the edge lengths or tetrad type variables $\rl$ and independent connection variables $\Omega$, SO(3,1) matrices defined on the 3-simplices, with some discrete gravity action $S_{\rm g}(\rl , \Omega  )$ in terms of both $\rl$ and $\Omega$ and integrate over $\Omega$. Connection variables were proposed by Fr\"{o}hlich \cite{Fro}. Introducing these additional variables is motivated by the fact that using only edge lengths or tetrad type variables $\rl$ does not allow a non-singular passage to the continuous time limit and the Hamiltonian canonical formalism that are stages in the standard construction of a functional integral in the continuous time. The latter should also follow from the full discrete functional measure in the continuous time limit, and this could be used for restoring the full discrete functional measure. The gravitational action $S_{\rm g}(\rl , \Omega  )$ in the variables $\rl$, $\Omega$ should be such that on the equations of motion (for $\Omega$) we obtain exactly the Regge action $S_{\rm g}(\rl )$. Besides, it is tempting to possibly reduce the group SO(3,1) of $\Omega$ to a subgroup. Thus, we arrive at gravity actions using self-dual or anti-self-dual $\Omega$ \cite{Kha}, and ultimately, in the general case, at their combination, characterized by the discrete analogue of the Barbero-Immirzi parameter $\gamma$ \cite{Barb,Imm},
\begin{eqnarray}                                                            %1
& & \hspace{-10mm} S_{\rm g} [ v, \Omega ] = \frac{1}{2} \sum_{\sigma^2} \left ( 1 + \frac{i}{\gamma } \right ) \sqrt{ \bv^2_{\sigma^2} } \arcsin \frac{\bv_{\sigma^2} * \pR_{\sigma^2} ( \Omega )}{\sqrt{ \bv^2_{\sigma^2}}} + \mbox{ c.c.} , \quad \bv * R \stackrel{\rm def}{=} \frac{1}{2}v^i R^{kl} \epsilon_{ikl} .
\end{eqnarray}

\noindent Here $R_{\sigma^2} ( \Omega )$ are the curvature matrices $\prod_{{\sigma^3} \supset {\sigma^2}} \Omega^{\pm 1}_{\sigma^3}$ (holonomy of $\Omega$) on the triangles $\sigma^2$. $2 \bv_{\sigma^2} = i \bl_{\sigma^1_1} \times \bl_{\sigma^1_2} - \bl_{\sigma^1_1} l^0_{\sigma^1_2} + \bl_{\sigma^1_2} l^0_{\sigma^1_1}$ is the complex area 3-vector of the triangle (2-simplex) $\sigma^2$ formed by the edges $\sigma^1_1$, $\sigma^1_2$ with vectors $l_{\sigma^1_1}$, $l_{\sigma^1_2}$. $\pmOmega$ and $\pmR ( \Omega ) = R ( \pmOmega )$ are (anti-)self-dual parts of matrices $\Omega$, $R$ expanded like $\Omega = \pOmega \mOmega$, $\mOmega = ( \pOmega )^*$, as elements of SO(3,1), a subgroup of SO(3,C) $\times$ SO(3,C). "c.c." means "complex conjugate", $( \cdot )^*$.

The choice of the sum of the self-dual and anti-self-dual contributions as the action simplifies the analysis. This is partly similar to the procedure of introducing Ashtekar's variables \cite{Ash} used to formulate loop quantum gravity (LQG) in continuum theory, when only the (anti-)self-dual su(2) connection is used. Contrary to that, we use both self-dual and anti-self-dual connections. Besides that, the mechanism for eliminating UV divergences initially operates via a discrete area spectrum in LQG with a finite nonzero quantum \cite{RovSmo} to compare with a finite nonzero typical elementary area/length scale in RC.

Using these variables $\Omega$, we can find the canonical continuous time functional measure. For a discrete system, it is typical that the Jacobian of the Poisson brackets of the constraints has a singularity at zero fields. In discrete gravity this occurs due to a sharp violation of the diffeomorphism invariance when the metric becomes non-flat \cite{Loll}. However, the full discrete measure that reduces to the continuous time limit regardless of what direction is chosen as a time can be found in some extended configuration superspace of independent area tensors. This measure can then be projected onto the physical hypersurface, where the area tensors are constrained to be bivectors constructed from some edge vectors. This projection means inserting an appropriate delta-function factor into the measure. With taking into account symmetry requirements, this factor is defined up to some additional factors $V^{\eta + \mathrm{const} }_{\sigma^4}$, true analogs of the continuum factors $ ( - \det \| g_{\lambda \mu} \| )^{\eta / 2 + \mathrm{const} }$. Here $\sigma^4$ is a 4-simplex, $V_{\sigma^4}$ is its 4-volume, $\eta$ is a parameter that thus determines the quantum extension of the theory.

We arrive at the aforementioned functional integration over $\Omega$-defined part ${\cal D} \Omega$ (the product of the invariant Haar measures on $\Omega_{\sigma^3} \in \mathrm{SO(3,1)}$ over the 3-simplices $\sigma^3$) of the mentioned full discrete measure $\d \mu ( \rl ) {\cal D} \Omega$,
\begin{equation}\label{int-S D-Omega=F}                                     %2
\int \exp [ i S_{\rm g}(\rl , \Omega  ) ] ( \cdot ) \d \mu ( \rl ) {\cal D} \Omega = \int \exp [ i \tilS_{\rm g} ( \rl ) ] ( \cdot ) F ( \rl ) D \rl ,
\end{equation}

\noindent resulting in a phase $\tilS_{\rm g} ( \rl )$ and a new functional measure $F ( \rl ) D \rl$ in terms of $\rl$ only.

The continuum analogue is the (Gaussian) integration over the connection $\omega \in \mathrm{so(3,1)}$ of the exponential of $i$ times the tetrad-connection form \cite{Holst,Fat} of the Einstein action, which gives $\exp{ (i \frac{1}{2} \int R \sqrt{-g} \d^4 x ) }$ in terms of the tetrad/metric up to a factor of a power of $g$. Now, in the discrete case, we denote by $\Omega = \Omega_0 ( \rl )$ a solution of the equations of motion for $\Omega$ and parameterize by $\omega \in \mathrm{so(3,1)}$ the deviation of $\Omega$ from $\Omega_0$ as $\Omega = \Omega_0 \exp \omega$. To find the phase $\tilS_{\rm g} ( \rl )$, it is appropriate to expand the integrand over $\omega$, since this gives a nonzero result for $\tilS_{\rm g} ( \rl )$ already in the zeroth order, and this is exactly the Regge action $S_{\rm g}(\rl , \Omega_0 ( \rl ) ) = S_{\rm g}(\rl )$ by the construction of $S_{\rm g}(\rl , \Omega  )$. $S_{\rm g}(\rl , \Omega  )$ is linear in the edge areas. Subsequent corrections to $S_{\rm g}(\rl )$ in $\tilS_{\rm g} ( \rl )$ are small if typical values of the integration variable $\omega$ are small. The latter occurs if a typical edge length scale is large, in particular if the spacelike elementary edge length scale $b_{\rm s}$ introduced below is large. The value $b_{\rm s}$ is a characteristic of the initial point of the perturbative expansion and is chosen based on some requirement of maximizing the functional measure, that is, based on dynamic considerations. It is large for a large $\eta$. Therefore, $\eta$ should be a large parameter.

To find the modulus of the result of integration over $\Omega$ in (\ref{int-S D-Omega=F}) or the resulting functional measure $F ( \rl ) D \rl$, it is advisable to expand the integrand in discrete analogues of the ADM lapse-shift functions \cite{ADM1}, since this gives a nontrivial result for the modulus already in the zeroth order, that is, when some edge vectors, which we can call {\it $t$-like} or {\it temporal}, vanish. In this case, no temporal triangle (that is, one that has a temporal edge) contributes to the action. The matrices $R_{\sigma^2} ( \Omega )$ on the remaining ({\it spatial} and {\it diagonal}) triangles are independent and can be taken as a subset of independent connection variables. The functional integration over $\Omega$ is then factorized into factors associated with individual spatial and diagonal triangles and can be performed in closed form \cite{Kha3}.

This stresses the relative simplicity of analyzing the functional integral contribution of configurations in the superspace, singled out by fixing the temporal edge vector components (at a low level), i.e. by imposing (a discrete version of) the synchronous gauge.

We find \cite{our2} that the Regge action can be reduced in the leading order in metric variations to a finite-difference form of the continuum Hilbert-Einstein action. To obtain a finite-difference or "lattice" expression, a periodic structure is required, and we use there the simplest periodic simplicial structure with a hypercubic cell consisting of 4!=24 4-simplices whose edges are cubic edges or diagonals \cite{RocWil}. Thus, the finite-difference Hilbert-Einstein action follows from $S_{\rm g}(\rl , \Omega  )$ in two steps: first, finding the phase $\tilS_{\rm g} ( \rl ) = S_{\rm g}(\rl ) + \dots $ by functional integrating over $\Omega$; second, approximating the obtained Regge action $S_{\rm g}(\rl )$ by the finite-difference Hilbert-Einstein action. Although, on the one hand, for a comparatively simple construction of an algorithmizable non-singular diagram technique, we need to further reduce the set of variables describing the 24 4-simplices to a 4-vector tetrad per site; on the other hand, to immediately move on to finite-difference forms, it is easier to appropriately reduce both the tetrad and connection variables to certain $l_\lambda^a$, $\Omega_\lambda$ per site already at the $S_{\rm g}(\rl , \Omega  )$ level: besides, in this case it will be more visual to estimate the order of magnitude of the possible effect of unaccounted subsequent terms in the expansion in $\omega$. This reduction is achieved by imposing certain conditions on the edge vectors and matrices $\Omega_{\sigma^3}$ that still allow any configuration in the superspace to be approximated in some topology with arbitrarily high accuracy. We have a corresponding mini-superspace simplicial action,
\begin{eqnarray}                                                          %3,4
& & S_{\rm g} [ v, \Omega ] = \frac{1 }{4 } \sum_{ \lambda \mu \nu \rho , {\rm sites}} \left ( 1 + \frac{i}{\gamma } \right ) \epsilon^{\lambda \mu \nu \rho} \sqrt{ \bv^2_{\lambda \mu} } \arcsin \frac{\bv_{\lambda \mu} * \pR_{\nu \rho} ( \Omega )}{\sqrt{ \bv^2_{\lambda \mu}}} + {\rm c. c. } , \\ & & 2 \bv_{\lambda\mu} = i \bl_\lambda \times \bl_\mu + l^0_\lambda \bl_\mu - l^0_\mu \bl_\lambda, \quad R_{\lambda\mu} (\Omega ) = \overOm_\lambda (\overT_\lambda \overOm_\mu) (\overT_\mu \Omega_\lambda) \Omega_\mu ,
\end{eqnarray}

\noindent where $\overline{( \cdot )}$ means the Hermitian conjugation, $T_\lambda$ is the shift operator of the $x^\lambda$-coordinate by 1 (to the next site). In \cite{our1} we found that this action is equivalent to the finite-difference form of the tetrad-connection representation \cite{Holst,Fat} of the Hilbert-Einstein action in the $\Omega$-elimination aspect at leading order in $l_\lambda^a$-variations. Excluding $\omega$ from this form using the equations of motion, we operate with finite differences in the leading order over themselves just as with ordinary derivatives, and thus arrive at a finite-difference form of the Einstein action $\frac{1}{2} \int R \sqrt{-g} \d^4 x$, where the 4-vectors $l_\lambda^a$ play the role of the tetrad.

We can imply some parametrization of the variables $\rl = (l_1, \dots, l_n )$ through some another variables $u = (u_1, \dots, u_n )$ such that the measure becomes Lebesgue in these new variables: $F ( \rl ) D \rl = D u$. We can then develop a perturbative expansion for the functional integral (\ref{int-S D-Omega=F}) in the neighbourhood of some initial point $\rl_{(0)} = \rl ( u_{(0)} )$ and a certain diagram technique \cite{khat,khat2}. As usual, it is convenient to choose this point so that it is a solution of the equations of motion for $\tilS_{\rm g} ( \rl )$, which ensures that there is no linear order term in the expansion of the action around this point in $\Delta u = u - u_{(0)}$:
\begin{eqnarray}\label{ddS}                                               %5,6
\hspace{0mm} \tilS_{\rm g} ( \rl ) & = & \tilS_{\rm g} ( \rl_{(0)} ) + \frac{1}{2} \sum_{j, k, l, m} \frac{\partial^2 \tilS_{\rm g} ( \rl_{(0)} )}{\partial l_j \partial l_l} \frac{\partial l_j (u_{(0)} )}{\partial u_k} \frac{\partial l_l (u_{(0)} )}{\partial u_m} \Delta u_k \Delta  u_m + \dots , \\ \label{dS/dl=0} \frac{\partial \tilS_{\rm g} ( \rl_{(0)} )}{\partial l_j} & = & 0 .
\end{eqnarray}

\noindent Analogously to requiring an extremum for the zeroth order term via fulfilling the equations of motion at $\rl_{(0)} = \rl ( u_{(0)} )$, we can require (within the freedom allowed by the equations of motion) an extremum for the second order term by the minimization of the determinant of the second order form in the exponent,
\begin{equation}\label{def-l0}                                              %7
F (\rl_{(0)} )^{-2} \det \left \| \frac{\partial^2 \tilS_{\rm g} (\rl_{(0)} )}{\partial l_i \partial l_k} \right \| = \mbox{ minimum}.
\end{equation}

\noindent To be exact, gauge-breaking and ghost terms are implied to be added to $\tilS_{\rm g} (\rl )$ here.

The aforementioned closed form of the result of functional integration over $\Omega$ (in the factorization approximation) \cite{Kha3} includes an exponentially suppressing factor $\exp ( - 2 \pi V_{\sigma^2} )$ in $F ( \rl )$ with the triangle area $V_{\sigma^2}$ (for a spatial or diagonal triangle $\sigma^2$). This can be illustrated by a simplified qualitative example of an integral of the type of $\int \exp ( i v_{ a b } R^{ a b } ) {\cal D} R$ ($ v_{a b}$ is the area tensor, $\frac{1}{2} \epsilon_{abcd} l^c_1 l^d_2 $ for the triangle formed by the edge vectors $l^c_1$, $l^d_2 $). Integrating this function of $ v_{a b}$ as independent variables with a monomial of the components $ v_{a b}$ over $\d^6 v$ (the requirement of analytic extendibility to independent $ v_{a b}$ is in good agreement with the aforementioned concept of an extended superspace of independent area tensors), we get the integral over ${\cal D} R$ of the multiple partial derivative of $\delta^6 ( R - \overline{R} )$. This integral turns out to be finite; to ensure this for any monomial of $ v_{a b}$, the function $\int \exp ( i v_{ a b } R^{ a b } ) {\cal D} R$ should contain exponential suppression of large areas.

$F ( \rl )$ also contains the aforementioned factor $V^\eta_{\sigma^4}$ with the 4-simplex volume $V_{\sigma^4}$ and is bell-shaped. The maximization condition (for $F ( \rl )$) (\ref{def-l0}) gives for the aforementioned typical spacelike elementary edge length scale $b_{\rm s}$:
\begin{equation}\label{b=l-pl-sqrt-eta}                                     %8
b_{\rm s} = l_{\rm Pl} \sqrt{ \frac{ \eta - 9 }{ 2 \pi } },
\end{equation}

\noindent $l_{\rm Pl} = \sqrt{8 \pi G}$ in the usual units. (The subtraction of the number 9 from $\eta$ in this expression is determined by a certain definition of $\eta$.) In this case, the initial metric is $g^{(0 )}_{\lambda \mu} = {\rm diag} (-b_{\rm t}^2, b_{\rm s}^2, b_{\rm s}^2, b_{\rm s}^2)$. Here $b_{\rm t}$ in (the discrete analogue of) the synchronous gauge is chosen as a gauge parameter (we consider the details in \cite{khat2}).

The metric $g_{\lambda \mu}$ can be parameterized by some $\tg_{\nu \rho} = \eta_{\nu \rho} + \tw_{\nu \rho}$, $g_{\lambda \mu} = g_{\lambda \mu} ( \{ \tg_{\nu \rho} \} )$, so that the functional measure $F ( \rl ) D \rl$ is the Lebesgue measure $\prod_\mathrm{sites} \d^{1 0} \tg_{\lambda \mu}$ in terms of $\tg_{\nu \rho}$. The parametrization $\rl = \rl ( u )$ (in particular, $g_{\lambda \mu} = g_{\lambda \mu} ( \{ \tg_{\nu \rho} \} )$) is not unique. Since we have the functional integration over $\Omega$ factorized over the individual triangles, it is natural to take a part of $\rl$ as variables $\rv_1, \dots \rv_m$ which have the sense of areas of the type of $\sqrt{ g_{2 2} g_{3 3} -  g_{2 3}^2 }$, perm(1,2,3), etc., and write the full measure in the form factorized over $\rv_1, \dots \rv_m$: $F ( \rl ) D \rl = [ \prod^m_{k = 1} f_k ( \rv_k, l_{m + 1}, \dots l_n ) \d \rv_k ] \prod^n_{k = m + 1} \d l_k$. In terms of $\tg_{\lambda \mu}$ we have the Lebesgue measure and a similar form with the variables related to $\tg_{\lambda \mu}$ with tilde $\trv_k, \tl_k$ and some simpler functions with tilde $\tf_k$: $ D u = [ \prod^m_{k = 1} \tf_k ( \trv_k, \tl_{m + 1}, \dots \tl_n ) \d \trv_k ] \prod^n_{k = m + 1} \d \tl_k$. Then it is natural to put $f_k ( \rv_k, l_{m + 1}, \dots, l_n ) \d \rv_k = \tf_k ( \trv_k, \tl_{m + 1}, \dots, \tl_n ) \d \trv_k$, $k = 1, \dots, m$, $l_k = \tl_k$, $k = m + 1, \dots, n$, and the problem splits into one-dimensional problems of expressing $\rv_k$ through $\trv_k$; $\tl_j = l_j$, $j = m + 1, \dots, n$, play the role of parameters.

In the functional measure we have factors raised to a power that is a large parameter, $\eta$. In this case, it is quite common for powers of a large parameter to appear in the perturbative expansion. Let us assume for a moment that there is no exponential suppression of large areas, and we propose a power volume factor in the measure, like $ ( - \det \| g_{\lambda \mu} \| )^{\eta / 2 }$, in the hope of suppressing small distances (the UV contribution). For some variable $y$ having the meaning of some length, we write the related part of the measure as $y^{\eta - 1} \d y$ and parameterize $y = y ( x )$ to get the Lebesgue measure $C \d x$. We take $C = y_0^\eta$ and expand $y = y ( x )$ around some point $y = y_0$, $x = 0$:
\begin{equation}                                                            %9
y_0^{- 1} y = (1 + \eta x)^{1 / \eta} = 1 + x + \dots + O ( \eta^{j - 1} ) \cdot x^j + \dots .
\end{equation}

\noindent The bilinear form of the action is due to the linear term in $x$ in this expansion and dynamically forces the typical value of $x$ to be equal to $O ( 1 )$ with respect to $\eta$. The subsequent terms define new interaction vertices. They have coefficients proportional to $\eta$ and higher powers of $\eta$, and the new vertices have similar coefficients. Thus, powers of $\eta$ appear in the perturbative expansion.

However, in the presence of exponential suppression of large areas in the measure, the situation turns out to be completely different. (The example measure given in the previous paragraph becomes something like $y^{\eta - 1} \exp ( - y ) \d y$, where $y$ is some area.) We show in \cite{khat,khat2} that then the perturbative expansion does not contain increasing powers of $\eta$ when choosing the initial point of the perturbative expansion not too far from the maximum point of the measure (\ref{def-l0}). Namely, the chosen initial edge length scale $b$ must differ from the length scale $b_\mathrm{s}$ (\ref{b=l-pl-sqrt-eta}), which maximizes the measure, by no more than a value of the order of the Planck length,
\begin{equation}                                                           %10
| b - b_\mathrm{s} | \lesssim O ( 1 ) \cdot l_{\rm Pl} .
\end{equation}

Limiting ourselves to the linear part of $g_{\lambda \mu} = g_{\lambda \mu} ( \{ \tg_{\nu \rho} \} )$ in $\tw_{\nu \rho}$, we obtain a discrete form of the standard continuum Feynman diagrams for gravity. If a given continuum diagram (or its structure) converges, this means that its value is determined by loop momenta that are smaller than or of the order of the external momenta; if the external quasi-momenta are significantly smaller than 1 in absolute value, then this discrete form should reproduce the continuum diagram when the components of the external quasi-momenta $b_\mathrm{s} \tq_{j \alpha}$, $b_\mathrm{t} \tq_{j 0}$ are small in absolute value compared to 1 ($\tq_j$ are momenta with components $(\tq_{j0}, \tq_{j1}, \tq_{j2}, \tq_{j3})$).

As mentioned above, since $b$ is fixed and does not tend to zero, the notion of the continuum limit amounts to the limit of small quasi-momenta, both external and loop. As we see, for the considered diagrams with converging continuum analogues, this limit is achieved in a simple way: by taking only the external quasi-momenta to be small, the effective loop quasi-momenta automatically become small.

Nonlinear in $\tw_{\nu \rho}$ terms in $g_{\lambda \mu} = g_{\lambda \mu} ( \{ \tg_{\nu \rho} \} )$ lead to the appearance of new interaction vertices in the action and new diagrams.

Further, to avoid unnecessary cumbersomeness and since we do not touch upon the mechanism of forming $b_\mathrm{s}$, $b_\mathrm{t}$, we assume $b_\mathrm{s} = 1$, $b_\mathrm{t} = 1$. The case of common $b_\mathrm{s}$, $b_\mathrm{t}$, where necessary, is treated by redefining variables and is analyzed in Subsection \ref{detail-pole}.

\subsection{Discrete Feynman diagrams}\label{Feynman}

Above we mentioned the relative simplicity of performing functional integration over connection for configurations in superspace singled out by fixing the temporal edge vector components (at a low level), i.e. by imposing (a discrete version of) the synchronous gauge. This gauge can also be approached in a broader context when trying to construct a consistent discrete analogue of the continuum Feynman diagrams. This implies certain limitations on the possible choice of the "true" zero-approximation finite-difference form of the action. These limitations concern the terms of non-leading orders in metric variations from site to site. (In the leading order this form coincides with the continuum GR action.) An analogous situation takes place for the discrete (finite-difference) form of Yang-Mills fields in the temporal gauge.

Discrete Feynman diagrams in a simplicial framework were probably first addressed in the paper \cite{HamLiu}, and our present paper is close in spirit to it. There these diagrams for a scalar field were considered. For a discrete scalar field $\phi$ described by its values at the sites of a hypercubic structure and the action in the form of a sum over the sites $( - 1 / 2 ) \sum_\mathrm{sites} [ g^{\lambda \mu} ( \Delta_\lambda \phi )( \Delta_\mu \phi ) + m^2 \phi^2 ] \sqrt{ - g }$ ( $g^{\lambda \mu} = \eta^{\lambda \mu} = \mathrm{diag} ( - 1, 1, 1, 1 )$, $ \Delta_\lambda = \exp ( i p_\lambda ) - 1 $ is the usual advanced finite-difference derivative, $p_\lambda$ is a quasi-momentum, $ - \pi < p_\lambda \leq \pi $), the propagator $- i \langle \phi \phi \rangle$ is $\propto ( - \oDelta \Delta - m^2 + i 0 )^{- 1} = [ 4 \sin^2 ( p_0 / 2 ) - \sum^3_{\alpha = 1} 4 \sin^2 ( p_\alpha / 2 ) - m^2 + i 0 ]^{- 1}$. A simplifying circumstance in this case is that two finite-difference derivatives in the action are contracted with each other, they enter the propagator in the form $\oDelta \Delta$, which ensures its relatively simple form and analytical structure.

In the case of a field with a nonzero integer spin, in particular the graviton itself, $\Delta_\lambda$ can be contracted with the field itself over a vector index. Then the calculations become complicated because of the need to distinguish between $\Delta_\lambda$ and $\pm \oDelta_\lambda$. This leads to a complicated analytical structure of the propagator. A natural requirement for a discrete Feynman diagram (or its structure) is to be reduced to its continuum version at small external momenta if the continuum version gives a non-infinite result (as in the case of one-loop corrections to Newton's law, including graviton loops). Meanwhile, the mentioned complicated analytical structure of the propagator may prevent from such a correspondence with the continuum theory. This problem requires an analysis.

So far we have not specified the gauge-fixing term required in the cases of Yang-Mills fields and gravity.

In the discrete case, the diffeomorphism symmetry is preserved in the leading order over metric variations from site to site (or from simplex to simplex). Therefore, when these variations are small, there are degrees of freedom close to the gauge ones of the continuum theory. That is, in the functional integral approach we are faced with a set of physically almost equivalent configurations of infinite functional measure. To eliminate this set, we can introduce a gauge, which means that we restrict ourselves to only a subset of all configurations in the superspace.

Unlike the continuum case, the result of calculating a given physical quantity will depend on the gauge. But summing/averaging over all gauges should yield the value of the original functional integral and the exact physical value in question. Here, summation over gauges is understood in a broad sense and also includes taking into account all possible simplicial structures (which is precisely what is implied by the RC strategy). The simplest such summation over structures would be an averaging over lattice orientations; for example, in the case of studying the Newtonian potential, this is equivalent to an averaging over the orientations of the line connecting the gravitating masses relative to the lattice axes.

Instead of singling out a subset in the configuration superspace and restricting ourselves to it, we can add a gauge-fixing term to the action, thereby introducing a weight factor into the functional integral. In the continuum limit, when metric variations from site to site are small, such a gauge-fixing term should reduce to a continuum gauge-fixing term.

Having added a gauge-fixing term to the action, we can find the propagator. Analytical properties of the propagator depend on the specific finite-difference form of the action. This is one of the rare cases when adding terms of non-leading order over metric variations from site to site to the action can change the result of the diagram calculation significantly. Therefore, we need to find some true zeroth approximation for the action, which requires specifying not only the leading but also the non-leading terms. The discrete propagator has its simplest form if the discrete version of the action plus a gauge-fixing term is obtained by replacing the derivative $\partial_\lambda$ with a symmetrized finite-difference form of the derivative $\Delta^{(s )}_\lambda$. The latter is anti-Hermitian, like $\partial_\lambda$, and it is due to this property that the discrete propagator follows from the continuum one by replacing $\partial_\lambda$ with $\Delta^{(s )}_\lambda$. In the momentum representation $\Delta^{(s )}_\lambda = i \sin p_\lambda$, $p_\lambda \in ( - \pi , \pi ]$ is the quasi-momentum. The graviton pole should be a pole of the factor $( \sum^3_{\alpha = 1} \sin^2 p_\alpha - \sin^2 p_0 - i 0 )^{- 1}$. For small $| \bp | \ll 1$, the pole is located not only at small $p_0 \approx \pm ( | \bp | - i 0 )$, as in the continuum limit, but also at $p_0 \approx \pm ( \pi - | \bp | + i 0 )$.

This pole doubling leads to that when integrating over $\d p_0$ by calculating the residues at the corresponding poles, we obtain a result approximately twice as large as that obtained by such integration in the analogous continuum diagram (it is assumed that the spatial quasi-momenta are small, so that $ \sin p_\alpha \approx p_\alpha$). Moreover, such doubling occurs not only when integrating over $\d p_0$, but also when integrating over the spatial components of quasi-momenta. This refers to a diagram or structure whose continuous version is finite. In this case, in the discrete version obtained by replacing $ p_\lambda \Rightarrow 2 \sin ( p_\lambda / 2 ) $, this corresponds to a region of $ p_\lambda $ close to zero that makes the dominant contribution, and in the version obtained by replacing $ p_\lambda \Rightarrow \sin p_\lambda $, this corresponds to two regions with the same dominant contribution: $ p_\lambda $ close to 0 and $ p_\lambda $ close to $ \pm \pi $. As an example, we consider such a finite continuum Feynman integral
\begin{equation}                                                           %11
I ( q_1, q_2 ) = \int \frac{ \d^4 p }{ p^2 ( p - q_1 )^2 ( p - q_2 )^2 } .
\end{equation}

\noindent In the discrete version obtained by replacing $ p_\lambda \Rightarrow \sin p_\lambda $, we split the integration interval $(-\pi, \pi]$ into intervals $(-\pi, -\pi / 2]$, $(-\pi / 2, \pi / 2]$, $ ( \pi / 2, \pi ]$, shift them into a single $(-\pi / 2, \pi / 2]$ by shifting $p_\lambda$ by $\pm \pi$ and scale $(-\pi / 2, \pi / 2]$ to $(-\pi, \pi]$ by replacing $ p_\lambda \Rightarrow p_\lambda / 2 $. The result is proportional to the expression for the discrete version obtained by replacing $ p_\lambda $ with $ 2 \sin ( p_\lambda / 2 ) $ with doubled external quasi-momenta ($q_1 \Rightarrow 2 q_1$, $q_2 \Rightarrow 2 q_2$), which reproduces $I ( 2 q_1, 2 q_2 )$ with a relative accuracy of $O ( q^2 )$, where $q$ is a scale of $q_1$, $q_2$:
\begin{eqnarray}\label{int-sin-p=64int-sin-p/2}                            %12
& & \hspace{-5mm} \int^\pi_{- \pi} \frac{ \d^4 p }{ [ \sum_\alpha \sin^2 ( p_\alpha - q_{ 1 \alpha } ) - \sin^2 ( p_0 - q_{ 1 0 } ) ] [ \sum_\alpha \sin^2 ( p_\alpha - q_{ 2 \alpha } ) - \sin^2 ( p_0 - q_{ 2 0 } ) ] } \nonumber \\ & & \hspace{-5mm} \cdot \frac{ 1 }{ \sum_\alpha \sin^2 p_\alpha - \sin^2 p_0 } = 64 \int^\pi_{- \pi} \frac{ \d^4 p }{ ( \sum_\alpha 4 \sin^2 \frac{ p_\alpha }{ 2 } - 4 \sin^2 \frac{ p_0 }{ 2 } ) [ \sum_\alpha 4 \sin^2 \frac{ p_\alpha - 2 q_{ 1 \alpha } }{ 2 } - 4 \sin^2 \frac{ p_0 - 2 q_{ 1 0 } }{ 2 } ] } \nonumber \\ & & \hspace{-5mm} \cdot \frac{ 1 }{ \sum_\alpha 4 \sin^2 \frac{ p_\alpha - 2 q_{2 \alpha} }{ 2 } - 4 \sin^2 \frac{ p_0 - 2 q_{2 0} }{ 2 } } = 64 [ 1 + O ( q^2 ) ] I ( 2 q_1, 2 q_2 ) = 16 [ 1 + O ( q^2 ) ] I ( q_1, q_2 ) ,
\end{eqnarray}

\noindent since $ I ( 2 q_1, 2 q_2 )  = \frac{ 1 }{ 4 } I ( q_1, q_2 ) $, and we have the result in the theory with the discrete action with the finite-difference derivative $\Delta^{( s )}_\lambda$ 16 times larger per loop than that given by the continuum diagram. Evidently, such a rough mismatch between the discrete and continuum theories requires to be managed somehow.

And even if we use $\Delta_\lambda$ and/or $\oDelta_\lambda$ everywhere in the action and pass through the aforementioned bulkiness caused by the need to distinguish between $\Delta_\lambda$ and $\pm \oDelta_\lambda$ in calculations, it turns out that for the hard synchronous gauge $g_{0 \lambda} = - \delta_{0 \lambda}$ these factors have sixth-degree denominators with respect to $\Delta_\lambda$, $\oDelta_\lambda$ and poles at $\Im p_0 \neq 0$ if $\bp \neq 0$. Looking ahead, we will have a picture of the location of nonphysical poles that is qualitatively similar to Fig.~\ref{f1}(c). And, as discussed there, introducing the principal value type prescription with $\varepsilon \to 0$ does not work properly for such poles for almost all spatial quasi-momenta $\bp$ and does not correspond to the continuum result. This also speaks against using only $\Delta_\lambda$ (and $\oDelta_\lambda$) everywhere.

The procedure of calculating an integral $\int^{+ \pi}_{- \pi} ( \Delta^{(s ) 2})^{- 1} \dots \d p_0$ can be based on integrating along the contour $C$, consisting of segments $[ - \pi , + \pi ]$, $[ + \pi , + \pi +i L ]$, $[ + \pi + i L , - \pi + i L ]$, $[ - \pi +i L , - \pi ]$, $L \to \infty$, Fig.~\ref{f2,75}. On the segments
\begin{figure}[h]
\unitlength 0.4pt
\begin{picture}(342,165)(-613,-51)
\put(96,-43){\line(1,0){38}}
\put(-43,-61){\line(0,1){160}}
\put(-51,-38){\circle*{6}}
\put(-190,-38){\circle*{6}}
\put(88,-38){\circle*{6}}
\put(-35,-49){\circle*{6}}
\put(-174,-49){\circle*{6}}
\put(104,-49){\circle*{6}}
\put(-182,-43){\line(-1,0){26}}
\put(-214,77){$- \pi + i L$}
\put(-220,-60){$- \pi$}
\put(-37,93){$\Im p_0$}
\put(-58,-66){$0$}
\put(79,77){$\pi + i L$}
\put(32,77){$C$}
\put(80,-61){$\pi$}
\put(123,-35){$\Re p_0$}

\thicklines

\put(134,-43){\vector(1,0){5}}
\put(-43,99){\vector(0,1){5}}
\put(96,-43){\vector(0,1){58}}
\put(96,15){\line(0,1){59}}
\put(96,74){\vector(-1,0){139}}
\put(-43,74){\line(-1,0){139}}
\put(-182,74){\vector(0,-1){58}}
\put(-182,16){\line(0,-1){59}}
\put(-182,-43){\vector(1,0){119}}
\put(-63,-43){\line(1,0){159}}

\end{picture}
\caption{Integration contour for calculating $\int^{+ \pi}_{- \pi} ( \Delta^{(s ) 2} + i 0 )^{- 1} \dots \d p_0$; \raisebox{0.7mm}{ \circle*{6}} \hspace{0mm} are poles.
\label{f2,75}}

\end{figure}
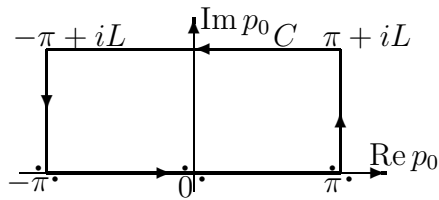
$[ + \pi , + \pi +i L ]$ and $[- \pi, - \pi +i L ]$, the values of the integrand coincide due to its periodicity in $\Re p_0$, and their contributions cancel each other out, since these segments are traversed in opposite directions. Therefore, the integral of interest is the sum of the contour integral and the integral over the remote segment $[ - \pi + i L , + \pi + i L ]$, $\int^\pi_{- \pi} = \oint_C + \int^{ \pi + i L}_{- \pi + i L}$; in turn, $\oint_C$ is equal to the sum of the residues at the poles inside the contour. The integral over the remote segment is an analogue of the integral over the remote semicircle in the continuum case.

An attempt to use instead of $\Delta^{(s )}_\lambda$ somehow the trigonometric functions of the half quasi-momentum $\Delta^{(s / 2 )}_\lambda = 2 i \sin ( p_\lambda / 2 ) $ would lead to a violation of the periodicity by $2 \pi$, the contributions from the segments $[ + \pi , + \pi +i L ]$ and $[- \pi, - \pi +i L ]$ would cease to compensate each other and would not correspond to the continuum case. These contributions themselves are unusual for the continuum and have non-standard i-ness.

A natural approach is to use both forms of the derivative, $\Delta^{(s )}_\lambda$ and $\Delta_\lambda$. In this case, $\Delta_\lambda$ is used only in those terms where the contraction of two derivatives occurs, $ \dots g^{\lambda \mu} ( \Delta_\lambda \dots )( \Delta_\mu \dots )$. The terms to which spin-2 components contribute are also of this type. The remaining terms use the $\Delta^{(s )}_\lambda$ form. When calculating the propagator in this approach, we encounter a dependence on one new value of scalar type $\cA = - \oDelta \Delta - \Delta^{(s ) 2} = O ( [ \Delta ]^4 )$ (a fourth-order quantity in metric variations).

Particular attention should be paid to the gauge-fixing term. If it contains derivatives, then the natural such term is the de Donder-Fock gauge-fixing term. With this approach, it should contain the derivative form $\Delta^{(s )}_\lambda$ and be equal to $ - \frac{ 1 }{ 4 } \sum_{\rm sites} \rlambda ( \Delta^{(s ) }_\nu h^{ \lambda \nu } ) \eta_{\lambda \mu} ( \Delta^{(s ) }_\rho h^{ \mu \rho } )$, $h^{\lambda \mu} = g^{\lambda \mu} \sqrt{- g}$, where $\rlambda$ is a parameter. In the special case $\rlambda = \infty$, adding this term is equivalent to imposing the condition $\Delta^{(s ) }_\mu h^{ \lambda \mu } = 0$. This suggests some general expression for $h^{\lambda \mu}$ using the transverse projector, $h^{\lambda \mu} = ( \Delta^{(s ) 2 } + i 0 )^{- 1} [ \delta^\lambda_\nu \Delta^{(s ) 2 } - \Delta^{(s ) \lambda } \Delta^{(s ) }_\nu ] X^{\nu \mu}$. This projector and, in particular, the factor $( \Delta^{(s ) 2 } + i 0 )^{- 1}$ should also appear in the propagator. And indeed, we know that the continuum propagator in such a gauge contains the transverse projector and an additional inverse d'Alembertian, and we now see that in the discrete case this inverse d'Alembertian is $( - \Delta^{(s ) 2 } - i 0 )^{- 1}$, leading to a doubling of the poles compared to the continuum limit when $p_\lambda$ is small. Though, there is a possibility that the terms with this additional factor may cancel each other out, as is the case in the Feynman gauge analogue at $\rlambda = 1$. But at $\cA \neq 0$, there are two inverse d'Alembertians, $( \oDelta \Delta )^{- 1}$ and $( - \Delta^{(s ) 2 } )^{- 1}$, and one should check that it is the latter that cancels out.

But even if we are left with the true inverse d'Alembertian, $( \oDelta \Delta )^{- 1}$, in the graviton propagator, there is also a ghost field propagator. The quadratic part of the ghost action is obtained by subjecting $\Delta^{(s ) }_\mu h^{ \lambda \mu }$ to the diffeomorphism transformation that also uses the symmetric form of the derivative $\Delta^{(s ) }_\lambda$ (eq. (\ref{gXi-g}) below). Then the ghost propagator uses the inverse d'Alembertian $( - \Delta^{(s ) 2 } )^{- 1}$ with the pole doubling. Thus, entire subdiagrams (ghost loops) do not match the continuum case, and hence neither do complete diagrams.

Thus we arrive at a gauge-fixing term, typically having no derivatives, the usual one being the (discrete version of the) synchronous gauge-fixing term, as being appropriate to the problem at hand. This is a gauge on $n^\lambda g_{\lambda \mu}$, where normally $n^\lambda = (1, 0, 0, 0)$, i.e. a gauge on the values $g_{0 \lambda}$, which are related to the Lagrange multipliers in the canonical Hamiltonian formalism for gravity. Correspondingly, this gauge refers to the so-called Hamiltonian gauges.

The synchronous gauge in the continuum theory leads to singularities of the propagator at $n p = 0$. These singularities are of infrared type and are therefore common to both continuum and discrete theories. They can be eliminated in analogy with Yang-Mills theories, where Landshoff proposed the prescription $p_0^{- 2} \Rightarrow (p_0^2 + \varepsilon^2)^{- 1}$ \cite{Land} for the gauge field propagator in the temporal gauge $A^a_0 = 0$, and this was confirmed by considering the "softened" gauge $n^\lambda A^a_\lambda = 0$ with $n^\lambda = (1, - \varepsilon (\partial^\alpha \partial_\alpha )^{- 1} \partial^\beta ) $ at $\varepsilon \to 0$ \cite{Ste}. Then the gauge field propagator contains terms with factors $( n p )^{- 1} = (p_0 + i \varepsilon )^{- 1}$, $( \on p )^{- 1} = (p_0 - i \varepsilon )^{- 1}$. We used such $n^\lambda$ for the gauge $n^\lambda w_{\lambda \mu} = 0$, $w_{\lambda \mu} = g_{\lambda \mu} - g_{\lambda \mu }^{(0 ) }$, in gravity \cite{khat0}. The graviton propagator is found in this gauge, the $p^{- j}_0 $ singularities ($j$ from 1 to 4) are replaced by products of $( n p )^{- 1}$ and $( \on p )^{- 1}$ factors, and the Faddeev-Popov ghost contribution is found to go to zero as $\varepsilon$ goes to 0.

Of interest is some modification of this prescription. In this prescription, the propagator $G ( n, \on )$ for the gauge-fixing term $ - \frac{1}{4} \int (n^\mu w_{\mu \lambda}) \rlambda^{\lambda \sigma } (n^\tau w_{\tau \sigma}) \d^4 x$ depends on mutually Hermitian conjugate operators $n$ and $\on$ (which are mutually conjugate complex numbers in the momentum representation). $G ( n, \on )$ is a Hermitian operator. Hermitianity is achieved mainly by multiplying mutually conjugate factors in individual terms: $( n p )^{- j} ( \on p )^{- j} = (p_0^2 + \varepsilon^2)^{- j}$. The terms $\sim (p_0^2 + \varepsilon^2)^{- j}$ are finite $\forall p_0$, but integrating over $p_0$ can yield negative powers of $\varepsilon$ that diverge at $\varepsilon \to 0$. Such a situation occurs for the non-pole terms of the graviton propagator. These terms at the one-loop level do not contribute to the absorptive part of the S-matrix, as mentioned in \cite{hooft}. But having in view multi-loop applications, we can look at another way of forming the propagator as a Hermitian operator by taking the half-sum of mutually conjugate functions of $n$ and $\on$: $[ G ( n, n ) + G ( \on, \on ) ] / 2$. $G ( n, n )$ and $G ( \on, \on )$ are analytic continuations of $G ( n, n ) = G ( \on, \on ) = G ( n, \on )$ for real (Hermitian) $n = \on$ to complex (neither Hermitian nor anti-Hermitian) $n$. The peculiarity of such an expression is that each term has poles only on one side of the integration path $\Im p_0 = 0$. Therefore, this path can be deformed to lie at a distance $O ( 1 )$ from the poles, and the passage to the limit as $\varepsilon$ tends to 0 is clearly non-singular. Up to terms with coefficients vanishing at $\varepsilon \to 0$, this reduces to the prescription $p_0^{-j} \Rightarrow [ (p_0 + i \varepsilon )^{-j} + (p_0 - i \varepsilon )^{-j} ] / 2$ for negative powers of $p_0$. It is often called a principal value prescription, although not in Cauchy's original sense, and we consider its matrix analogue or a principal value type prescription for the propagator. We can find the gauge-fixing term required for this form of $G$ \cite{khat0.5}. It is a function of $G ( \rnu, \rnu )$, $\rnu^\lambda = (1, 0, 0, 0) $. Remarkably, this term has the form $ - \frac{ 1 }{ 4 } \int f_\lambda \Lambda^{\lambda \mu} f_\mu \d^4 x $ with an operator $\Lambda^{\lambda \mu} = O ( \varepsilon^{- 2} )$ depending on $\partial$ and not on coordinates and a metric functional $f_\lambda$, that is, it indeed imposes a gauge on four values $f_\lambda$ or on four degrees of freedom. We can also find that the ghost contribution vanishes in the limit $\varepsilon \to 0$.

In the present paper, we analyze the soft synchronous gauge in the form of such a principal value type prescription in the discrete framework, using the above mentioned refined finite-difference form of the action $\cS_{\rm g}$, which uses both $\Delta^{(s ) }_\lambda$ and $\Delta_\lambda$ derivative forms. The required gauge-fixing term is a function of the propagator $\cG ( \rnu, \rnu )$, $\rnu^\lambda = (1, 0, 0, 0) $, for such an action and for the gauge-fixing term $ - \frac{1}{4} \sum_{\rm sites} (n^\mu w_{\mu \lambda}) \rlambda^{\lambda \sigma } (n^\tau w_{\tau \sigma})$.

Further we analyze the most general form of the action $\cS_{\rm g}$ for our purposes, where a part $k$ of some term uses $\Delta_\lambda$, and the remaining part $1 - k$ uses the $\Delta^{(s ) }_\lambda$ derivative form, which still results in the factor $( \oDelta \Delta )^{- 1}$ in the propagator without pole doubling. We find that it is precisely for $k = 1$ that the considered principal value type prescription actually works, and a smooth approach to the continuum limit (small quasi-momenta $\bp$) is ensured. Then this value $k = 1$ is used. The required gauge-fixing term for the considered principal value type prescription and its finiteness are analyzed; it is found that the ghost contribution tends to zero at $\varepsilon \to 0$. If a continuum diagram (or some of its structures) has no UV divergences, then this means that its discrete analogue is dominated by metric fields with small site-to-site variations or small quasi-momenta, and we can consider the leading order in the site-to-site variations of the metric and use the effective propagator $G^{\rm eff} (n, n)$ obtained from $\cG (n, n)$ by equating $- \oDelta \Delta$ to $\Delta^{(s) 2}$ in the leading order in finite differences, except for $- \oDelta \Delta$ appearing in the denominator. This certainly looks considerably less bulky than the full $\cG (n, n)$. With this propagator, one can perform all the machinery to form the gauge-fixing term for the considered principal value type prescription and find the ghost contribution. Another relatively simple expression for the propagator is obtained if we neglect $\varepsilon^2$ and limit ourselves to spatial indices (the temporal index leads to smallness at $\varepsilon \to 0$). The discussion is illustrated by electromagnetic (Yang-Mills) analogs, where the formulas are much simpler.

The paper is organized as follows. In Section \ref{ng-gauge} we consider the general form of the action $\cS_{\rm g}$ with $\Delta^{(s ) }_\lambda$ and $\Delta_\lambda$, characterized by the parameter $k$, and analyze the analytic continuation $\cG ( n, n )$ of the propagator $\cG ( n, \on )$ for the gauge-fixing term $ - \frac{1}{4} \sum_{\rm sites} (n^\mu w_{\mu \lambda}) \rlambda^{\lambda \sigma } (n^\tau w_{\tau \sigma})$ for real $n = \on$ to complex $n$. Prior to this, in Subsection \ref{nA-elmag}, the situation is illustrated by the electromagnetic case and the temporal gauge, where an action is considered with similarly introduced $\Delta^{(s ) }_\lambda$ and $\Delta_\lambda$, as well as a term fixing $nA$, and a similarly continued propagator $\cD ( n, n )$ is found. In Subsection \ref{ng-grav}, gravity itself is considered and the optimal value of the parameter $k$, equal to 1, is found. In Subsection \ref{detail-pole}, we introduce some spacelike length $b_{\rm s} \neq 1$ and some timelike length $b_{\rm t} \neq 1$. The fact is discussed that for sufficiently large $b_{\rm t} b_{\rm s}^{- 1}$ ($3 b_{\rm t}^2 b_{\rm s}^{- 2} > 1$) there exist spatial quasi-momenta for which the poles of the same term in the propagator are located on both sides of the integration path $\Im p_0 = 0$, and the considered principal value type prescription does not work properly. Under the same conditions, another inconvenience arises, namely, some spatial quasi-momenta appear for which the graviton pole does not exist at a real $p_0$. Therefore, although here we assume that $b_{\rm s} = 1$ and $b_{\rm t} = 1$ outside this Subsection to avoid bulkiness, one should keep in mind the eventual transition to physical $b_{\rm s}$, $b_{\rm t}$. In particular, in our papers \cite{khat,khat2} we have $3 b_{\rm t}^2 b_{\rm s}^{- 2} < 1$. Section \ref{principal} discusses the principal value type prescription itself. In Subsection \ref{gauge-fixing} we consider the general form of the gauge-fixing term for the principal value type prescription. Subsection \ref{A=0finite} considers such a term and its finiteness in the case where we can work in the limit $\cA \to 0$, or in the leading order over finite differences and use the effective propagator $G^{\rm eff} (n, n)$. In Subsection \ref{finiteAnot0}, we analyze the gauge-fixing term for the principal value type prescription and its finiteness for the full propagator $\cG (n, n)$. Subsection \ref{em_finite} discusses electromagnetic illustration. In Section \ref{ghost}, we consider the ghost contribution for the gauge-fixing term for the principal value type prescription. In Subsection \ref{ghost-general} we analyze the general expression for the effective ghost factor introduced into the functional integral simultaneously with the addition of the gauge-fixing term of the considered form to the action. In Subsection \ref{A=0ghost}, we establish that the effective ghost action vanishes at $\varepsilon \to 0$ for the $G^{\rm eff} (n, n)$-based gauge fixing term. In Subsection \ref{Aneq0ghost}, we find that the effective ghost action vanishes at $\varepsilon \to 0$ for the gauge-fixing term based on the full propagator $\cG (n, n)$. In the analysis of Subsections \ref{A=0ghost}, \ref{Aneq0ghost}, we use some freedom in choosing non-leading orders over metric/field variations from site to site in a certain expression that cannot be captured by a continuum analogue but influence the computation of its functional determinant. This choice is formulated as an improvement of the finite-difference diffeomorphism formula for the metric (known in the leading order over finite differences from the continuum analogue) at the expense of non-leading corrections in Subsection \ref{diff-improve}. In Subsection \ref{(-g)^alpha*g} we introduce some non-Gaussianity (trilinear and higher order terms) into the gauge-fixing term for the principal value type prescription. The considered modification of the gauge-fixing term from bilinear in $w_{\lambda \mu} = g_{\lambda \mu} - g_{0 \lambda}^{(0 ) }$ is made by replacing $w_{\lambda \mu} \Rightarrow ( - g )^\alpha w_{\lambda \mu}$. (We use such a gauge-fixing term in our paper \cite{khat2}.) We find that the effective ghost contribution vanishes up to a simple power volume factor in the functional measure at $\varepsilon \to 0$, and we also find that new interaction vertices associated with this term give a vanishing contribution to the diagram technique in this limit. Subsection \ref{YM-ghost} discusses the Yang-Mills illustration for determining the ghost contribution. Then the Conclusion follows.

\section{Discrete action and Hamiltonian gauge (\texorpdfstring{$n^\lambda g_{\lambda \mu}$}{nλ gλμ}) fixing term}\label{ng-gauge}

\subsection{The case of electromagnetic field}\label{nA-elmag}

The discrete action takes the form
\begin{eqnarray}                                                           %13
& & S_{\rm em} [ A ] = - \frac{1}{4} \sum_{\rm sites} \left[ \left( \Delta^{(s )\lambda} A^\mu - \Delta^{(s )\mu} A^\lambda \right) \left( \Delta^{(s )}_\lambda A_\mu - \Delta^{(s )}_\mu A_\lambda \right) \right], ~ \Delta^{(s )}_\lambda = \frac{ T_\lambda - \oT_\lambda }{2} .
\end{eqnarray}

\noindent Here we assume the metric $g_{\lambda \mu} = \eta_{\lambda \mu} = {\rm diag} (-1, 1, 1, 1)$, $T_\lambda f(\dots , x^\lambda , \dots ) = f(\dots , x^\lambda + 1 , \dots )$ (shift operator), $f$ is a function. Overlining means Hermitian conjugation. Although discreteness violates gauge symmetry, for small variations of the field from site to site there are degrees of freedom close to the gauge degrees of freedom of continuum theory. To exclude from the functional integral a set of physically almost equivalent configurations of infinite measure, we introduce into this integral an averaging factor equivalent to adding to the action a term fixing the gauge. Such a term for averaging over the temporal gauge would be proportional to $(\rnu^\lambda A_\lambda )^2$, where $\rnu^\lambda = (1, 0, 0, 0)$. To "soften" the gauge singularities at $p_0 = \rnu p = 0$, it is proposed in Ref \cite{Ste} to replace $\rnu^\lambda$ here by $n^\lambda$, where $n^\lambda$ is some differential operator infinitely close to $\rnu^\lambda$ that is neither Hermitian nor anti-Hermitian. We can add a discrete version of such a gauge-fixing term and a source term to the discrete action and, by varying it with respect to $A$, find the propagator,
\begin{eqnarray}\label{Sem[AJ]}                                            %14
& & S_{\rm em}^\prime [ A, J ] = S_{\rm em} - \sum_{\rm sites} \left[ J^\lambda A_\lambda + \frac{\rlambda}{2} \left( n^\lambda A_\lambda \right)^2 \right] , \nonumber \\ & & n^\lambda = \rnu^\lambda - \varepsilon \frac{ \Delta^{(s ) \lambda}_\perp }{ \Delta^{(s ) 2}_\perp } , \quad \Delta^{(s ) \lambda}_\perp = \Delta^{(s ) \lambda} - \frac{ \rnu^\lambda }{ \rnu^2 } (\rnu \Delta^{(s )} ) , \quad \rlambda^{- 1} \stackrel{\rm def}{=} \ralpha , \quad A_\lambda = D_{\lambda \mu} ( n, \on ) J^\mu .
\end{eqnarray}

\noindent The symmetrized finite-difference derivative $\Delta^{(s )}_\lambda$ is anti-Hermitian, as is the continuum derivative. Therefore, the calculation of the discrete propagator $D_{\lambda \mu} ( n, \on )$ is the same as the calculation of the continuum one, and the result should follow from the continuum one \cite{Ste} by replacing the continuum derivative $\partial_\lambda$ with the discrete one $\Delta^{(s ) }_\lambda$.

However, we are interested in the (non-Hermitian) operator $D_{\lambda \mu} ( n, n )$ (which, when summed with the Hermitian conjugate $D_{\lambda \mu} ( \on, \on )$, will yield a Hermitian one). In the momentum representation, this can be seen as an analytical continuation from $n = \rnu$ (the usual "hard" synchronous gauge) to complex $n$, or as the result of a formal replacement $\on \Rightarrow n$ in $D_{\lambda \mu} ( n, \on )$. So, on the one hand, it looks simpler than $D_{\lambda \mu} ( n, \on )$.

On the other hand, we consider some complication of the action at the level of non-leading orders over field variations from site to site. Due to the properties of $\Delta^{(s )}_\lambda = i \sin p_\lambda$, the quasi-momentum configurations related by the transformation $p_\lambda \to p_\lambda + \pi$, for any given component of any given loop quasi-momentum $p_\lambda$, contribute equally (up to a sign, perhaps) to a diagram based on the original discrete action $S_{\rm em}$ (to introduce an interaction, we can think of the Yang-Mills action, which has the same bilinear form, up to a trivial insertion of color structure). In particular, the propagator has a pole part $\propto ( \Delta^{(s ) 2} + i 0 )^{- 1} = ( \sin^2 p_0 - \sum^3_{\alpha = 1} \sin^2 p_\alpha + i 0 )^{- 1}$, and no matter how small the quasi-momentum $\bp$ is, poles are present both at small $p_0$ and at $p_0$ close to $\pm \pi$. Integrating over $\d p_0$ by calculating residues at the corresponding poles, we obtain a result approximately twice as large as that obtained by such an integration in the analogous continuum diagram.

The key to fixing the situation lies in changing the term in the action where two derivatives are contracted with each other, by replacing $\Delta^{(s )}_\lambda$ with $\Delta_\lambda = T_\lambda - 1$ there. Then the pole part becomes $\propto ( - \oDelta \Delta + i 0 )^{- 1} = ( 4 \sin^2 ( p_0 / 2 ) - \sum^3_{\alpha = 1} 4 \sin^2 ( p_\alpha / 2 ) + i 0 )^{- 1}$, and for small $\bp$ the poles are found only for small $p_0$ in the period $(- \pi , \pi ]$. Thus, the refined action is
\begin{eqnarray}                                                           %15
& & \cS_{\rm em} = \frac{1}{2} \sum_{\rm sites} \left[ \left( \Delta^{(s ) \lambda} A^\mu \right) \left( \Delta^{(s )}_\mu A_\lambda \right) - \left( \vphantom{\Delta^{(s ) \lambda}} \Delta^\lambda A^\mu \right) \left( \vphantom{\Delta^{(s ) \lambda}} \Delta_\lambda A_\mu \right) \right] .
\end{eqnarray}

\noindent As discussed above, we add the gauge-fixing term written in (\ref{Sem[AJ]}), but $n$ is formally treated as real in the momentum representation (Hermitian), and after evaluation it extends to complex $n$. The result reads
\begin{eqnarray}\label{cD(nn)}                                             %16
& & \hspace{-5mm} \cD_{\lambda \mu} ( n, n ) = \frac{- 1}{ \oDelta \Delta } \left[ \eta_{\lambda \mu} - \frac{ - ( n^2 + \ralpha \oDelta \Delta ) \Delta^{(s )}_\lambda \Delta^{(s )}_\mu + ( n \Delta^{(s )} ) ( \Delta^{(s )}_\lambda n_\mu + \Delta^{(s )}_\mu n_\lambda ) + \cA n_\lambda n_\mu }{ ( n \Delta^{(s )} )^2 + \cA ( n^2 + \ralpha \oDelta \Delta ) } \right] , \nonumber \\ & & \cA \stackrel{\rm def}{=} - \oDelta \Delta - \Delta^{(s ) 2} .
\end{eqnarray}

\noindent To denote the refined action and the corresponding propagator-related quantities, we use symbols marked with a check mark at the top.

In the following we want to pass to the limit $\ralpha \to 0$, $\varepsilon \to 0$. Both of these parameters are responsible for the typical value of $\rnu^\lambda A_\lambda$ through the gauge-fixing term: the effect of $\ralpha$ is $O( 1 / \sqrt{ | \rlambda | } ) = O( \sqrt{ | \ralpha | })$, the effect of $\varepsilon $ is $O ( \varepsilon )$. The question is the relationship between these effects. If $\sqrt{ | \ralpha | }$ is of higher order than $\varepsilon$, this means that at first the gauge-fixing factor actually becomes the $\delta$-function factor $\prod_{\rm sites} \delta ( \rnu^\lambda A_\lambda )$. Here we should continue $\rnu$ to complex $n$ or $\on$, but the $\delta$-function of a complex argument is not defined. Thus, $\ralpha$ cannot be of higher order than $\varepsilon^2$. Therefore, to minimize the impact of a nonzero value of $\ralpha$, we should take minimally $\ralpha = O( \varepsilon^2 )$.

\subsection{The case of gravity itself}\label{ng-grav}

The discrete action takes the form
\begin{eqnarray}\label{Sg}                                                 %17
& & S_{\rm g} [ g ] = \frac{1}{ 8 } \sum_{\rm sites} g^{\lambda \rho} g^{\mu \sigma} g^{\nu \tau} \left[ 2 \left( \Delta^{(s )}_\lambda g_{\mu \nu} \right) \left( \Delta^{(s )}_\tau g_{\sigma \rho} \right) - \left( \Delta^{(s )}_\nu g_{\mu \lambda} \right) \left( \Delta^{(s )}_\tau g_{\sigma \rho} \right) \right. \nonumber \\ & & \left. - 2 \left( \Delta^{(s )}_\lambda g_{\mu \rho} \right) \left( \Delta^{(s )}_\sigma g_{\nu \tau} \right) + \left( \Delta^{(s )}_\nu g_{\lambda \rho} \right) \left( \Delta^{(s )}_\tau g_{\mu \sigma} \right) \right] \sqrt{- g} .
\end{eqnarray}

\noindent This can be combined with the "soft" gauge-fixing term and a source term,
\begin{eqnarray}\label{S_g'}                                               %18
& & S_{\rm g}^\prime [ g , J ] = S_{\rm g} - \sum_{\rm sites} \left[ J^{\lambda \mu} w_{\lambda \mu} + \frac{1}{4} (n^\mu w_{\mu \lambda}) \rlambda^{\lambda \sigma } (n^\tau w_{\tau \sigma}) \right] , \quad g_{\lambda \mu} = \eta_{\lambda \mu} + w_{\lambda \mu} , \nonumber \\ & & \eta^{\lambda \mu } = {\rm diag} (-1, 1, 1, 1) , ~ ( \| \rlambda^{\lambda \mu} \|^{- 1} )_{\sigma \tau} \stackrel{\rm def}{=} \ralpha_{\sigma \tau} , ~ w_{\lambda \mu} = G_{\lambda \mu \sigma \tau} ( n, \on ) J^{\sigma \tau} .
\end{eqnarray}

It would be physically natural that if a continuum diagram converges, then for ordinary, non-Planckian external momenta, or for distances much larger than the typical edge length scale that plays the role of a lattice spacing, it should be reproduced with high accuracy by its discrete version. However, as in the electromagnetic case, the discrete diagram has additional contributions compared to its continuous counterpart, in particular additional poles that must be taken into account when integrating over $\d p_0$. To deal with this problem, we refine the terms in the action where two derivatives are contracted with each other by replacing $\Delta^{(s )}_\lambda$ with $\Delta_\lambda = T_\lambda - 1$ there. There are two such terms in $S_{\rm g}$ (\ref{Sg}). One of them, the second term in square brackets, controls the dynamics of the tensor structure of $w_{\lambda \mu}$ and must be refined in any case if we want to change the denominator of the propagator. The other, fourth term in square brackets, governs the dynamics of only a scalar part of $w_{\lambda \mu}$ -- ${\rm tr} w$ and a priori does not require its complete replacement with an analogue with $\Delta_\lambda$. So first we replace its $k$ part from the total number 1, leaving the other part $1 - k$ unchanged. Thus, the refined action is
\begin{eqnarray}\label{cSg}                                                %19
& & \cS_{\rm g} = \frac{1}{ 8 } \sum_{\rm sites} g^{\lambda \rho} g^{\mu \sigma} g^{\nu \tau} \left[ 2 \left( \Delta^{(s )}_\lambda g_{\mu \nu} \right) \left( \Delta^{(s )}_\tau g_{\sigma \rho} \right) - \left( \Delta_\nu g_{\mu \lambda} \right) \left( \Delta_\tau g_{\sigma \rho} \right) - 2 \left( \Delta^{(s )}_\lambda g_{\mu \rho} \right) \left( \Delta^{(s )}_\sigma g_{\nu \tau} \right) \right. \nonumber \\ & & \left. + k \left( \Delta_\nu g_{\lambda \rho} \right) \left( \Delta_\tau g_{\mu \sigma} \right) + ( 1 - k ) \left( \Delta^{(s )}_\nu g_{\lambda \rho} \right) \left( \Delta^{(s )}_\tau g_{\mu \sigma} \right) \right] \sqrt{- g} .
\end{eqnarray}

A priori, there are some reference values of $k$. For example, one can choose $k$ such that the considered combination of squares of finite-difference derivatives best approximates the square of the continuum derivative at small quasi-momenta. In the momentum representation,
\begin{equation}                                                           %20
k \oDelta_1 \Delta_1 + ( 1 - k ) \overline{ \Delta^{(s )}_1} \Delta^{(s )}_1 = p_1^2 + \left( \frac{k}{4} - \frac{1}{3} \right) p_1^4 + O ( p_1^6 ) ,
\end{equation}

\noindent and the maximal relative accuracy $O ( p_1^4 )$ instead of the typical $O ( p_1^2 )$ is achieved with $k = 4 / 3$.

It is convenient to first calculate the propagator for the simpler case of $S_{\rm g}$ (\ref{Sg}), when all finite differences in it have the form $\Delta^{(s) }$, and then trace the changes caused by the above replacement of some $\Delta^{(s) }$s by $\Delta$.
\begin{eqnarray}\label{ds2g=J+nf+ds2g}                                     %21
& & \Delta^{(s) 2} w_{\lambda \mu} = 4 J_{\lambda \mu} + \on_\mu F_\lambda + \on_\lambda F_\mu + \Delta^{(s) }_\mu \Delta^{(s) \nu} w_{\lambda \nu} + \Delta^{(s) }_\lambda \Delta^{(s) \nu} w_{\mu \nu} - \eta^{\nu \rho} \Delta^{(s) }_\lambda \Delta^{(s) }_\mu w_{\nu \rho} \nonumber \\ & & \hspace{-10mm} - \eta_{\lambda \mu} \Delta^{(s) \nu} \Delta^{(s) \rho} w_{\nu \rho} + \eta_{\lambda \mu} \eta^{\nu \rho} \Delta^{(s) 2} w_{\nu \rho} , ~ \text{where} ~ F^\lambda \stackrel{\rm def }{ = } \rlambda^{\lambda \rho} n^\nu w_{\rho \nu} , ~ n^\mu w_{\lambda \mu} = \ralpha_{\lambda \mu} F^\mu \stackrel{\rm def }{ = }  f_\lambda .
\end{eqnarray}

\noindent Here $F_\lambda$ can be found straightaway by applying the operator $\Delta^{(s) \mu}$ to both sides of (\ref{ds2g=J+nf+ds2g}),
\begin{eqnarray}                                                           %22
& & \on_\mu \Delta^{(s) \mu} F_\lambda + \on_\lambda \Delta^{(s) \mu} F_\mu = - 4 \Delta^{(s) \mu} J_{\lambda \mu} , \nonumber \\ & & F_\lambda = - 4 (\on \Delta^{(s)} )^{- 1} \Delta^{(s) \mu} J_{\lambda \mu} + 2 \on_\lambda (\on \Delta^{(s)} )^{- 2} \Delta^{(s) \mu} \Delta^{(s) \nu} J_{\mu \nu} .
\end{eqnarray}

\noindent Knowing $F_\lambda$, we find $h \stackrel{\rm def }{ = } \Delta^{(s) \lambda} \Delta^{(s) \mu} w_{\lambda \mu} - \eta^{\lambda \mu} \Delta^{(s) 2} w_{\lambda \mu}$ by taking the trace of (\ref{ds2g=J+nf+ds2g}),
\begin{equation}                                                           %23
h = 2 \eta^{\lambda \mu} J_{\lambda \mu} + \on^\lambda F_\lambda .
\end{equation}

\noindent Contracting (\ref{ds2g=J+nf+ds2g}) with $n^\mu$, we find $r_\lambda \stackrel{\rm def }{ = } \Delta^{(s) \mu} w_{\lambda \mu} - \eta^{\mu \nu} \Delta^{(s) }_\lambda w_{\mu \nu}$ in terms of the found $F_\lambda$, $f_\lambda = \ralpha_{\lambda \mu} F^\mu$ and $h$,
\begin{eqnarray}                                                           %24
( n \Delta^{(s) } ) r_\lambda & = & - 4 n^\mu J_{\lambda \mu} + \Delta^{(s) 2} f_\lambda - \Delta^{(s) }_\lambda \Delta^{(s) \mu} f_\mu - ( n \on ) F_\lambda - \on_\lambda n^\mu F_\mu + n_\lambda h .
\end{eqnarray}

\noindent Then we can contract this $r_\lambda$ with $n^\lambda$ and, also knowing $F_\lambda$, find $\eta^{\lambda \mu} w_{\lambda \mu}$. Substituting the latter back into $r_\lambda$ gives $\Delta^{(s) \mu} w_{\lambda \mu}$. Finally, we can substitute the found terms containing $w_{\lambda \mu}$ into the RHS of (\ref{ds2g=J+nf+ds2g}) and find the propagator, $w_{\lambda \mu} = G_{\lambda \mu \sigma \tau} J^{\sigma \tau}$,
\begin{eqnarray}\label{G}                                                  %25
& & \hspace{-5mm} \frac{1}{2} G_{\lambda \mu \sigma \tau} ( n, \on ) = - \frac{i}{2} \langle w_{\lambda \mu} w_{\sigma \tau} \rangle = \frac{1}{\Delta^{(s) 2} } [ L_{\lambda \sigma} ( n, \on ) L_{\mu \tau} ( n, \on ) + L_{\mu \sigma} ( n, \on ) L_{\lambda \tau} ( n, \on ) \nonumber \\ & & \hspace{-5mm} - L_{\lambda \mu} ( n, n ) L_{\sigma \tau} ( \on, \on ) ] - \frac{ ( \ralpha_{\lambda \sigma} \Delta^{(s) }_\tau + \ralpha_{\lambda \tau} \Delta^{(s) }_\sigma) \Delta^{(s) }_\mu + ( \ralpha_{\mu \sigma} \Delta^{(s) }_\tau + \ralpha_{\mu \tau} \Delta^{(s) }_\sigma) \Delta^{(s) }_\lambda }{(n \Delta^{(s) })(\on \Delta^{(s) })} \nonumber \\ & & \hspace{-5mm} + \Delta^{(s) }_\lambda \Delta^{(s) }_\mu \frac{ n^\nu \ralpha_{\nu \sigma} \Delta^{(s) }_\tau + n^\nu \ralpha_{\nu \tau} \Delta^{(s) }_\sigma }{(n \Delta^{(s) })^2 (\on \Delta^{(s) } ) } + \frac{ \ralpha_{\lambda \nu} \on^\nu \Delta^{(s) }_\mu + \ralpha_{\mu \nu} \on^\nu \Delta^{(s) }_\lambda }{(n \Delta^{(s) }) (\on \Delta^{(s) })^2 } \hspace{0mm} \Delta^{(s) }_\sigma \Delta^{(s) }_\tau - \frac{ n^\nu \ralpha_{\nu \rho} \on^\rho }{(n \Delta^{(s) })^2 (\on \Delta^{(s) })^2 } \nonumber \\ & & \hspace{-5mm} \cdot \Delta^{(s) }_\lambda \Delta^{(s) }_\mu \Delta^{(s) }_\sigma \Delta^{(s) }_\tau ; ~
L_{\lambda \mu} ( m, n ) \stackrel{\rm def }{=} \eta_{\lambda \mu} - \Delta^{(s) }_\lambda \frac{m_\mu }{ m \Delta^{(s) } } - \frac{n_\lambda }{ n \Delta^{(s) } } \Delta^{(s) }_\mu + \frac{(m n) \Delta^{(s) }_\lambda \Delta^{(s) }_\mu}{(m \Delta^{(s) } )( n \Delta^{(s) } )} .
\end{eqnarray}

Next we will consider the action of interest to us $\cS_{\rm g}$ (\ref{cSg}) with both $\Delta$ and $\Delta^{(s) }$ used. We denote the corresponding propagator as $\cG_{\lambda \mu \sigma \tau}$, in contrast to $G_{\lambda \mu \sigma \tau}$ (equation (\ref{G})) considered above for the action containing only $\Delta^{(s) }$. In what follows, we are interested in the (non-Hermitian) operator $\cG_{\lambda \mu \sigma \tau} ( n, n )$ (the summation of which with the Hermitian conjugate operator $\cG_{\lambda \mu \sigma \tau} ( \on, \on )$ leads to a Hermitian one). In the momentum representation this can be viewed as an analytical continuation from the usual "hard" synchronous gauge at $n = \rnu$ to complex $n$ and is obtained by formally substituting $\on = n$ into the formulas. This simplification is favorable for the present rather balky calculations. Equation (\ref{ds2g=J+nf+ds2g}) is modified,
\begin{eqnarray}\label{d2g=J+nf+d2g+ds2g}                                  %26
& & \hspace{-10mm} - \oDelta \Delta w_{\lambda \mu} = 4 J_{\lambda \mu} + n_\mu F_\lambda + n_\lambda F_\mu + \Delta^{(s) }_\mu \Delta^{(s) \nu} w_{\lambda \nu} + \Delta^{(s) }_\lambda \Delta^{(s) \nu} w_{\mu \nu} - \eta^{\nu \rho} \Delta^{(s) }_\lambda \Delta^{(s) }_\mu w_{\nu \rho} \nonumber \\ & & \hspace{-10mm} - \eta_{\lambda \mu} \Delta^{(s) \nu} \Delta^{(s) \rho} w_{\nu \rho} + \eta_{\lambda \mu} \eta^{\nu \rho} \left[ - k \oDelta \Delta + (1 - k) \Delta^{(s) 2} \right] w_{\nu \rho} = 4 J_{\lambda \mu} + n_\mu F_\lambda + n_\lambda F_\mu \nonumber \\ & & \hspace{-15mm} + \Delta^{(s )}_\mu r_\lambda + \Delta^{(s )}_\lambda r_\mu + \eta^{\nu \rho} \Delta^{(s) }_\lambda \Delta^{(s) }_\mu w_{\nu \rho} + \eta_{\lambda \mu } ( - h + k \cA \eta^{\nu \rho} w_{\nu \rho} ) , \quad w_{\lambda \mu} \stackrel{\rm def}{=} \cG_{\lambda \mu \sigma \tau} ( n, n ) J^{\sigma \tau}.
\end{eqnarray}

\noindent As in the case of eq. (\ref{ds2g=J+nf+ds2g}), we apply the operations $\Delta^{(s) \mu} ( \cdot )$, $\eta^{\lambda \mu}( \cdot )$, $n^\mu ( \cdot )$ to (\ref{d2g=J+nf+d2g+ds2g}), $n^\lambda ( \cdot )$ to the found $r_\lambda$ in order to find $F_\lambda$, $h$, $r_\lambda$, $\eta^{\lambda \mu} w_{\lambda \mu}$, respectively; but now on the RHS there are terms $O ( \cA )$, linear functionals of these functions. Then it is more convenient to first express $F_\lambda$, $r_\lambda$ through $\eta^{\lambda \mu} w_{\lambda \mu}$, $h$,
\begin{eqnarray}\label{Dw}                                                 %27
& & \hspace{-5mm} r^\lambda = \left\{ \left[ \eta_{\sigma \tau} + \cA ( n \Delta^{(s) } )^{- 2} \left( n^2 \eta_{\sigma \tau} + (\oDelta \Delta ) \ralpha_{\sigma \tau} + \Delta^{(s) }_\sigma \Delta^{(s) \kappa} \ralpha_{\kappa \tau} \right) \right]^{- 1} \right\}^{\lambda \mu} ( n \Delta^{(s) } )^{- 1} \left[ \vphantom{\left( \frac{1}{2} \right)} - 4 n^\nu J_{\mu \nu} \right. \nonumber \\ & & \left. + 2 n_\mu \eta^{\nu \rho} J_{\nu \rho} - \left( n^2 \eta_{\mu \nu} + (\oDelta \Delta) \ralpha_{\mu \nu} + \Delta^{(s) }_\mu \Delta^{(s) \rho} \ralpha_{\rho \nu} \right) K^\nu + \left( k - \frac{1}{2} \right) \cA n_\mu \eta^{\nu \rho} w_{\nu \rho} \right] , \nonumber \\
& & K_\lambda \stackrel{\rm def }{=} - 4 ( n \Delta^{(s) } )^{- 1} \Delta^{(s) \mu} J_{\lambda \mu} + 2 n_\lambda ( n \Delta^{(s) } )^{- 2} \Delta^{(s) \mu} \Delta^{(s) \nu} J_{\mu \nu} \nonumber \\ & & + ( k - 1 ) \cA \left[ - ( n \Delta^{(s) } )^{- 1} \Delta^{(s) }_\lambda + \frac{1}{2} ( n \Delta^{(s) } )^{- 2} \Delta^{(s) 2} n_\lambda \right] \eta^{\mu \nu} w_{\mu \nu} - \frac{1}{2} \cA n_\lambda ( n \Delta^{(s) } )^{- 2} h
\end{eqnarray}

\noindent and
\begin{eqnarray}\label{F}                                                  %28
& & \hspace{-7mm} F^\lambda = \cA ( n \Delta^{(s) } )^{- 1} r^\lambda + K^\lambda = \left\{ \left[ \eta_{\sigma \tau} + \cA ( n \Delta^{(s) } )^{- 2} \left( n^2 \eta_{\sigma \tau} + (\oDelta \Delta ) \ralpha_{\sigma \tau} + \Delta^{(s) }_\sigma \Delta^{(s) \kappa} \ralpha_{\kappa \tau} \right) \right]^{- 1} \right\}^{\lambda \mu} \nonumber \\ & & \cdot \left[ \vphantom{\left( \frac{1}{2} \right)} \cA ( n \Delta^{(s) } )^{- 2} \left( - 4 n^\nu J_{\mu \nu} + 2 n_\mu \eta^{\nu \rho} J_{\nu \rho} \right) + K_\mu + \left( k - \frac{1}{2} \right) \cA^2 ( n \Delta^{(s) } )^{- 2} n_\mu \eta^{\nu \rho} w_{\nu \rho} \right] ,
\end{eqnarray}

\noindent where $\eta^{\lambda \mu} w_{\lambda \mu}$, $h$ are subject to a system of two equations,
\begin{eqnarray}                                                        %29,30
& & \label{trw} \hspace{-7mm} \left\{ 1 + \left( 2 k - \frac{1}{2} \right) \cA ( n \Delta^{(s) } )^{- 2} n^2 - \frac{k - 1}{2} \cA ( n \Delta^{(s) } )^{- 4} \Delta^{(s) 2} n^4 - \cA \left\{ \left[ 2 + \cA ( n \Delta^{(s) } )^{- 2} n^2 \right] \Delta^{(s) \nu} \right. \right. \nonumber \\ & & \hspace{-8mm} \left. \left. + ( n \Delta^{(s) } )^{- 1} (\oDelta \Delta ) n^\nu \right\} \rbeta_{\nu \rho} \left\{ (1 - k) ( n \Delta^{(s) } )^{- 2} \Delta^{(s) \rho} + \left[ \left( k - \frac{1}{2} \right) \cA + \frac{k - 1}{2} \Delta^{(s) 2} \right] ( n \Delta^{(s) } )^{- 3} n^\rho \right. \right. \nonumber \\ & & \left. \left. \hspace{-7mm} \vphantom{\frac{1}{2}} \right\} \right\} \eta^{\lambda \mu} w_{\lambda \mu} + \frac{1}{2} \cA \left\{ ( n \Delta^{(s) } )^{- 4} n^4 + \left\{ \left[ 2 + \cA ( n \Delta^{(s) } )^{- 2} n^2 \right] \Delta^{(s) \nu} + ( n \Delta^{(s) } )^{- 1} (\oDelta \Delta ) n^\nu \right\} \rbeta_{\nu \rho} n^\rho \right. \nonumber \\ & & \left. \hspace{-7mm} \cdot ( n \Delta^{(s) } )^{- 3} \right\} h = 4 ( n \Delta^{(s) } )^{- 2} n^\lambda n^\mu J_{\lambda \mu} - 2 ( n \Delta^{(s) } )^{- 2} n^2 \eta^{\lambda \mu} J_{\lambda \mu} - 4 ( n \Delta^{(s) } )^{- 3} n^2 n^\lambda \Delta^{(s) \mu} J_{\lambda \mu} \nonumber \\ & & \hspace{-7mm} + 2 ( n \Delta^{(s) } )^{- 4} n^4 \Delta^{(s) \lambda} \Delta^{(s) \mu} J_{\lambda \mu} + \left\{ \left[ 2 + \cA ( n \Delta^{(s) } )^{- 2} n^2 \right] \Delta^{(s) \nu} + ( n \Delta^{(s) } )^{- 1} (\oDelta \Delta ) n^\nu \right\} \rbeta_{\nu \rho} \nonumber \\ & & \hspace{-7mm} \cdot \! \left[ \cA ( n \Delta^{(s) } )^{- 3} \left( - 4 \eta^{\rho \lambda} n^\mu + 2 n^\rho \eta^{\lambda \mu} \right) - 4 ( n \Delta^{(s) } )^{- 2} \eta^{\rho \lambda} \Delta^{(s) \mu} + 2 ( n \Delta^{(s) } )^{- 3} n^\rho \Delta^{(s) \lambda} \Delta^{(s) \mu} \right] J_{\lambda \mu} , \\
& & \label{DDw} \hspace{-7mm} \left\{ \left( \frac{1}{2} - k \right) \cA + \frac{1 - k}{2} \cA ( n \Delta^{(s) } )^{- 2} \Delta^{(s) 2} n^2 - \cA^2 \Delta^{(s) \nu} \rbeta_{\nu \rho} \left\{ \vphantom{\frac{1}{2}} (1 - k) ( n \Delta^{(s) } )^{- 2} \Delta^{(s) \rho} \right. \right. \nonumber \\ & & \left. \left. \hspace{-7mm} + \left[ \left( k - \frac{1}{2} \right) \cA + \frac{k - 1}{2} \Delta^{(s) 2} \right] ( n \Delta^{(s) } )^{- 3} n^\rho \right\} \right\} \eta^{\lambda \mu} w_{\lambda \mu} + \left[ 1 + \frac{1}{2} \cA ( n \Delta^{(s) } )^{- 2} n^2 \right. \nonumber \\ & & \left. \hspace{-7mm} + \frac{1}{2} \cA^2 \Delta^{(s) \nu} \rbeta_{\nu \rho} n^\rho ( n \Delta^{(s) } )^{- 3} \right] h = 2 \eta^{\lambda \mu} J_{\lambda \mu} - 4 ( n \Delta^{(s) } )^{- 1} n^\lambda \Delta^{(s) \mu} J_{\lambda \mu} \nonumber \\ & & \hspace{-7mm} + 2 ( n \Delta^{(s) } )^{- 2} n^2 \Delta^{(s) \lambda} \Delta^{(s) \mu} J_{\lambda \mu} + \cA \Delta^{(s) \nu} \rbeta_{\nu \rho} \left[ \cA ( n \Delta^{(s) } )^{- 3} \left( - 4 \eta^{\rho \lambda} n^\mu + 2 n^\rho \eta^{\lambda \mu} \right) \right. \nonumber \\ & & \left. \hspace{-7mm} - 4 ( n \Delta^{(s) } )^{- 2} \eta^{\rho \lambda} \Delta^{(s) \mu} + 2 ( n \Delta^{(s) } )^{- 3} n^\rho \Delta^{(s) \lambda} \Delta^{(s) \mu} \right] J_{\lambda \mu} .
\end{eqnarray}

\noindent Here the dependence on $\ralpha$ is realized through
\begin{eqnarray}\label{beta=f(alpha)}                                      %31
& & \hspace{-15mm} \rbeta_\lambda{}^\mu \stackrel{\rm def }{ = } \ralpha_{\lambda \nu} \left\{ \left[ \eta_{\sigma \tau} + \cA ( n \Delta^{(s) } )^{- 2} \left( n^2 \eta_{\sigma \tau} + (\oDelta \Delta ) \ralpha_{\sigma \tau} + \Delta^{(s) }_\sigma \Delta^{(s) \kappa} \ralpha_{\kappa \tau} \right) \right]^{- 1} \right\}^{\nu \mu} , \nonumber \\ & & \hspace{-15mm} \ralpha_\lambda{}^\mu = \left[ 1 + \cA ( n \Delta^{(s) } )^{- 2} n^2 \right] \rbeta_{\lambda \nu} \left\{ \left[ \eta_{\sigma \tau} - \cA ( n \Delta^{(s) } )^{- 2} \left( (\oDelta \Delta ) \rbeta_{\sigma \tau} + \Delta^{(s) }_\sigma \Delta^{(s) \kappa} \rbeta_{\kappa \tau} \right) \right]^{- 1} \right\}^{\nu \mu} ;
\end{eqnarray}

\noindent in particular,
\begin{eqnarray}\label{1/[1+A(1+dda)]}                                     %32
& & \left\{ \left[ \eta_{\sigma \tau} + \cA ( n \Delta^{(s) } )^{- 2} \left( n^2 \eta_{\sigma \tau} + (\oDelta \Delta ) \ralpha_{\sigma \tau} + \Delta^{(s) }_\sigma \Delta^{(s) \kappa} \ralpha_{\kappa \tau} \right) \right]^{- 1} \right\}^{\lambda \mu} \nonumber \\ & & = \left[ 1 + \cA ( n \Delta^{(s) } )^{- 2} n^2 \right]^{- 1} \left[ \eta^{\lambda \mu} - \cA ( n \Delta^{(s) } )^{- 2} \left( (\oDelta \Delta ) \rbeta^{\lambda \mu} + \Delta^{(s) \lambda} \Delta^{(s) }_\nu \rbeta^{\nu \mu} \right) \right] .
\end{eqnarray}

\noindent The determinant of the system (\ref{trw},\ref{DDw}) takes the form
\begin{eqnarray}                                                           %33
& & \hspace{-5mm} \det{}_{(w,h)} = \left[ 1 + \cA ( n \Delta^{(s) } )^{- 2} n^2 \right]^2 + (k - 1) \cA \left\{ \frac{1}{2} ( n \Delta^{(s) } )^{- 4} n^2 \left[ - 3 ( \oDelta \Delta ) n^2 + 4 \left( ( n \Delta^{(s) } )^2 \right. \right. \right. \nonumber \\ & & \left. \left. \left. \hspace{-5mm} - \Delta^{(s) 2} n^2 \right) \right] + 2 ( n \Delta^{(s) } )^{- 2} \left[ 1 + \cA ( n \Delta^{(s) } )^{- 2} n^2 \right] \Delta^{(s) \lambda} \rbeta_{\lambda \mu} \Delta^{(s) \mu} + ( \oDelta \Delta ) ( n \Delta^{(s) } )^{- 3} \left[ 1 \vphantom{\frac{1}{2}} \right. \right. \nonumber \\ & & \left. \left. \hspace{-5mm} + \frac{1}{2} \cA ( n \Delta^{(s) } )^{- 2} n^2 \right] \left( \Delta^{(s) \lambda} \rbeta_{\lambda \mu} n^\mu + n^\mu \rbeta_{\mu \lambda} \Delta^{(s) \lambda} \right) - \frac{1}{2} ( \oDelta \Delta ) ( n \Delta^{(s) } )^{- 4} \Delta^{(s) 2} n^\lambda \rbeta_{\lambda \mu} n^\mu \right\} \nonumber \\ & & \hspace{-5mm} - \frac{1}{2} ( k - 1 ) \cA^3 ( \oDelta \Delta ) ( n \Delta^{(s) } )^{- 6} \left[ ( \Delta^{(s) \lambda} \rbeta_{\lambda \mu} \Delta^{(s) \mu} ) ( n^\nu \rbeta_{\nu \rho} n^\rho ) - ( \Delta^{(s) \lambda} \rbeta_{\lambda \rho} n^\rho ) ( n^\nu \rbeta_{\nu \mu} \Delta^{(s) \mu} ) \right] \nonumber \\ & & \hspace{-5mm} - \frac{1}{2} \left( k - \frac{1}{2} \right) \cA^2 \left[ 1 + \cA ( n \Delta^{(s) } )^{- 2} n^2 \right] ( n \Delta^{(s) } )^{- 4} ( \oDelta \Delta ) n^\lambda \rbeta_{\lambda \mu} n^\mu .
\end{eqnarray}

\noindent Taking into account the dependence on $\varepsilon$ in the linear approximation, we omit the terms $O ( \rbeta ) = O ( \varepsilon^2 )$ and write
\begin{eqnarray}                                                           %34
& & \hspace{-15mm} ( n \Delta^{(s) } )^4 \det{}_{(w,h)} = \left[ ( n \Delta^{(s) } )^2 + \cA n^2 \right]^2 - \frac{k - 1}{2} \cA n^2 \left\{ 3 \oDelta \Delta n^2 + 4 \left[ \Delta^{(s) 2 } n^2 - ( n \Delta^{(s) } )^2 \right] \right\} .
\end{eqnarray}

\noindent Here we set $n \Delta^{(s) } = \rnu \Delta^{(s) } - \varepsilon$, $n^2 \approx \rnu^2$, $\rnu^\lambda = (1, 0, 0, 0)$ and in the momentum representation we obtain that the zeroes of $\det{}_{(w,h)}$ and, consequently, the poles of the functions $w$, $h$ associated with the propagator are located for small $\bp$ at
\begin{eqnarray}                                                           %35
& & 2 \sin \frac{ p_0 }{ 2 } = - i \varepsilon \left[ 1 + O ( | \bp | ) \right] \nonumber \\ & & \hspace{-10mm} + \sqrt{ \sum_\alpha \left( 2 \sin^2 \frac{ p_\alpha }{ 2 } \right)^2 \left[ 1 + O ( | \bp | ) \right] + \sqrt{\frac{k - 1}{ 2 }} \left[ \sum_{\alpha,\beta} \left( 2 \sin^2 \frac{ p_\alpha }{ 2 } \right)^2 \sin^2 p_\beta \right]^{1 / 2} \hspace{-5mm} \left[ 1 + O ( | \bp | ) \right] } .
\end{eqnarray}

\begin{figure}[h]
	\centering
	\includegraphics[scale=1]{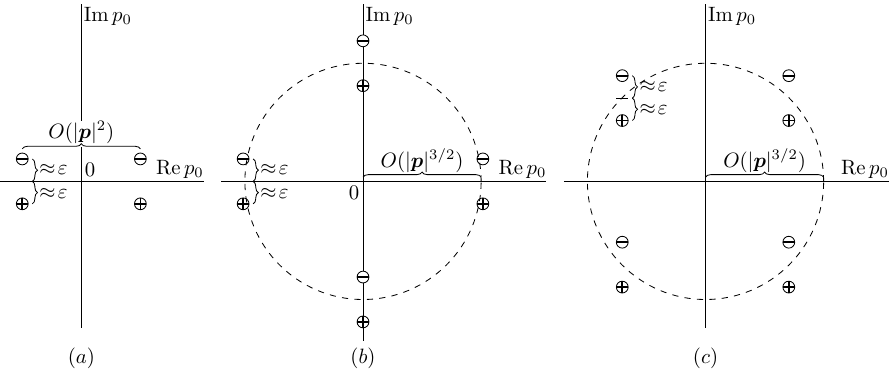}
	\caption{Location of nonphysical poles of the propagators $\oplus$ (of the quantities like $\cG (n, n)$) and $\ominus$ (of the quantities like $\cG (\on, \on)$) for $k = 1$ or electromagnetic case (a), $k > 1$ (b) and $k < 1$ (c).}
	\label{f1}
\end{figure}

\noindent The location of these poles in the complex plane of $p_0$ is shown in Fig.~\ref{f1} for different $k$.
It is seen that either for $k = 1$ or in the electromagnetic case these poles are initially at $\varepsilon = 0$ on the real axis. Then introducing an arbitrarily small $\varepsilon$ shifts each of these poles (of the two) to the upper or each to the lower half-plane, depending on the sign of $\varepsilon$. The propagator-related quantities $r^\lambda$ (\ref{Dw}), $F^\lambda$ (\ref{F}) can acquire the additional denominator $[ 1 + \cA ( n \Delta^{(s )} )^{- 2} n^2 ]$ (\ref{1/[1+A(1+dda)]}) with almost the same, up to $O ( \varepsilon^2 )$, positions of zeros with the same properties. The value $\cG (n, n)$ (or $\cG (\on, \on)$) required to form the principal value type propagator $[ \cG (n, n) + \cG (\on, \on) ]/2$ is defined through (\ref{d2g=J+nf+d2g+ds2g}) by a combination of $w$, $h$, $r^\lambda$, $F^\lambda$. These quantities have poles of this $\cG (n, n)$ except for the physical poles ($( \oDelta \Delta )^{- 1} $), i. e. nonphysical poles. Working with small quasi-momenta means approaching the continuum limit; as $| \bp |$ tends to zero, the two poles that differ in the sign of $\Re p_0 = O ( \bp^2 )$ smoothly merge, without crossing the integration path $\Im p_0 = 0$, into one pole $p_0 = - i \varepsilon$ (or $p_0 = i \varepsilon$), characteristic of the continuum theory.

On the contrary, for $k \neq 1$, the poles of the same term $\cG (n, n)$ (or $\cG (\on, \on)$) are located on both sides of the integration path $\Im p_0 = 0$ for almost all (for $\varepsilon \to 0$) spatial quasi-momenta $\bp$: $ | \bp | \gtrsim | \bp |_0 = O ( \varepsilon^{2 / 3} ) $, Fig.~\ref{f1}(b,c). Then the result of integration over $\d p_0$ is contributed by the residues at the poles on either side, is generally nonzero, and does not match the continuum result.

Thus, it is precisely $k = 1$ that allows the considered principal value type prescription to work and to ensure a smooth approach to the continuum limit (small quasi-momenta $\bp$). In what follows we take $k = 1$. The behaviour of the denominators responsible for the nonphysical poles in $\cG (n, n)$ is similar to $[ ( n \Delta^{(s )} )^2 + n^2 \cA ]^j$, where
\begin{eqnarray}\label{(nd)^2+n^2A^2}                                      %36
& & ( n \Delta^{(s )} )^2 + n^2 \cA = - ( \rnu \oDelta ) ( \rnu \Delta ) - 2 \varepsilon ( \rnu \Delta^{(s )} ) - \rnu^2 ( \oDelta_\perp \Delta_\perp + \Delta^{(s ) 2}_\perp ) + O ( \varepsilon^2 ) \nonumber \\ & & = - 4 \sin^2 \frac{ p_0 + i \varepsilon }{2} + 4 \sum_{\alpha = 1}^3 \sin^4 \frac{ p_\alpha }{2} + O ( \varepsilon^2 ) ,
\end{eqnarray}

\noindent and the complex conjugate of this expression for $\cG (\on, \on)$. Here $\rnu^\lambda = (1, 0, 0, 0)$.

For $k = 1$, it is more convenient to operate with the values $h - \cA \eta^{\lambda \mu} w_{\lambda \mu}$ and $\eta^{\lambda \mu} w_{\lambda \mu} + n^2 ( n \Delta^{(s) } )^{- 2} h$. In the expressions for these quantities, the numerators and denominator $Det$ can be reduced by a common factor $1 + \cA ( n \Delta^{(s) } )^{- 2} n^2$,
\begin{eqnarray}\label{h-Aw}                                         %37,38,39
& & \hspace{-5mm} ( h - \cA \eta^{\lambda \mu} w_{\lambda \mu} ) Det_0 = 2 [ 1 + \cA ( n \Delta^{(s) } )^{- 2} n^2 ] \eta^{\lambda \mu} J_{\lambda \mu} - 2 \cA ( n \Delta^{(s) } )^{- 2} n^\lambda n^\mu J_{\lambda \mu} \nonumber \\ & & \hspace{-5mm} + 2 ( n \Delta^{(s) } )^{- 2} n^2 \Delta^{(s) \lambda} \Delta^{(s) \mu} J_{\lambda \mu} - 4 ( n \Delta^{(s) } )^{- 1} n^\lambda \Delta^{(s) \mu} J_{\lambda \mu} + \cA ( n \Delta^{(s) } )^{- 1} (\oDelta \Delta ) n^\nu \rbeta_{\nu \rho} \nonumber \\ & & \hspace{-5mm} \cdot \left[ \cA ( n \Delta^{(s) } )^{- 3} \left( 2 \eta^{\rho \lambda} n^\mu - n^\rho \eta^{\lambda \mu} \right) + 2 ( n \Delta^{(s) } )^{- 2} \eta^{\rho \lambda} \Delta^{(s) \mu} - ( n \Delta^{(s) } )^{- 3} n^\rho \Delta^{(s) \lambda} \Delta^{(s) \mu} \right] J_{\lambda \mu} , \\ & & \hspace{-5mm} [ \eta^{\lambda \mu} w_{\lambda \mu} + n^2 ( n \Delta^{(s) } )^{- 2} h ] Det_0 = 4 ( n \Delta^{(s) } )^{- 4} \left[ n^2 \Delta^{(s) \lambda} - ( n \Delta^{(s) } ) n^\lambda \right] \left[ n^2 \Delta^{(s) \mu} \right. \nonumber \\ & & \left. \hspace{-5mm} - ( n \Delta^{(s) } ) n^\mu \right] J_{\lambda \mu} + \left\{ 2 [ 1 + \cA ( n \Delta^{(s) } )^{- 2} n^2 ] \Delta^{(s) \nu} + ( n \Delta^{(s) } )^{- 1} (\oDelta \Delta ) n^\nu \right\} \rbeta_{\nu \rho} \left[ \cA ( n \Delta^{(s) } )^{- 3} \right. \nonumber \\ & & \left. \hspace{-5mm} \cdot \left( - 4 \eta^{\rho \lambda} n^\mu + n^\rho \eta^{\lambda \mu} \right) - 4 ( n \Delta^{(s) } )^{- 2} \eta^{\rho \lambda} \Delta^{(s) \mu} + 2 ( n \Delta^{(s) } )^{- 3} n^\rho \Delta^{(s) \lambda} \Delta^{(s) \mu} \right] J_{\lambda \mu} \nonumber \\ & & \hspace{-5mm} + 2 \cA^2 \Delta^{(s) \nu} \rbeta_{\nu \rho} n^\rho ( n \Delta^{(s) } )^{- 5} n^\lambda n^\mu J_{\lambda \mu} + \cA \left[ 2 \Delta^{(s) \nu} + ( n \Delta^{(s) } )^{- 1} (\oDelta \Delta ) n^\nu \right] \rbeta_{\nu \rho} n^\rho \left[ 2 ( n \Delta^{(s) } )^{- 4} \right. \nonumber \\ & & \left. \hspace{-5mm} \cdot n^\lambda \Delta^{(s) \mu} - n^2 ( n \Delta^{(s) } )^{- 5} \Delta^{(s) \lambda} \Delta^{(s) \mu} \right] J_{\lambda \mu} - 2 \cA^2 ( n \Delta^{(s) } )^{- 4} (\oDelta \Delta ) \left( \Delta^{(s) \nu} n^\sigma - n^\nu \Delta^{(s) \sigma} \right) \rbeta_{\nu \rho} n^\rho \nonumber \\ & & \hspace{-5mm} \cdot \rbeta_{\sigma \tau} \eta^{\tau \lambda} \left[ \cA ( n \Delta^{(s) } )^{- 3} n^\mu + ( n \Delta^{(s) } )^{- 2} \Delta^{(s) \mu} \right] J_{\lambda \mu} , \\ & & Det_0 \stackrel{\rm def }{ = } 1 + \cA ( n \Delta^{(s) } )^{- 2} n^2 - \frac{1}{4} \cA^2 (\oDelta \Delta ) ( n \Delta^{(s) } )^{- 4} n^\lambda \rbeta_{\lambda \mu} n^\mu .
\end{eqnarray}

For the other two values $F_\lambda$, $r_\lambda$ defining the propagator by equation (\ref{d2g=J+nf+d2g+ds2g}), we have in terms of $h - \cA \eta^{\lambda \mu} w_{\lambda \mu}$, $\eta^{\lambda \mu} w_{\lambda \mu} + n^2 ( n \Delta^{(s) } )^{- 2} h$ and $\ralpha$, expressed in terms of $\rbeta$ by solving (\ref{beta=f(alpha)}):
\begin{eqnarray}\label{F_k=1}                                           %40,41
& & \hspace{-5mm} F^\lambda = \left[ 1 + \cA ( n \Delta^{(s) } )^{- 2} n^2 \right]^{- 1} \left[ \eta^{\lambda \mu} - \cA ( n \Delta^{(s) } )^{- 2} ( \oDelta \Delta \rbeta^{\lambda \mu} + \Delta^{(s) \lambda } \Delta^{(s) }_\sigma \rbeta^{\sigma \mu } ) \right] \left[ \cA ( n \Delta^{(s) } )^{- 2} \vphantom{\frac{1}{2}} \right. \nonumber \\ & & \left. \hspace{-5mm} \cdot \left( - 4 n^\nu J_{ \mu \nu } + 2 n_\mu \eta^{\nu \rho} J_{\nu \rho} \right) - 4 ( n \Delta^{(s) } )^{- 1} \Delta^{(s) \nu } J_{ \mu \nu } + 2 n_\mu ( n \Delta^{(s) } )^{- 2} \Delta^{(s) \nu } \Delta^{(s) \rho } J_{\nu \rho} \right. \nonumber \\ & & \left. \hspace{-5mm} - \frac{1}{2} \cA n_\mu ( n \Delta^{(s) } )^{- 2} \left( h - \cA \eta^{\nu \rho} w_{\nu \rho} \right) \right], \\
& & \hspace{-5mm} \label{r} r^\lambda = \left[ 1 + \cA ( n \Delta^{(s) } )^{- 2} n^2 \right]^{- 1} ( n \Delta^{(s) } )^{- 1} \left\{ \left[ \eta^{\lambda \mu} - \cA ( n \Delta^{(s) } )^{- 2} ( \oDelta \Delta \rbeta^{\lambda \mu} + \Delta^{(s) \lambda } \Delta^{(s) }_\sigma \rbeta^{\sigma \mu } ) \right] \vphantom{\frac{1}{2}} \right. \nonumber \\ & & \left. \hspace{-5mm} \cdot \left( - 4 n^\nu J_{ \mu \nu } + 2 n_\mu \eta^{\nu \rho} J_{\nu \rho} \right) - \left( n^2 \eta^{ \lambda \mu } + \oDelta \Delta \rbeta^{\lambda \mu} + \Delta^{(s) \lambda } \Delta^{(s) }_\sigma \rbeta^{\sigma \mu } \right) \left[ - 4 ( n \Delta^{(s) } )^{- 1} \Delta^{(s) \nu } J_{ \mu \nu } \right. \right. \nonumber \\ & & \left. \left. \hspace{-5mm} + 2 n_\mu ( n \Delta^{(s) } )^{- 2} \Delta^{(s) \nu } \Delta^{(s) \rho } J_{\nu \rho} \right] + \frac{1}{2} \cA n^\lambda \left[ \eta^{\nu \rho} w_{\nu \rho} + n^2 ( n \Delta^{(s) } )^{- 2} h \right] \right. \nonumber \\ & & \left. \hspace{-5mm} + \frac{1}{2} \cA ( n \Delta^{(s) } )^{- 2} \left( \oDelta \Delta \rbeta^{\lambda \mu} + \Delta^{(s) \lambda } \Delta^{(s) }_\sigma \rbeta^{\sigma \mu } \right) n_\mu \left( h - \cA \eta^{\nu \rho} w_{\nu \rho} \right) \right\}.
\end{eqnarray}

The propagator is given by
\begin{eqnarray}\label{d2g=J+nf+dr+d2(g+h/d^2)+(h-Ag)}                     %42
& & - \oDelta \Delta w_{\lambda \mu} = 4 J_{\lambda \mu} + n_\mu F_\lambda + n_\lambda F_\mu + \Delta^{(s )}_\mu r_\lambda + \Delta^{(s )}_\lambda r_\mu + \left[ 1 + \cA ( n \Delta^{(s )} )^{- 2} n^2 \right]^{- 1} \nonumber \\ & & \cdot \Delta^{(s )}_\lambda \Delta^{(s )}_\mu \left\{ \left[ \eta^{\nu \rho} w_{\nu \rho} + n^2 ( n \Delta^{(s) } )^{- 2} h \right] - n^2 ( n \Delta^{(s) } )^{- 2} ( h - \cA \eta^{\nu \rho} w_{\nu \rho} ) \right\} \nonumber \\ & & - \eta_{\lambda \mu } ( h - \cA \eta^{\nu \rho} w_{\nu \rho} ) , \quad w_{\lambda \mu} \stackrel{\rm def}{=} \cG_{\lambda \mu \sigma \tau} ( n, n ) J^{\sigma \tau}.
\end{eqnarray}

\noindent When analyzing the pole structure of the propagator, one should remember that $\rbeta$ is actually a function of $\ralpha$, $n$, $\Delta$, $\Delta^{(s )}$ (\ref{beta=f(alpha)}) and that $\rbeta = O ( ( n \Delta^{( s )} )^2 )$ in the neighborhood of $n \Delta^{( s )} = 0$. Compared to the case $\cA = 0$, the pole factors of the type $( n \Delta^{(s )} )^{- j}$ in terms are mostly replaced by factors of the type $[ ( n \Delta^{(s )} )^2 + n^2 \cA ]^{- l}$, but not all. Namely, the factor $( n \Delta^{(s )} )^{- 1}$ enters the propagator (\ref{d2g=J+nf+dr+d2(g+h/d^2)+(h-Ag)}) through $r$ (\ref{r}), where it enters into the product with another pole factor,$( n \Delta^{(s )} )^{- 1} [ ( n \Delta^{(s )} )^2 + n^2 \cA ]^{- 1}$. Obviously, the predecessor of this product in the case of $\cA = 0$ is the pole factor $( n \Delta^{(s )} )^{- 3}$. In addition to the pole at $p_0 \approx - i \varepsilon$, the factor $( n \Delta^{(s )} )^{- 1} \propto ( \sin p_0 + i \varepsilon )^{- 1} $ also has poles at $p_0 \approx \pm \pi + i \varepsilon$, but since the sign of $\Im p_0$ changes, there is no doubling of the poles compared to the continuum case, as can be seen if the integration contour is closed to cover $p_0 \approx - i \varepsilon$.

If a continuum diagram converges, then it is contributed mainly by the loop momenta of the order of the external momenta. For habitual external momenta, much smaller than the Planck scale, this allows us to write down for this diagram its expansion over typical variations of the external fields from site to site. To obtain the leading order over these variations, it suffices to use the effective propagator $G^{\rm eff} (n, n)$ obtained from $\cG (n, n)$ by equating $- \oDelta \Delta$ to $\Delta^{(s) 2}$ in the leading order over finite differences, except for $- \oDelta \Delta$ appearing in the denominator. That is, neglecting $\cA$ (which is $O ( \Delta^4 )$) on the right side of the equation $- \oDelta \Delta w_{\lambda \mu} = 4 J_{\lambda \mu} + \dots$ (\ref{d2g=J+nf+d2g+ds2g}). This $G^{\rm eff} $ differs from $G$ (\ref{G}) by having $- \oDelta \Delta$ instead of $\Delta^{(s) 2}$ in the denominator. If we add the designation of the functional dependence of $G$ on $\ralpha$, for example, as $G (n, n, \ralpha )$, then
\begin{equation}\label{Geff}                                               %43
G^{\rm eff} (n, n, \ralpha ) = \frac{\Delta^{(s) 2}}{- \oDelta \Delta} G \left(n, n, \frac{- \oDelta \Delta}{\Delta^{(s) 2}} \ralpha \right) .
\end{equation}

\noindent Obviously, there are no pole factors $[ ( n \Delta^{(s )} )^2 + n^2 \cA ]^{- l}$ in terms, and the factors regularizing $( \rnu \Delta^{(s )} )^{- j}$ are $( n \Delta^{(s )} )^{- j}$.

Another relatively simple expression for the propagator is obtained if we neglect the value of $\ralpha = O ( \varepsilon^2 )$ as compared to $O ( 1 )$ and restrict ourselves to the spatial-spatial components of the metric,
\begin{eqnarray}\label{Gabgd}                                              %44
& & \cG_{\alpha \beta \gamma \delta} ( n , n ) = \frac{- 2}{ \oDelta \Delta } [ L_{\alpha \gamma} ( n, n ) L_{\beta \delta} ( n, n ) + L_{\beta \gamma} ( n, n ) L_{\alpha \delta} ( n, n ) - L_{\alpha \beta} ( n, n ) L_{\gamma \delta} ( n, n ) ] , \nonumber \\ & & L_{\alpha \beta} ( n, n ) \stackrel{\rm def }{=} \eta_{\alpha \beta} + \frac{ n^2 }{ ( n \Delta^{(s )} )^2 + \cA n^2 } \Delta^{(s )}_\alpha \Delta^{(s )}_\beta
\end{eqnarray}

\noindent (in fact, here $\eta_{\alpha \beta} = \delta_{\alpha \beta}$). Neglecting the value $\alpha$ means, in particular, that keeping $\varepsilon^2$ in $n^2$ is an excess of precision. But keeping $\varepsilon$ in $ n \Delta^{(s )} = \rnu \Delta^{(s )} - \varepsilon$  makes sense and allows us to bypass the singularity.

\subsubsection{More detailed arrangement of nonphysical poles at \texorpdfstring{$k = 1$}{k=1}}\label{detail-pole}

Of interest is also a more detailed consideration of nonphysical poles, for small and especially for large quasi-momenta $\bp$. To analyze the latter consistently, we need to consider a more general case of the starting point of the perturbative expansion $g^{(0 )}_{\lambda \mu} \neq \eta_{\lambda \mu}$. The necessary generality is provided by some spacelike lengths $b_{\rm s}$ and timelike lengths $b_{\rm t}$ (lattice spacings),
\begin{eqnarray}                                                           %45
& & \hspace{-15mm} g^{(0 )}_{\lambda \mu} = {\rm diag} (-b_{\rm t}^2, b_{\rm s}^2, b_{\rm s}^2, b_{\rm s}^2) = l^{(0) a}_\lambda \eta_{a b} l^{(0) b}_\mu , \quad l^{(0) a}_\lambda = {\rm diag} (b_{\rm t}, b_{\rm s}, b_{\rm s}, b_{\rm s}) , \quad l^{(0) a}_\mu l^{(0) \lambda}_a \equiv \delta^\lambda_\mu ,
\end{eqnarray}

\noindent so that we can introduce a new scaled metric tensor variable $\tg_{a b}$ and other field variables,
\begin{equation}                                                           %46
g_{\lambda \mu} = l^{(0) a}_\lambda \tg_{a b} l^{(0) b}_\mu, ~ A_\lambda = l^{(0) a}_\lambda \tA_a .
\end{equation}

\noindent This induces a transition to scaled finite differences $\tDelta^{(s )}_a$, $\tDelta_a$ and gauge parameters $\tn^a$, $\trnu^a$, $\tvareps$, $\trlambda^{a b}$,
\begin{eqnarray}                                                           %47
& & \tDelta^{(s )}_a = \tl^{(0) \lambda}_a \Delta^{(s )}_\lambda , ~ \tDelta_a = \tl^{(0) \lambda}_a \Delta_\lambda , ~ \tl^{(0) \lambda}_a \stackrel{\rm def}{=} l^{(0) \lambda}_a \sqrt{ \det \| l^{(0) b}_\mu \|} , ~ \tvareps = \varepsilon \sqrt{ \det \| l^{(0) b}_\mu \|} , \nonumber \\ & & \tn^a = n^\lambda l^{(0 ) a}_\lambda , ~ \trnu^a = \rnu^\lambda l^{(0 ) a}_\lambda , ~ \trlambda^{a b} = l^{(0 ) a}_\lambda \rlambda^{\lambda \mu} l^{(0 ) b}_\mu ,
\end{eqnarray}

\noindent so that formulas for the action and the gauge-fixing term in terms of tilde values could be obtained simply by replacing non-tilde values with tilde values. (It should be taken into account that the factor $\sqrt{- g}$ in the case of a general metric is also present in $S_{\rm em}$, $\cS_{\rm em}$, thereby ensuring the natural emergence of the same $\tDelta^{(s )}_a$, $\tDelta_a$ there.) The indices of the tilde values are raised (lowered) with the help of $\eta^{a b}$ ($\eta_{a b}$). In particular, nonphysical poles (apart from those determined by $\tn \tDelta^{(s )} = 0$) are defined by (omitting the terms $O ( \varepsilon^2 )$)
\begin{eqnarray}                                                           %48
& & \hspace{-15mm} \left( \tn \tDelta^{(s )} \right)^2 - \tn^2 \left( \otDelta \tDelta + \tDelta^{(s )} \tDelta^{(s ) } \right) = 0, ~ \tn^a = \trnu^a - \tvareps \frac{\tDelta^{(s ) a}_\perp}{\tDelta^{(s ) 2}_\perp} , ~ \tDelta^{(s ) a}_\perp = \tDelta^{(s ) a} - \frac{ \trnu^a }{ \trnu^2 } (\trnu \tDelta^{(s )} ) .
\end{eqnarray}

\noindent Here $\tn \tDelta^{(s) } = \trnu \tDelta^{(s) } - \tvareps$, and we set $\tn^2 \approx \trnu^2$ and in the momentum representation we obtain
\begin{equation}                                                           %49
\sin^2 \frac{p_0 }{2 } = \sigma - \frac{\varepsilon }{2 } i \sin p_0 , ~ \sigma \stackrel{\rm def}{= } \frac{b_{\rm t}^2 }{b_{\rm s}^2 } \sum_{\alpha = 1}^3 \sin^4 \frac{p_\alpha }{2 } .
\end{equation}

\noindent Up to $O ( \varepsilon^2 )$ this is compatible with the form of the denominators (\ref{(nd)^2+n^2A^2}); because of the periodicity of $p_0$, it is more convenient to consider the situation in the plane of $\exp ( i p_0 )$ (with complex $p_0$), for which we have two solutions,
\begin{equation}                                                           %50
\exp ( i p_{0 \pm} ) = \frac{1 - 2 \sigma \pm \sqrt{4 ( \sigma^2 - \sigma ) + \varepsilon^2 }}{1 - \varepsilon } .
\end{equation}

\begin{figure}[h]
	\centering
	\includegraphics[scale=1.00]{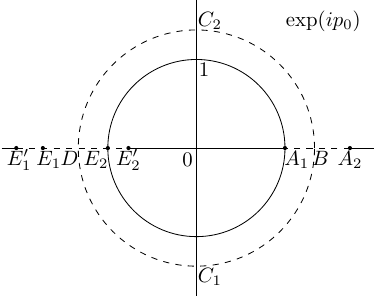}
	\caption{Location of nonphysical poles for $k = 1$.}
	\label{f2}
\end{figure}

\noindent $\sigma$ increases as $p_j$s change from $0$ to $\pi$, and this pair of poles describes the curves
shown in Fig.~\ref{f2} by dashed lines.
If $3b_{\rm t}^2 / b_{\rm s}^2 \leq 1$, then $\sigma$ does not exceed 1 and these curves $A_1BC_1DE_1 \cup A_2BC_2DE_2$ (shown for the limiting case $3b_{\rm t}^2 / b_{\rm s}^2 = 1$) lie on one side of the integration contour (a circle of unit radius with center at the origin), while on this contour there lies only one pole $A_1$ at $\bp = 0$ and $E_2$ at $p_1 = p_2 = p_3 = \pi$ (only if $3b_{\rm t}^2 / b_{\rm s}^2 = 1$). For definiteness, we take $\varepsilon > 0$, and the considered curves lie outside the contour (for $\varepsilon < 0$, they would lie inside the contour).

If $3b_{\rm t}^2 / b_{\rm s}^2 > 1$, then $\sigma$ can exceed 1, and the segment $E_2 E^\prime_2$ of the corresponding curves $A_1BC_1DE^\prime_1 \cup A_2BC_2DE^\prime_2$ is inside the integration contour.

The idea of the principal value prescription is to place the poles of each of the two terms on one side of the integration path. But if the integration path is clamped between the poles of the same term, as in the latter case, then the integration result is contributed by the residue at any of these poles and cannot be zero and does not match the continuum result.

The question of whether $3b_{\rm t}^2 / b_{\rm s}^2$ is greater or less than 1 is also important for the physical poles themselves, located at the zeroes of
\begin{equation}                                                           %51
\otDelta_a \tDelta^a - i 0 \propto \frac{b_{\rm t}^2 }{b_{\rm s}^2 } \sum_{\alpha = 1}^3 \sin^2 \frac{p_\alpha}{2} - \sin^2 \frac{p_0 }{2 } -i 0 ,
\end{equation}

\noindent and for $3 b_{\rm t}^2 / b_{\rm s}^2 > 1$ we are faced with a situation where for some quasi-momenta $\bp$ these poles acquire a (finite) imaginary part and become nonphysical. We consider this fact in \cite{khat} and note that if both spacelike and timelike elementary lengths are determined from the maximum point of the functional measure, then $b_{\rm t} / b_{\rm s} = \gamma $, the Barbero-Immirzi parameter, which was estimated from calculations of the black hole entropy using the area operator spectrum in LQG in a number of papers \cite{AshBaeCorKra,LewDom,Meis,Khr}, and these estimates satisfy $3 \gamma^2 < 1$ with a margin.

Besides that, the result of constructing the perturbative expansion depends on the gauge, that is, on the subset of the configuration superspace over which the functional integral is evaluated (in the perspective, some averaging should be performed over all gauges). The estimate of $b_{\rm t}$, $b_{\rm s}$ from the maximum point of the functional measure is based on some model assumptions about the measure extended to the entire configuration superspace. We can return to a smaller region where the measure can be found in the "factorization approximation" of functional integration over the connection. This is achieved by considering the zeroth order over discrete analogues of the ADM lapse-shift functions \cite{ADM1}, some {\it temporal} edge vectors, which for this purpose we fix at a low level so that $b_{\rm t} b_{\rm s}^{- 1} << 1$.

Thus, the inequality
\begin{equation}                                                           %52
3 b_{\rm t}^2 b_{\rm s}^{- 2} < 1
\end{equation}

\noindent looks quite natural. This condition allows the principal value prescription in the discrete case to match the continuum limit. The trivial choice $b_{\rm t} = 1$, $b_{\rm s} = 1$ does not satisfy this condition, but we shall continue to use this choice for simplicity of notation, taking into account that in the real case the metric $g_{\lambda \mu}$ and finite differences $\Delta^{(s )}_\lambda$, $\Delta_\lambda$ should be replaced by $\tg_{a b}$ and $\tDelta^{(s )}_a$, $\tDelta_a$ (as well as components of other fields).

Thus, if for some quasi-momenta $\bp$ there are no physical poles, then the principal value prescription for the nonphysical poles in the discrete case leads to extra contributions for such $\bp$ compared to the continuum limit. If there are physical poles for all $\bp$, then the principal value type prescription scheme in the discrete case matches the continuum result.

\section{Principal value type prescription and the correspon\-ding gauge-fixing term}\label{principal}

\subsection{The form of the gauge-fixing term}\label{gauge-fixing}

For the general bilinear action form
\begin{equation}                                                           %53
\cS_{\rm g} = \frac{1}{2} \sum_{\rm sites} w_{\lambda \mu} \ccM^{\lambda \mu \sigma \tau} w_{\sigma \tau} + O ( w^3 )
\end{equation}

\noindent we have the usual propagator form corresponding to the "soft" gauge-fixing term (\ref{S_g'}),
\begin{equation}                                                           %54
\hspace{0mm} \cG_{\lambda \mu \sigma \tau} ( n, \on ) = \left[ \left\| \ccM^{\nu \rho \kappa \varphi} - \frac{1}{2} \on^{( \nu } \rlambda^{\rho ) ( \kappa } n^{\varphi )} \right\|^{-1} \right]_{\lambda \mu \sigma \tau} \! \! \! .
\end{equation}

\noindent Here, $\on$ means the Hermitian conjugate of $n$, and then $\cG_{\lambda \mu \sigma \tau} ( n, \on )$ is Hermitian symmetric (if $\ccM$, $\rlambda$ are Hermitian symmetric), which is obtained by the standard calculation process starting from varying the action, but there is another way to "soften" the gauge and get a Hermitian symmetric value with a complex $n$. We define
\begin{equation}                                                           %55
\hspace{0mm} \cG_{\lambda \mu \sigma \tau} ( n, n ) = \left[ \left\| \ccM^{\nu \rho \kappa \varphi} - \frac{1}{2} n^{( \nu } \rlambda^{\rho ) ( \kappa } n^{\varphi )} \right\|^{-1} \right]_{\lambda \mu \sigma \tau} \! \! \! ,
\end{equation}

\noindent a non-Hermitian operator, and the same for $n \Rightarrow \on$ and form their half-sum
\begin{equation}\label{vpG=}                                               %56
\cG_{\lambda \mu \sigma \tau} = \frac{1}{2} \cG_{\lambda \mu \sigma \tau} ( n, n ) + \frac{1}{2} \cG_{\lambda \mu \sigma \tau} ( \on, \on ) .
\end{equation}

\noindent The operator (\ref{vpG=}) can be written as
\begin{equation}                                                           %57
\frac{1}{2} ( {\cal C } + i {\cal E } )^{-1} + \frac{1}{2} ( {\cal C } - i {\cal E } )^{-1} = ( { \cal C } + { \cal E } { \cal C }^{-1} { \cal E } )^{-1} ,
\end{equation}

\noindent where ${\cal C }$ and $\pm i {\cal E }$ are the Hermitian and anti-Hermitian parts of the operators $\ccM - n \rlambda n /2$ and $\ccM - \on \rlambda \on /2$.

Thus, the principal value type propagator $\cG$ is the inverse of ${ \cal C } + { \cal E } { \cal C }^{-1} { \cal E }$, which can be written as the original bilinear form plus a correction, $\ccM + \Delta \ccM$. In the momentum representation, the Hermitian/anti-Hermitian parts are the $\Re$/$i \Im$ parts, and the correction takes the form
\begin{eqnarray}\label{DM}                                                 %58
\Delta \ccM^{\lambda \mu \sigma \tau} = - \frac{1}{2} \Re \left( n^{( \mu } \rlambda^{\lambda ) ( \sigma } n^{\tau )}\right) + \frac{1}{4} \left[ \Im \left( n^{( \mu } \rlambda^{\lambda ) \zeta } n^{\pi }\right) \right] \cG^{\rm aux}_{\zeta \pi \chi \psi} \left[ \Im \left( n^{ \chi } \rlambda^{\psi ( \sigma } n^{\tau )}\right) \right] ,
\end{eqnarray}

\noindent where the auxiliary propagator is
\begin{equation}\label{Gaux}                                               %59
\cG^{\rm aux}_{\zeta \pi \chi \psi} = \left[ \left\| \ccM^{\nu \rho \kappa \varphi} - \frac{1}{2} \Re \left( n^{( \nu } \rlambda^{\rho ) ( \kappa } n^{\varphi )}\right) \right\|^{-1} \right]_{\zeta \pi \chi \psi}.
\end{equation}

\noindent The $\Re$- and $i \Im$-parts are
\begin{eqnarray}                                                        %60,61
& & \label{Re(nln)} \Re \left( n^{ \nu } \rlambda^{\rho \kappa } n^{\varphi }\right) = \rnu^{ \nu } \rlambda^{\rho \kappa } \rnu^{\varphi } + \varepsilon^2 \frac{ \Delta_{ \perp}^{(s) \nu } }{ \Delta_{ \perp}^{(s) 2 } } \rlambda^{\rho \kappa } \frac{ \Delta_{ \perp}^{(s) \varphi } }{ \Delta_{ \perp}^{(s) 2 } } , \\
& & \label{Im(nln)} i \Im \left( n^\mu \rlambda^{\lambda \zeta } n^\pi \right) = - \varepsilon \frac{ \rnu^\mu \rlambda^{\lambda \zeta} \Delta_\perp^{(s) \pi } + \Delta_\perp^{(s) \mu } \rlambda^{\lambda \zeta} \rnu^\pi }{\Delta_\perp^{(s) 2}} .
\end{eqnarray}

In terms of the "hard" synchronous gauge propagator
\begin{eqnarray}                                                           %62
\cG^{(0)}_{\zeta \pi \chi \psi} = \left[ \left\| \ccM^{\nu \rho \kappa \varphi} - \frac{1}{2} \rnu^{( \nu } \rlambda^{\rho ) ( \kappa } \rnu^{\varphi )} \right\|^{-1} \right]_{\zeta \pi \chi \psi} = \cG_{\zeta \pi \chi \psi} ( \rnu , \rnu ) ,
\end{eqnarray}

\noindent $\cG^{\rm aux}_{\zeta \pi \chi \psi}$ can be expressed as
\begin{equation}\label{Gaux=f(M)}                                          %63
\cG^{\rm aux}_{\zeta \pi \chi \psi} = \cG^{(0)}_{\zeta \pi \chi \psi} + \frac{1}{2} \cG^{(0)}_{\zeta \pi \nu \rho } \frac{ \Delta_{ \perp}^{(s) \nu } }{ \Delta_{ \perp}^{(s) 2 } } \left( \cmM^{ - 1 } \right)^{\rho \kappa } \frac{ \Delta_{ \perp}^{(s) \varphi } }{ \Delta_{ \perp}^{(s) 2 } } \cG^{(0)}_{\kappa \varphi \chi \psi} ,
\end{equation}

\noindent where we introduce the notation
\begin{equation}\label{M=f(DGD)}                                           %64
\cmM_{\lambda \tau} = \frac{\ralpha_{\lambda \tau}}{\varepsilon^2 }  - \frac{1}{2} \frac{ \Delta_{ \perp}^{(s) \mu } }{ \Delta_{ \perp}^{(s) 2 } } \cG^{(0)}_{\lambda \mu \sigma \tau} \frac{ \Delta_{ \perp}^{(s) \sigma } }{ \Delta_{ \perp}^{(s) 2 } } .
\end{equation}

In eq. (\ref{DM}), $\cG^{(0)}_{\lambda \mu \sigma \tau}$ enters in the form $\Delta_{ \perp}^{(s) \mu } \cG^{(0)}_{\lambda \mu \sigma \tau} \Delta_{ \perp}^{(s) \sigma }$, $\Delta_{ \perp}^{(s) \mu } \cG^{(0)}_{\lambda \mu \sigma \tau} \rnu^\sigma $, $\rnu^\mu \cG^{(0)}_{\lambda \mu \sigma \tau} \Delta_{ \perp}^{(s) \sigma }$, and $\rnu^\mu \cG^{(0)}_{\lambda \mu \sigma \tau} \rnu^\sigma $. Expanding (\ref{DM}) using (\ref{Re(nln)}), (\ref{Im(nln)}) and (\ref{Gaux=f(M)}) and expressing $\Delta_{ \perp}^{(s) \mu } \cG^{(0)}_{\lambda \mu \sigma \tau} \Delta_{ \perp}^{(s) \sigma }$ in terms of $\cmM_{\lambda \tau}$ using (\ref{M=f(DGD)}), we find for the required correction to the bilinear form of the action:
\begin{eqnarray}\label{DM=OM^(-1)O+}                                       %65
& & \hspace{-10mm} \Delta \ccM^{\lambda \mu \sigma \tau} = - \frac{1}{2 \varepsilon^2 } \left( \rnu^\mu \delta^\lambda_\rho + \frac{ \varepsilon^2 }{2} \frac{ \Delta_{ \perp}^{(s) \mu } }{ \Delta_{ \perp}^{(s) 2 } } \rlambda^{\lambda \zeta } \rnu^\pi \cG^{(0)}_{\zeta \pi \nu \rho} \frac{ \Delta_{ \perp}^{(s) \nu } }{ \Delta_{ \perp}^{(s) 2 } } \right) \left( \cmM^{ - 1 } \right)^{\rho \kappa } \nonumber \\ & & \hspace{-10mm} \cdot \left( \delta^\sigma_\kappa \rnu^\tau + \frac{ \varepsilon^2 }{2} \frac{ \Delta_{ \perp}^{(s) \varphi } }{ \Delta_{ \perp}^{(s) 2 } } \cG^{(0)}_{\kappa \varphi \chi \psi} \rnu^\chi \rlambda^{\psi \sigma} \frac{ \Delta_{ \perp}^{(s) \tau } }{ \Delta_{ \perp}^{(s) 2 } } \right) - \frac{ \varepsilon^2 }{2} \frac{ \Delta_{ \perp}^{(s) \lambda } }{ \Delta_{ \perp}^{(s) 2 } } \left( \rlambda^{\mu \sigma} + \frac{1}{2} \rlambda^{\mu \zeta } \rnu^\pi \cG^{(0)}_{\zeta \pi \chi \psi} \rnu^\chi \rlambda^{\psi \sigma} \right) \frac{ \Delta_{ \perp}^{(s) \tau } }{ \Delta_{ \perp}^{(s) 2 } } .
\end{eqnarray}

\noindent Here we mean symmetrization with respect to permutations $\lambda \leftrightarrow \mu$ and $\sigma \leftrightarrow \tau$. That is, the gauge-fixing term that provides the principal value type form of the propagator takes the form
\begin{eqnarray}\label{F[g]}                                               %66
& & \hspace{-5mm} \ccF [ g ] = \frac{1}{2} \sum_{\rm sites} w_{\lambda \mu} \Delta \ccM^{\lambda \mu \sigma \tau} w_{\sigma \tau} = \sum_{\rm sites} \left( - \frac{1}{4} \cmf_\rho [ g ] \cLambda^{\rho \kappa} \cmf_\kappa [ g ] + \frac{1}{2} w_{\lambda \mu} \mm^{\lambda \mu \sigma \tau} w_{\sigma \tau} \right) , ~ \cLambda = \frac{1}{\varepsilon^2} \cmM^{- 1} , \nonumber \\ & & \hspace{-5mm} \cmf_\rho = \crO^{\lambda \mu}_\rho w_{\lambda \mu} , \quad \mm^{\lambda \mu \sigma \tau} = - \frac{ \varepsilon^2 }{2} \frac{ \Delta_{ \perp}^{(s) \lambda } }{ \Delta_{ \perp}^{(s) 2 } } \left( \rlambda^{\mu \sigma} + \frac{1}{2} \rlambda^{\mu \zeta } \rnu^\pi \cG^{(0)}_{\zeta \pi \chi \psi} \rnu^\chi \rlambda^{\psi \sigma} \right) \frac{ \Delta_{ \perp}^{(s) \tau } }{ \Delta_{ \perp}^{(s) 2 } }, \nonumber \\ & & \label{trO} \hspace{-5mm} \crO^{\lambda \mu}_\rho = \delta^\lambda_\rho \rnu^\mu + \frac{ \varepsilon^2 }{ 2 } \frac{ \Delta^{( s ) \nu }_\perp }{ \Delta^{( s ) 2 }_\perp } \cG^{(0)}_{\rho \nu \pi \zeta} \rnu^\pi \rlambda^{\zeta \lambda} \frac{ \Delta^{( s ) \mu }_\perp }{ \Delta^{( s ) 2 }_\perp } .
\end{eqnarray}

Thus, $\Delta \ccM^{\lambda \mu \sigma \tau}$ is represented in the form $\propto \overline{ \crO } \cmM^{- 1} \crO$ plus a potentially small term $\mm \propto \varepsilon^2$. On the other hand, this term contains the matrix $\rlambda$, which we assumed above to be large ($O ( \varepsilon^{-2} ) $). Compare this with the case $\cA = 0$, that is, with the same construction for the action containing only $\Delta^{(s )}$ and the propagator $G$ instead of $\cG$,
\begin{eqnarray}                                                           %67
G^{(0)}_{\zeta \pi \chi \psi} = \left[ \left\| \cM^{\nu \rho \kappa \varphi} - \frac{1}{2} \rnu^{( \nu } \rlambda^{\rho ) ( \kappa } \rnu^{\varphi )} \right\|^{-1} \right]_{\zeta \pi \chi \psi} = G_{\zeta \pi \chi \psi} ( \rnu , \rnu ) .
\end{eqnarray}

\noindent $G_{\lambda \mu \sigma \tau} ( n, \on )$ (\ref{G}) contains $\| \ralpha \| = \| \rlambda \|^{- 1}$ linearly, and $G^{(0)}_{\lambda \mu \sigma \tau}$ is $G_{\lambda \mu \sigma \tau} ( n, \on )$ for $n = \on = \rnu$. When contracting $G^{(0)}$ with $\rnu$ over any index, only $O ( \ralpha )$ (non-pole) part survives. The term we are interested in turns out to be equal to zero,
\begin{equation}\label{l+lng0nl=0}                                         %68
\rlambda^{\mu \sigma} + \frac{1}{2} \rlambda^{\mu \zeta } \rnu^\pi G^{(0)}_{\zeta \pi \chi \psi} \rnu^\chi \rlambda^{\psi \sigma} = 0.
\end{equation}

\noindent For $\cG^{(0)}$ we can write the expansion in terms of $G^{(0)}$ over
\begin{equation}                                                           %69
(\ccM - \cM )^{\lambda \mu \sigma \tau} = \frac{1}{2} ( \eta^{\lambda \sigma} \eta^{\mu \tau} + \eta^{\lambda \tau} \eta^{\mu \sigma} - 2 k \eta^{\lambda \mu} \eta^{\sigma \tau} ) \cA .
\end{equation}

\noindent Its validity is ensured by the existence of a region in the momentum space (small quasi-momenta compared to their limiting values) in which $\cA$ is a small perturbation, with subsequent analytical continuation from this region. This $\cG^{(0)}$ appears in the desired expression as
\begin{eqnarray}\label{ntGn=nGn+}                                          %70
\rnu^\pi \cG^{(0)}_{\zeta \pi \chi \psi} \rnu^\chi & = & \rnu^\pi G^{(0)}_{\zeta \pi \chi \psi} \rnu^\chi + \rnu^\pi G^{(0)}_{\zeta \pi \lambda \mu} ( \cM - \ccM )^{\lambda \mu \sigma \tau} G^{(0)}_{\sigma \tau \chi \psi} \rnu^\chi + \dots \nonumber \\ & & + \rnu^\pi G^{(0)}_{\zeta \pi \lambda \mu} ( \cM - \ccM )^{\lambda \mu \nu \rho} G^{(0)}_{\nu \rho \eta \xi} \dots ( \cM - \ccM )^{\kappa \varphi \sigma \tau} G^{(0)}_{\sigma \tau \chi \psi} \rnu^\chi + \dots .
\end{eqnarray}

\noindent It is important that $\cA$ enters (\ref{ntGn=nGn+}) starting from the second term, and all terms starting from the second are of order $O ( \ralpha^2 )$, since $\rnu G^{(0)} = O ( \ralpha )$. That is, the dependence of $\rnu \cG^{(0)} \rnu$ on $\cA$ arises in the order $O ( \ralpha^2 ) = O ( \rlambda^{-2} )$. In $\Delta \ccM$ this is multiplied by two $\rlambda$, resulting in order $O ( \rlambda^0 ) = O(1)$, while the first term $\rnu G^{(0)} \rnu$ is cancelled ((\ref{l+lng0nl=0})),
\begin{equation}\label{l+lntg0nl=O(A)}                                     %71
\rlambda^{\mu \sigma} + \frac{1}{2} \rlambda^{\mu \zeta } \rnu^\pi \cG^{(0)}_{\zeta \pi \chi \psi} \rnu^\chi \rlambda^{\psi \sigma} = O ( \cA ).
\end{equation}

\noindent In total, this contributes $\varepsilon^2 O ( \cA )$ to $\Delta \ccM$. We can also get this by directly studying $F^\mu = \rlambda^{\mu \zeta } \rnu^\pi \cG^{(0)}_{\zeta \pi \chi \psi} J^{\chi \psi}$ (\ref{F}) (using (\ref{Dw}), (\ref{trw}), (\ref{DDw}); $n = \rnu$), from which we read off and find $\rlambda^{\mu \zeta } \rnu^\pi \cG^{(0)}_{\zeta \pi \chi \psi} \rnu^\chi$ and find that to order zero in $\ralpha$, $O(1)$, it is independent of $\cA$.

\subsection{Finiteness of the gauge-fixing term in the limiting case \texorpdfstring{$\cA = - \oDelta \Delta - \Delta^{(s ) 2} \to 0$}{A → 0}}\label{A=0finite}

When calculating diagrams, from the bilinear form of action we need only the propagator, possible singularities of the coefficients of this form are inessential. But if we parameterize the metric nonlinearly, these singularities will move to new interaction vertices and will be undesirable.

The question of the finiteness of the coefficients in the gauge-fixing term for the principal value prescription arises because it is determined by $\cG^{\rm aux}$ (\ref{Gaux}), the propagator corresponding to the gauge-fixing term with the matrix $\Re \left( n^{ \nu } \rlambda^{\rho \kappa } n^{\varphi }\right)$ (\ref{Re(nln)}). In the case of the full complex matrix $ n^{ \nu } \rlambda^{\rho \kappa } n^{\varphi }$, the values $\left( n \Delta^{(s)} \right)^{- j}$ replacing the singularities $\left( \rnu \Delta^{(s)} \right)^{- j}$ are finite, but now this matrix is real, and by analogy one can expect the appearance of contributions $\propto \left( \Re \left[ \left( n \Delta^{(s)} \right)^2 \right] \right)^{- j} = \left[ \left( \rnu \Delta^{(s)} \right)^2 + \varepsilon^2 \right]^{- j}$, which are equal to $(-\sin^2 p_0 + \varepsilon^2 )^{- j}$ (for $\rnu = (1, 0, 0, 0)$), defined not for each $p_0$. Or, in other words, $\Delta \ccM$ (\ref{DM=OM^(-1)O+}) uses a "hard" synchronous gauge propagator $\cG^{(0)}$, which is singular, and one must check whether these singularities are cancelled out in the final answer. In fact, we find that the gauge-fixing term can be defined in a finite way.

In the limiting case $\cA \to 0$ for the gauge-fixing term, we have the same expressions (\ref{M=f(DGD)}), (\ref{DM=OM^(-1)O+}), (\ref{F[g]}), but with $\cG$ replaced by $G^{\rm eff}$, $\cmM$ by $\mM$ and other quantities marked with a check mark by those without a check mark. The $G^{\rm eff(0)}_{\zeta \pi \chi \psi}$ appearing there (see (\ref{Geff}), (\ref{G}) at $n = \rnu$) is simpler than $\cG^{(0)}_{\zeta \pi \chi \psi}$. Since $\rnu^\pi G^{\rm eff(0)}_{\zeta \pi \chi \psi} = \rnu^\pi G^{(0)}_{\zeta \pi \chi \psi}$, then ${\mathrm O}^{\lambda \mu}_\rho$ and $\mm^{\lambda \mu \sigma \tau}$ for $G^{\rm eff}$ are the same as for $G$, in particular, $\mm = 0$ ((\ref{l+lng0nl=0})), and we have
\begin{eqnarray}\label{OforA=0}                                            %72
& & \hspace{-5mm} \Delta \cM^{\lambda \mu \sigma \tau} = - \frac{1 }{2 \varepsilon^2 } \overline{\mathrm O}^{\lambda \mu}_\rho \left( \mM^{ - 1 } \right)^{\rho \kappa } {\mathrm O}^{\sigma \tau}_\kappa , \quad {\mathrm O}^{\lambda \mu}_\rho = \delta^{ ( \lambda }_\rho \rnu^{ \mu ) } + \frac{ \varepsilon^2 }{ 2 } \frac{ \Delta_{ \perp}^{(s) \nu } }{ \Delta_{ \perp}^{(s) 2 } } G^{\rm eff(0)}_{\rho \nu \pi \zeta} \rnu^\pi \rlambda^{ \zeta ( \lambda} \frac{ \Delta_{ \perp}^{(s) \mu ) } }{ \Delta_{ \perp}^{(s) 2 } } \nonumber \\ & & = \delta^{ ( \lambda }_\rho \rnu^{ \mu ) } - \frac{ \varepsilon^2 }{ \rnu^2 } \rnu_\rho \frac{ \Delta_{ \perp}^{(s) \lambda } \Delta_{ \perp}^{(s) \mu } }{ ( \Delta_{ \perp}^{(s) 2 } )^2 } - \frac{ \varepsilon^2 }{ \rnu \Delta^{(s) } } \left( \delta^{ ( \lambda }_\rho - \frac{ \rnu_\rho \rnu^{ ( \lambda } }{ \rnu^2 } + \Delta_{ \perp \rho }^{(s) } \frac{ \Delta_{ \perp}^{(s) ( \lambda } }{ \Delta_{ \perp}^{(s) 2 } } \right) \frac{ \Delta_{ \perp}^{(s) \mu ) } }{ \Delta_{ \perp}^{(s) 2 } } \nonumber \\ & & + \frac{ \varepsilon^2 }{ ( \rnu \Delta^{(s) } )^2 } \Delta_{ \perp \rho }^{(s) } \frac{ \rnu^{ ( \lambda } \Delta_{ \perp}^{(s) \mu ) } }{ \Delta_{ \perp}^{(s) 2 } } .
\end{eqnarray}

The $\mM$ considered here for $G^{\rm eff}$ differs from that calculated directly for $G$ by scaling its $O ( \ralpha^0 ) = O ( 1 )$ part by a factor of $\Delta^{(s) 2} ( - \oDelta \Delta)^{- 1} $. $G^{(0)}_{\rho \nu \pi \zeta} \rnu^\pi \rlambda^{ \zeta \lambda}$ does not depend on $\ralpha$ (a property of the case $\cA = 0$ inherited from the continuum theory due to some symmetry of the finite difference $\Delta^{(s)}$), so ${\mathrm O}^{\lambda \mu}_\rho$ does not, but $\mM_{\lambda \tau}$ does, and we should specify $\ralpha$. The most general such $\ralpha$ respecting the symmetry defined by the two singled out 4-vectors $\Delta^{( s )}_\lambda$, $\rnu_\lambda$ or $\Delta^{( s )}_{ \perp \lambda }$, $\rnu_\lambda$ is a combination of five structures,
\begin{eqnarray}\label{ralpha}                                             %73
& & \ralpha_{\lambda \mu} = \ralpha_P P_{\lambda \mu} + \ralpha_\rnu \frac{ \rnu_\lambda \rnu_\mu }{ \rnu^2 } + \ralpha_\Delta \frac{ \Delta^{(s) }_{ \perp \lambda} \Delta^{(s) }_{ \perp \mu } }{ \Delta^{(s) 2 }_{ \perp } } + \ralpha_{\rnu \Delta } \rnu_\lambda \Delta^{(s) }_{ \perp \mu } + \ralpha_{ \Delta \rnu } \Delta^{(s) }_{ \perp \lambda } \rnu_\mu, \nonumber \\ & & P_{\lambda \mu} = \eta_{\lambda \mu} - \frac{ \rnu_\lambda \rnu_\mu }{ \rnu^2 } - \frac{ \Delta^{(s) }_{ \perp \lambda} \Delta^{(s) }_{ \perp \mu } }{ \Delta^{(s) 2 }_{ \perp } }.
\end{eqnarray}

\noindent Then $\mM$ is also a combination of these structures,
\begin{eqnarray}                                                           %74
& & \hspace{-10mm} \mM_{\lambda \mu} = \imu_P P_{\lambda \mu} + \imu_\rnu \frac{ \rnu_\lambda \rnu_\mu }{ \rnu^2 } + \imu_\Delta \frac{ \Delta^{(s) }_{ \perp \lambda} \Delta^{(s) }_{ \perp \mu } }{ \Delta^{(s) 2 }_{ \perp } } + \imu_{\rnu \Delta } \rnu_\lambda \Delta^{(s) }_{ \perp \mu } + \imu_{ \Delta \rnu } \Delta^{(s) }_{ \perp \lambda } \rnu_\mu, \nonumber \\ & & \hspace{-10mm} \imu_P = \frac{1}{ ( \rnu \Delta^{(s) } )^2 } \left[ \frac{ \rnu^2 \Delta^{(s) 2} }{ \Delta^{(s) 2 }_{ \perp } ( \oDelta \Delta ) } + \ralpha_P \right] +  \frac{ \ralpha_P }{ \varepsilon^2 }, ~ \imu_\rnu = \frac{ \ralpha_\rnu }{ \varepsilon^2 } + \frac{ \ralpha_\Delta }{ \rnu^2 \Delta^{(s) 2 }_{ \perp } }, \nonumber \\ & & \hspace{-10mm} \imu_\Delta = \frac{ \rnu^4 }{ ( \rnu \Delta^{(s) } )^4 } \left[ \frac{ ( \Delta^{(s) 2 } )^2 }{ \Delta^{(s) 2 }_{ \perp } ( \oDelta \Delta ) } + \ralpha_\rnu \frac{ \Delta^{(s) 2 }_{ \perp } }{ \rnu^2 } \right] - 2 ( \ralpha_{\rnu \Delta } + \ralpha_{ \Delta \rnu } ) \frac{ \rnu^2 \Delta^{(s) 2 }_{ \perp } }{ ( \rnu \Delta^{(s) } )^3 } + 4 \frac{ \ralpha_\Delta }{ ( \rnu \Delta^{(s) } )^2 } + \frac{ \ralpha_\Delta }{ \varepsilon^2 }, \nonumber \\ & & \hspace{-10mm} \imu_{\rnu \Delta } = \frac{ \ralpha_{\rnu \Delta } }{ \varepsilon^2 } - \frac{ \ralpha_{ \Delta \rnu } }{ ( \rnu \Delta^{(s) } )^2 } + 2 \frac{ \ralpha_\Delta }{ \rnu^2 \Delta^{(s) 2 }_{ \perp } ( \rnu \Delta^{(s) } ) }, ~ \imu_{ \Delta \rnu } = \frac{ \ralpha_{ \Delta \rnu } }{ \varepsilon^2 } - \frac{ \ralpha_{\rnu \Delta } }{ ( \rnu \Delta^{(s) } )^2 } + 2 \frac{ \ralpha_\Delta }{ \rnu^2 \Delta^{(s) 2 }_{ \perp } ( \rnu \Delta^{(s) } ) }.
\end{eqnarray}

\noindent The reciprocal matrix reads
\begin{eqnarray}                                                           %75
& & ( \mM^{- 1} )_{\lambda \mu} = \bimu_P P_{\lambda \mu} + \bimu_\rnu \frac{ \rnu_\lambda \rnu_\mu }{ \rnu^2 } + \bimu_\Delta \frac{ \Delta^{(s) }_{ \perp \lambda} \Delta^{(s) }_{ \perp \mu } }{ \Delta^{(s) 2 }_{ \perp } } + \bimu_{\rnu \Delta } \rnu_\lambda \Delta^{(s) }_{ \perp \mu } + \bimu_{ \Delta \rnu } \Delta^{(s) }_{ \perp \lambda } \rnu_\mu, \nonumber \\ & & \bimu_P = \frac{1}{ \imu_P }, ~ \bimu_\rnu = \frac{ \imu_\Delta }{ \mathrm{det}_{\textstyle \mM} }, ~ \bimu_\Delta = \frac{ \imu_\rnu }{ \mathrm{det}_{\textstyle \mM} }, ~ \bimu_{ \rnu \Delta } = \frac{ - \imu_{ \rnu \Delta } }{ \mathrm{det}_{\textstyle \mM} }, ~ \bimu_{ \Delta \rnu } = \frac{ - \imu_{ \Delta \rnu } }{ \mathrm{det}_{\textstyle \mM} }, \nonumber \\ & & \mathrm{det}_{\textstyle \mM} = \imu_\rnu \imu_\Delta - \rnu^2 \Delta^{(s) 2 }_{ \perp } \imu_{\rnu \Delta} \imu_{\Delta \rnu } = \left\{\frac{ \rnu^4 }{ ( \rnu \Delta^{(s) } )^4 } \left[ \frac{ ( \Delta^{(s) 2 } )^2 }{ \Delta^{(s) 2 }_{ \perp } ( \oDelta \Delta ) } + \ralpha_\rnu \frac{ \Delta^{(s) 2 }_{ \perp } }{ \rnu^2 } \right]\right\} \nonumber \\ & & \cdot \left( \frac{ \ralpha_\Delta }{ \rnu^2 \Delta^{(s) 2 }_{ \perp } } + \frac{ \ralpha_\rnu }{ \varepsilon^2 } \right) + \frac{2}{ \varepsilon^2 } \left\{ 2 \frac{ \ralpha_\rnu \ralpha_\Delta }{ ( \rnu \Delta^{(s) } )^2 } - \left( \ralpha_{ \rnu \Delta } + \ralpha_{ \Delta \rnu } \right) \left[ \ralpha_\rnu \frac{ \rnu^2 \Delta^{(s) 2 }_{ \perp } }{ ( \rnu \Delta^{(s) } )^3 } + \frac{ \ralpha_\Delta }{ \rnu \Delta^{(s) } } \right] \right\} \nonumber \\ & & + \rnu^2 \Delta^{(s) 2 }_{ \perp } \left\{ \frac{ \ralpha_{\rnu \Delta}^2 + \ralpha_{ \Delta \rnu }^2 }{ \varepsilon^2 ( \rnu \Delta^{(s) } )^2 } - \ralpha_{\rnu \Delta} \ralpha_{ \Delta \rnu } \left[ \frac{1}{ \varepsilon^4 } + \frac{1}{ ( \rnu \Delta^{(s) } )^4 } \right] \right\}.
\end{eqnarray}

\noindent It is seen that there is a singularity in ${\mathrm O}^{\lambda \mu}_\rho$ at $\rnu \Delta^{(s) } \to 0$, which shows up differently when contracted with different structures over $\rho$:
\begin{eqnarray}\label{(n,P,D)O}                                           %76
& & ( \rnu^\rho, P^{ \kappa \rho }, \Delta^{( s ) \rho }_\perp ) {\mathrm O}^{\lambda \mu}_\rho = ( O(1), O( ( \rnu \Delta^{(s) } )^{- 1} ), O( ( \rnu \Delta^{(s) } )^{- 2} ).
\end{eqnarray}

\noindent But this singularity in $\overline{\mathrm O}^{\lambda \mu}_\rho$, ${\mathrm O}^{\sigma \tau}_\kappa$ is cancelled by the smallness at $\rnu \Delta^{(s) } \to 0$ of the coefficients at each structure over $\rho$ and over $\kappa$ in $\left( \mM^{ - 1 } \right)^{\rho \kappa }$:
\begin{eqnarray}                                                           %77
& & \bimu_\rnu = O(1), ~ ( \bimu_P, \bimu_{\rnu \Delta}, \bimu_{\Delta \rnu} ) = O( ( \rnu \Delta^{(s) } )^2 ), ~ \bimu_\Delta = O( ( \rnu \Delta^{(s) } )^4 ).
\end{eqnarray}

The possible singularity we are left with may be at the zeros of $\mathrm{det}_{\textstyle \mM}$. (A singularity at $\Delta^{(s) 2 }_{ \perp } = 0$ is also possible, but this is an integrable singularity.) The form of $\mathrm{det}_{\textstyle \mM}$ assumes that everywhere, except for a set of measure zero in the configuration space of variables other than $\rnu \Delta^{( s )}$, all its zeros are simple roots. That is, in the vicinity of the $n$-th zero at $\rnu \Delta^{( s )} = ( \rnu \Delta^{( s )} )_n$, there may be a simple pole $\propto ( \rnu \Delta^{( s )} - ( \rnu \Delta^{( s )} )_n )^{- 1}$ in the coefficients of the gauge-fixing term. This is an integrable singularity in the sense of Cauchy's principal value.

\subsection{Finiteness of the gauge-fixing term at \texorpdfstring{$\cA \neq 0$}{A ̸= 0}}\label{finiteAnot0}

In this case, the term $\mm$ in $\Delta \ccM$, which violates the factorization of $\Delta \ccM$ into factors $\overline{ \crO }$, $\cmM^{- 1}$, $\crO$, is not equal to zero. The value of $ \rlambda^{\mu \zeta } \rnu^\pi \cG^{(0)}_{\zeta \pi \chi \psi} \rnu^\chi$, which determines $\mm$, is determined by $F^\mu = \rlambda^{\mu \zeta } \rnu^\pi \cG^{(0)}_{\zeta \pi \chi \psi} J^{\chi \psi} = 2 \rlambda^{\mu \zeta } \rnu^\pi \cG^{(0)}_{\zeta \pi \chi \psi} \rnu^\chi \jmath^\psi$, obtained by substituting $J^{\chi \psi} = \rnu^\chi \jmath^\psi + \rnu^\psi \jmath^\chi$ into $F^\mu$ (\ref{F_k=1}). This gives
\begin{eqnarray}\label{l+lnG(0)nl}                                         %78
& & \hspace{-5mm} \rlambda^{\mu \sigma} + \frac{1}{2} \rlambda^{\mu \zeta } \rnu^\pi \cG^{(0)}_{\zeta \pi \chi \psi} \rnu^\chi \rlambda^{\psi \sigma} = \cA ( \rnu \Delta^{(s)} )^{- 2} \left[ 1 + \cA ( \rnu \Delta^{(s)} )^{- 2} \rnu^2 \right]^{- 1} \left\{ ( \oDelta \Delta ) \eta^{\mu \sigma} + \Delta^{(s) \mu} \Delta^{(s) \sigma} \right. \nonumber \\ & & \left. \hspace{-5mm} - \cA ( \rnu \Delta^{(s)} )^{- 2} \left[ ( \oDelta \Delta ) \eta^{\mu \lambda} + \Delta^{(s) \mu} \Delta^{(s) \lambda} \right] \rbeta_{\lambda \tau} \left[ ( \oDelta \Delta ) \eta^{\tau \sigma} + \Delta^{(s) \tau} \Delta^{(s) \sigma} \right] \right\} \nonumber \\ & & \hspace{-5mm} - \frac{1}{4} \cA^2 ( \oDelta \Delta ) ( \rnu \Delta^{(s)} )^{- 4} \left[ 1 + \cA ( \rnu \Delta^{(s)} )^{- 2} \rnu^2 - \frac{1}{4} \cA^2 ( \oDelta \Delta ) ( \rnu \Delta^{(s)} )^{- 4} ( \rnu \beta \rnu ) \right]^{- 1} \nonumber \\ & & \cdot \left[ 1 + \cA ( \rnu \Delta^{(s)} )^{- 2} \rnu^2 \right]^{- 1} \left\{ \eta^{\mu \lambda} - \cA ( \rnu \Delta^{(s)} )^{- 2} \left[ ( \oDelta \Delta ) \delta^\mu_\nu + \Delta^{(s) \mu} \Delta^{(s) }_\nu \right] \rbeta^{\nu \lambda} \right\} \rnu_\lambda \nonumber \\ & & \cdot \rnu_\tau \left\{ \eta^{\tau \sigma} - \cA ( \rnu \Delta^{(s)} )^{- 2} \beta^{\tau \rho} \left[ ( \oDelta \Delta ) \delta^\sigma_\rho + \Delta^{(s) }_\rho \Delta^{(s) \sigma} \right] \right\} ,
\end{eqnarray}

\noindent which defines $\mm$ (\ref{F[g]}). Substituting $J^{\nu \rho} = \Delta^{(s) \nu}_\perp \jmath^\rho + \Delta^{(s) \rho}_\perp \jmath^\nu$ into $F^\lambda = \rlambda^{\lambda \zeta } \rnu^\pi \cG^{(0)}_{\zeta \pi \nu \rho} J^{\nu \rho} = 2 \rlambda^{\lambda \zeta } \rnu^\pi \cG^{(0)}_{\zeta \pi \nu \rho} \Delta^{(s) \nu}_\perp \jmath^\rho$, we find $\rlambda^{\lambda \zeta } \rnu^\pi \cG^{(0)}_{\zeta \pi \nu \rho} \Delta^{(s) \nu}_\perp$, which defines $\overline{ \crO }^{ \lambda \mu }_\rho$. This gives for $ \crO^{ \lambda \mu }_\rho$:
\begin{eqnarray}\label{OatAnot0}                                           %79
& & \hspace{-5mm} \crO^{ \lambda \mu }_\rho = \delta_\rho^\lambda \rnu^\mu + \frac{ \varepsilon^2 }{ ( \Delta^{(s) 2}_\perp )^2 } \left[ 1 + \cA ( \rnu \Delta^{(s)} )^{- 2} \rnu^2 \right]^{- 1} \left\{ \cA ( \rnu \Delta^{(s)} )^{- 2} ( \Delta^{(s)}_{\perp \rho} \rnu_\sigma - \rnu_\rho \Delta^{(s)}_{\perp \sigma} ) - ( \rnu \Delta^{(s)} )^{- 1} \right. \nonumber \\ & & \left. \hspace{-5mm} \cdot ( \Delta^{(s) }_\rho \Delta^{(s)}_{\perp \sigma} + \Delta^{(s) 2}_\perp \eta_{\rho \sigma} ) + ( \rnu \Delta^{(s)} )^{- 2} \Delta^{(s) 2}_\perp \Delta^{(s) }_\rho \rnu_\sigma - \frac{1}{2} \cA ( \rnu \Delta^{(s)} )^{- 2} \left[ 1 + \cA ( \rnu \Delta^{(s)} )^{- 2} \rnu^2 \right. \right. \nonumber \\ & & \left. \left. \hspace{-5mm} - \frac{1}{4} \cA^2 ( \oDelta \Delta ) ( \rnu \Delta^{(s)} )^{- 4} ( \rnu \beta \rnu ) \right]^{- 1} \left\{ \left[ 1 + \cA ( \rnu \Delta^{(s)} )^{- 2} \rnu^2 + ( \rnu \Delta^{(s)} )^{- 2} \Delta^{(s) 2}_\perp \rnu^2 \right] \Delta^{(s)}_{\perp \rho} \right. \right. \nonumber \\ & & \left. \left. \hspace{-5mm} - \frac{1}{2} \cA ( \rnu \Delta^{(s)} )^{- 3} ( \oDelta \Delta ) \left[ \cA ( \rnu \Delta^{(s)} )^{- 1} ( \Delta^{(s)}_{\perp \rho} \rnu_\tau - \rnu_\rho \Delta^{(s)}_{\perp \tau} ) - \Delta^{(s) }_\rho \Delta^{(s)}_{\perp \tau} - \Delta^{(s) 2}_\perp \eta_{\rho \tau} + ( \rnu \Delta^{(s)} )^{- 1} \right. \right. \right. \nonumber \\ & & \left. \left. \left. \hspace{-5mm} \cdot \Delta^{(s) 2}_\perp \Delta^{(s) }_\rho \rnu_\tau \right] \rbeta^{\tau \nu} \rnu_\nu \right\} \rnu_\sigma \right\} \left\{ \eta^{\sigma \lambda} - \cA ( \rnu \Delta^{(s)} )^{- 2} \beta^{\sigma \chi} \left[ ( \oDelta \Delta ) \delta^\lambda_\chi + \Delta^{(s) }_\chi \Delta^{(s) \lambda} \right] \right\} \Delta^{(s) \mu }_\perp .
\end{eqnarray}

\noindent Symmetrization under the permutation $\lambda \leftrightarrow \mu$ is implied.

The $\cmM_{\lambda \tau}$ looks similar but more bulky. It contains $ \Delta_{ \perp}^{(s) \mu } \cG^{(0)}_{\lambda \mu \sigma \tau} \Delta_{ \perp}^{(s) \sigma }$ that follows from the found $h - \cA \eta^{\lambda \mu} w_{\lambda \mu}$, $\eta^{\lambda \mu} w_{\lambda \mu} + n^2 ( n \Delta^{(s) } )^{- 2} h$, $F^\lambda$, $r^\lambda$ (\ref{h-Aw}-\ref{r}) (at $n = \rnu$ and $J^{\sigma \tau} = \Delta^{(s) \sigma}_\perp \jmath^\tau + \Delta^{(s) \tau}_\perp \jmath^\sigma$) substituted to
\begin{eqnarray}                                                           %80
& & \hspace{-5mm} \Delta_{ \perp}^{(s) \mu } w_{\lambda \mu} = \Delta^{(s) \mu } w_{\lambda \mu} - (\rnu^2)^{- 1} ( \rnu \Delta^{(s )} ) \rnu^\mu w_{\lambda \mu} = \Delta^{(s )}_\lambda \eta^{\mu \nu} w_{\lambda \mu} + r_\lambda - (\rnu^2)^{- 1} ( \rnu \Delta^{(s )} ) \ralpha_{\lambda \mu} F^\mu \nonumber \\ & & \hspace{-5mm} = \left[ 1 + \cA ( \rnu \Delta^{(s )} )^{- 2} \rnu^2 \right]^{- 1} \Delta^{(s )}_\lambda \left\{ \left[ \eta^{\mu \nu} w_{\mu \nu} + \rnu^2 ( \rnu \Delta^{(s) } )^{- 2} h \right] - \rnu^2 ( \rnu \Delta^{(s) } )^{- 2} ( h - \cA \eta^{\mu \nu} w_{\mu \nu} ) \right\} \nonumber \\ & & \hspace{-5mm} + r_\lambda - (\rnu^2)^{- 1} ( \rnu \Delta^{(s )} ) \ralpha_{\lambda \mu} F^\mu
\end{eqnarray}

\noindent while $ \Delta_{ \perp}^{(s) \mu } w_{\lambda \mu} = 2 \Delta_{ \perp}^{(s) \mu } \cG^{(0)}_{\lambda \mu \sigma \tau} \Delta_{ \perp}^{(s) \sigma } \jmath^\tau$.

The found $\rnu^\mu \cG^{(0)}_{\lambda \mu \sigma \tau} \rnu^\sigma $, $\Delta_{ \perp}^{(s) \mu } \cG^{(0)}_{\lambda \mu \sigma \tau} \rnu^\sigma $, $\rnu^\mu \cG^{(0)}_{\lambda \mu \sigma \tau} \Delta_{ \perp}^{(s) \sigma }$ and $\Delta_{ \perp}^{(s) \mu } \cG^{(0)}_{\lambda \mu \sigma \tau} \Delta_{ \perp}^{(s) \sigma }$, defining $\mm$, $\crO$, $\ocrO$, $\cmM$, respectively, have such properties that, firstly, $\rnu \Delta^{( s )} = 0$ is no longer a singular point for $\mm$, $\crO$ and $\ocrO$ (at least with the exception of a set of measure zero in the three-dimensional space of $\Delta^{( s )}_{\perp \lambda}$) and $\cmM$ does not appear degenerate in it. Secondly, in these matrices for all five structures (listed, for example, in (\ref{ralpha})) a singular factor $\propto [ 1 + \cA ( \rnu \Delta^{(s )} )^{- 2} \rnu^2 ]^{- 1}$ arises. Thirdly, for all four structures except $P_{\lambda \tau}$, there is an additional singular factor $[ 1 + \cA ( \rnu \Delta^{(s)} )^{- 2} \rnu^2 - \frac{1}{4} \cA^2 ( \oDelta \Delta ) ( \rnu \Delta^{(s)} )^{- 4} ( \rnu \beta \rnu ) ]^{- 1}$.

In particular, instead of (\ref{(n,P,D)O}) we have for $\crO$ contracted with different structures near the singularities:
\begin{eqnarray}                                                           %81
& & \hspace{-10mm} ( \rnu^\rho, P^{ \kappa \rho }, \Delta^{( s ) \rho }_\perp ) {\mathrm O}^{\lambda \mu}_\rho = \left[ 1 + \cA ( \rnu \Delta^{(s )} )^{- 2} \rnu^2 \right]^{- 1} \nonumber \\ & & \hspace{-10mm} \cdot \left\{ \left[ 1 + \cA ( \rnu \Delta^{(s )} )^{- 2} \rnu^2 - \frac{1}{4} \cA^2 ( \oDelta \Delta ) ( \rnu \Delta^{(s)} )^{- 4} ( \rnu \beta \rnu ) \right]^{- 1} ( O (1), 0, O(1 ) ) + ( 0, O(1 ), 0 ) \right\},
\end{eqnarray}

\noindent and for $\cmM$ written as a combination of the aforementioned five structures,
\begin{eqnarray}                                                           %82
& & \cmM_{\lambda \mu} = \cimu_P P_{\lambda \mu} + \cimu_\rnu \frac{ \rnu_\lambda \rnu_\mu }{ \rnu^2 } + \cimu_\Delta \frac{ \Delta^{(s) }_{ \perp \lambda} \Delta^{(s) }_{ \perp \mu } }{ \Delta^{(s) 2 }_{ \perp } } + \cimu_{\rnu \Delta } \rnu_\lambda \Delta^{(s) }_{ \perp \mu } + \cimu_{ \Delta \rnu } \Delta^{(s) }_{ \perp \lambda } \rnu_\mu,
\end{eqnarray}

\noindent we have near the aforementioned singularities
\begin{eqnarray}                                                           %83
& & \hspace{-10mm} ( \cimu_P, \cimu_\rnu, \cimu_\Delta, \cimu_{\rnu \Delta }, \cimu_{ \Delta \rnu } ) = \left[ 1 + \cA ( \rnu \Delta^{(s )} )^{- 2} \rnu^2 \right]^{- 1} \nonumber \\ & & \hspace{-10mm} \cdot \left\{ \left[ 1 + \cA ( \rnu \Delta^{(s )} )^{- 2} \rnu^2 - \frac{1}{4} \cA^2 ( \oDelta \Delta ) ( \rnu \Delta^{(s)} )^{- 4} ( \rnu \beta \rnu ) \right]^{- 1} (0, O (1), O (1), O(1 ), O (1) ) \right. \nonumber \\ & & \left. + ( O(1 ), 0, 0, 0, 0 ) \vphantom{\frac{1}{4}} \right\}.
\end{eqnarray}

\noindent The expression $\Delta \ccM$ is bilinear in $\ocrO$, $\crO$ and may contain squares of singular factors, namely $[ 1 + \cA ( \rnu \Delta^{(s )} )^{- 2} \rnu^2 ]^{- 2}$ and $[ 1 + \cA ( \rnu \Delta^{(s)} )^{- 2} \rnu^2 - \frac{1}{4} \cA^2 ( \oDelta \Delta ) ( \rnu \Delta^{(s)} )^{- 4} ( \rnu \beta \rnu ) ]^{- 2}$. This would lead to singularities of constant sign of the type of $( \rnu \Delta^{( s )} - ( \rnu \Delta^{( s )} )_n )^{- 2}$, the integral of which diverges in the sense of the Cauchy principal value. But this singularity in $\ocrO \otimes \crO$ is partially cancelled by the smallness near the singularities of the coefficients at each structure over $\rho$, $\kappa$ in $( \cmM^{- 1} )^{\rho \kappa}$:
\begin{eqnarray}                                                           %84
& & \hspace{-10mm} ( \cmM^{- 1} )_{\lambda \mu} = \bcimu_P P_{\lambda \mu} + \bcimu_\rnu \frac{ \rnu_\lambda \rnu_\mu }{ \rnu^2 } + \bcimu_\Delta \frac{ \Delta^{(s) }_{ \perp \lambda} \Delta^{(s) }_{ \perp \mu } }{ \Delta^{(s) 2 }_{ \perp } } + \bcimu_{\rnu \Delta } \rnu_\lambda \Delta^{(s) }_{ \perp \mu } + \bcimu_{ \Delta \rnu } \Delta^{(s) }_{ \perp \lambda } \rnu_\mu, \nonumber \\
& & \hspace{-10mm} ( \bcimu_P, \bcimu_\rnu, \bcimu_\Delta, \bcimu_{\rnu \Delta }, \bcimu_{ \Delta \rnu } ) = \left[ 1 + \cA ( \rnu \Delta^{(s )} )^{- 2} \rnu^2 \right] \nonumber \\ & & \hspace{-10mm} \cdot \left\{ \left[ 1 + \cA ( \rnu \Delta^{(s )} )^{- 2} \rnu^2 - \frac{1}{4} \cA^2 ( \oDelta \Delta ) ( \rnu \Delta^{(s)} )^{- 4} ( \rnu \beta \rnu ) \right] (0, O (1), O (1), O(1 ), O (1) ) \right. \nonumber \\ & & \left. + ( O(1 ), 0, 0, 0, 0 ) \vphantom{\frac{1}{4}} \right\}.
\end{eqnarray}

\noindent As a result, we have the factors proportional to products of simple poles: $[ 1 + \cA ( \rnu \Delta^{(s )} )^{- 1} \rnu^2 ]^{- 1}$ and $[ 1 + \cA ( \rnu \Delta^{(s)} )^{- 2} \rnu^2 - \frac{1}{4} \cA^2 ( \oDelta \Delta ) ( \rnu \Delta^{(s)} )^{- 4} ( \rnu \beta \rnu ) ]^{- 1}$ in $\overline{ \crO } \cmM^{- 1} \crO$. The same singularities are present in $\mm$ (\ref{l+lnG(0)nl}).

The form of $\mathrm{det}_{\textstyle \cmM}$ generalizes the form of $\mathrm{det}_{\textstyle \mM}$, and its zeros are also typically simple roots. In overall, singularities appearing in $\Delta \ccM$ are usually of the type of simple poles $( \rnu \Delta^{( s )} - ( \rnu \Delta^{( s )} )_n )^{- 1}$, which are integrable in the Cauchy principal value sense.

The squares of the singular factors may a priori arise in a more symmetric case, for example in the limiting case $\cA \to 0$ ($( \rnu \Delta^{( s )} )^{- 2}$ and $( \rnu \Delta^{( s )} )^{- 4}$), but as we discussed in the previous Subsection \ref{A=0finite}, they also cancel out.

It is also important to note that we have taken $\rbeta = O ( ( \rnu \Delta^{( s )} )^2 )$ in the vicinity of $\rnu \Delta^{( s )} = 0$ given its origin from the matrix $\ralpha$ parameterizing the original $n^\lambda g_{\lambda \mu}$-gauge-fixing term, although nothing prevents us from taking $\rbeta$ as a more fundamental parameter and considering the more general case $\rbeta = O ( 1 )$. However, we then found a quadratic singularity $( \rnu \Delta^{( s )} )^{-2}$ in $\Delta \ccM$, whose interpretation in the sense of the Cauchy principal value does not give a finite answer.

\subsection{Electromagnetic illustration}\label{em_finite}

For the action
\begin{equation}                                                           %85
\cS_{\rm em} = \frac{1}{2} \sum_{\rm sites} A_\lambda \ccM^{\lambda \mu} A_\mu
\end{equation}

\noindent and the "soft" gauge-fixing term (\ref{Sem[AJ]}), the principal value type gauge-fixing term is described by the following correction to $\ccM$:
\begin{eqnarray}\label{ccM-em}                                             %86
& & \hspace{-10mm} \Delta \ccM^{\lambda \mu} = - \frac{1 }{ \varepsilon^2 } \overline{\widecheck{\mathrm O}^\lambda} \chM^{- 1} \crO^\mu - \varepsilon^2 \frac{ \Delta^{(s ) \lambda }_\perp }{ \Delta^{(s ) 2 }_\perp } \left( \rlambda + \rlambda^2 \rnu^\nu \cD^{(0 )}_{\nu \rho} \rnu^\rho \right) \frac{ \Delta^{(s ) \mu }_\perp }{ \Delta^{(s ) 2 }_\perp } , \nonumber \\ & & \hspace{-15mm} \cD^{(0 )}_{\lambda \mu} \stackrel{\rm def}{=} \cD_{\lambda \mu} ( \rnu , \rnu ) , \quad \crO^\lambda \stackrel{\rm def}{=} \rnu^\lambda + \varepsilon^2 \frac{ \Delta^{(s ) \sigma }_\perp }{ \Delta^{(s ) 2 }_\perp } \cD^{(0 )}_{\sigma \nu} \rnu^\nu \rlambda \frac{ \Delta^{(s ) \lambda }_\perp }{ \Delta^{(s ) 2 }_\perp } , \quad \chM \stackrel {\rm def}{=} \frac{ \ralpha }{ \varepsilon^2 } - \frac{ \Delta^{(s ) \sigma }_\perp }{ \Delta^{(s ) 2 }_\perp } \cD^{(0 )}_{\sigma \tau} \frac{ \Delta^{(s ) \tau }_\perp }{ \Delta^{(s ) 2 }_\perp } .
\end{eqnarray}

\noindent Using $\cD_{\lambda \mu} ( n, n )$ (\ref{cD(nn)}), we obtain the propagator-related quantities entering (\ref{ccM-em}):
\begin{eqnarray}                                                           %87
& & \crO^\lambda = \rnu^\lambda - \varepsilon^2 \frac{ \Delta^{(s ) \lambda }_\perp }{ \Delta^{(s ) 2 }_\perp } \frac{ \rnu \Delta^{(s )} }{ ( \rnu \Delta^{(s )} )^2 + \cA ( \rnu^2 + \ralpha \oDelta \Delta ) } , \quad \chM^{- 1} = \nonumber \\ & & = \frac{ \Delta^{(s ) 2 }_\perp [ ( \rnu \Delta^{(s )} )^2 + \cA ( \rnu^2 + \ralpha \oDelta \Delta ) ] }{ \displaystyle \frac{ \mathstrut \ralpha }{ \mathstrut \varepsilon^2 } \Delta^{(s ) 2 }_\perp [ ( \rnu \Delta^{(s )} )^2 + \cA ( \rnu^2 + \ralpha \oDelta \Delta ) + \varepsilon^2 ] - \rnu^2 + \ralpha \cA } , \nonumber \\ & & \rlambda + \rlambda^2 \rnu^\nu \cD^{(0 )}_{\nu \rho} \rnu^\rho = \frac{ \cA \oDelta \Delta }{ ( \rnu \Delta^{(s )} )^2 + \cA ( \rnu^2 + \ralpha \oDelta \Delta ) } .
\end{eqnarray}

\noindent Two operators $\overline{\widecheck{\mathrm O}^\lambda}$, $\crO^\mu$ can give a singularity squared, $[ ( \rnu \Delta^{(s )} )^2 + \cA ( \rnu^2 + \ralpha \oDelta \Delta ) ]^{- 2}$, but $\chM^{- 1}$ softens this singularity to a product of simple poles, which under the integral sign over $\d p_0$ can be considered as leading to a finite value in the sense of the Cauchy principal value.

In the limiting case $\cA \to 0$, the corresponding values are indicated without a check mark at the top, and we have $\Delta \cM^{\lambda \mu} \propto \overline{{\mathrm O}^\lambda} M^{- 1} \rO^\mu$. Two operators $\overline{{\mathrm O}^\lambda}$ and $\rO^\mu$ might give a singularity squared, $( \rnu \Delta^{(s )} )^{- 2}$, but $M^{- 1}$ cancels this singularity.

Strictly speaking, when passing to the limit $\cA \to 0$, it is physically justifiable to keep the actual $\cA$ in the denominators of the propagators; then, if the continuum counterpart of a considered diagram converges, then setting $\cA = 0$ in the nominators means omitting the terms of the non-leading order over typical variations of the external fields from site to site ($ \cA = O ( \Delta^4 ) $).

Therefore, we can consider some $D^{\rm eff} (n, n)$, which differs from $\cD (n, n)$ naively taken at $\cA = 0$ by scaling its $O ( \ralpha^0 ) = O ( 1 )$ part by a factor of $\Delta^{(s) 2} ( - \oDelta \Delta)^{- 1} $. The operator $\rO^\lambda$ is determined by the $O ( \ralpha )$ part of $D (\rnu, \rnu)$ and remains unchanged under this scaling. For $M$, we have
\begin{equation}\label{1/M_em}                                             %88
M^{- 1} = \frac{ \Delta^{ (s ) 2 }_\perp ( \rnu \Delta^{ (s ) } )^2 }{ \displaystyle \frac{\mathstrut \ralpha }{\mathstrut \varepsilon^2 } \Delta^{ (s ) 2 }_\perp [ ( \rnu \Delta^{ (s ) } )^2 + \varepsilon^2 ] + \rnu^2 \Delta^{(s ) 2 } ( \oDelta \Delta )^{- 1} } .
\end{equation}

\noindent with the same property to cancel the singularity squared $( \rnu \Delta^{(s )} )^{- 2}$,                        which is infinite in the sense of the Cauchy  principal value.

Thus, in this more simple system we are faced with the same mechanisms of providing finiteness of the principal value type gauge-fixing term as in gravity, and these mechanisms are different for $\cA \neq 0$ and limiting $\cA \to 0$ cases: in the $\cA \to 0$ case, possible singularities at $\rnu \Delta^{(s )} \to 0$ are cancelled, and in the $\cA \neq 0$ case, singularities (like those at $ ( \rnu \Delta^{(s )} )^2 + \cA ( \rnu^2 + \ralpha \oDelta \Delta ) \to 0 $) are not cancelled completely, but appear as products of simple poles admitting finite definition in the sense of the Cauchy  principal value (in the momentum representation).

\section{Ghost contribution}\label{ghost}

\subsection{General expression}\label{ghost-general}

Together with the gauge-fixing multiplier $\exp ( i \ccF [ g ] )$, we also introduce the corresponding normalization factor $\cPhi [ g ]$ under the functional integral sign to ensure separating out degrees of freedom close to the gauge degrees of freedom of the continuum theory, when the field/metric variations from 4-simplex to 4-simplex are small,
\begin{eqnarray}\label{1/Phi=int-exp-F}                                    %89
& & \cPhi [ g ]^{- 1} = \int \exp{( i \ccF [ g^\Xi ] )} \prod_{\rm sites} \d \Xi \nonumber \\ & & = \int \exp{ \sum_{\rm sites} \left( - \frac{i}{4} \cmf_\rho [ g^\Xi ] \cLambda^{\rho \kappa} \cmf_\kappa [ g^\Xi ] + \frac{i}{2} w^\Xi_{\lambda \mu} \mm^{\lambda \mu \sigma \tau} w^\Xi_{\sigma \tau} \right)} \prod_{\rm sites} \d \Xi , \quad \d \Xi = \prod_\lambda \d \xi^\lambda .
\end{eqnarray}

\noindent Here $\Xi$ is the group of diffeomorphisms or coordinate transformations $\delta x^\lambda = \xi^\lambda ( x )$. The finite-difference action is invariant with respect to $\Xi$ only in the leading order over metric variations from site to site. In this order we have
\begin{eqnarray}\label{gXi-g}                                           %90,91
& & \delta^\Xi g_{\lambda \mu} = g^\Xi_{\lambda \mu} - g_{\lambda \mu} = - \Delta^{(s) }_\mu \xi_\lambda - \Delta^{(s) }_\lambda \xi_\mu + 2 \Gamma^\nu_{\lambda \mu} \xi_\nu + O ( (\xi )^2 ), \nonumber \\ & & \Gamma^\nu_{\lambda \mu} = \frac{1}{2} g^{\nu \rho} ( \Delta^{(s )}_\mu g_{\rho \lambda} + \Delta^{(s )}_\lambda g_{\rho \mu} - \Delta^{(s )}_\rho g_{\lambda \mu} ) , \\ & & \label{fgXi=fg+xi} \cmf_\rho [ g^\Xi ] = \cmf_\rho [ g ] - \ccO_\rho{}^\nu \xi_\nu + O ( (\xi )^2 )
\end{eqnarray}

\noindent ($O ( (\xi )^2 )$ terms mean that $\xi$ is not necessarily infinitesimal), where
\begin{equation}\label{ccO=crOD}                                           %92
\ccO_\rho{}^\nu \xi_\nu \stackrel{\rm def }{ = } \crO^{\lambda \mu}_\rho \left( \Delta^{(s) }_\mu \xi_\lambda + \Delta^{(s) }_\lambda \xi_\mu - 2 \Gamma^\nu_{\lambda \mu} \xi_\nu \right) .
\end{equation}

In the functional integral, the gauge-fixing multiplier $\exp ( i \ccF [ g ] )$ provides configurations with $\cmf_\rho [ g ] = O ( \varepsilon )$ to dominate. In the integral over $\Xi$ (\ref{1/Phi=int-exp-F}), the configurations with $\cmf_\rho [ g^\Xi ] = O ( \varepsilon )$ dominate. This means that the typical values of $\xi_\lambda$ are $O ( \varepsilon )$ (from (\ref{fgXi=fg+xi})). Then the integral (\ref{1/Phi=int-exp-F}) is an integral of the exponential of the sum of $O(\varepsilon^0)$ terms (bilinear, linear and constant with respect to $\xi$) and higher orders in $\varepsilon$ (beginning from trilinear in $\xi$). This integral can be expanded into a sum of Gaussian integrals by expanding the exponential over $O(( \xi )^3)$ part. This can be viewed as an expansion over diagrams with the internal lines of the field $\xi$. The term with $\mm$ can also be considered as a correction $O ( \varepsilon^2 )$.

At this stage, we consider which powers of $\varepsilon$ in the diagram contributions to the effective ghost action $\cS_{\rm ghost} = - i \ln \cPhi$ can be significant in the limit $\varepsilon \to 0$. (The considered effective action turns out to be non-pole and imaginary, that is, means some real factor in the functional integral measure.) Ghost loop diagrams can provide an estimate of the contribution to the density of the effective ghost action (in the continuum case) or, in the considered discrete case, the contribution from a single site. To get a finite estimate of $\cS_{\rm ghost}$ in the whole spacetime, we need to put the system for intermediate regularization in a box with a large but finite number of sites along each coordinate.

In particular, let there be $N$ sites in the direction of time $x^0$. Finiteness of $N$ provides an additional IR regularization to the effect of nonzero $\varepsilon$. Since we are aiming to end up with the effect of $\varepsilon$, the effect of $N$ should be relatively small. The latter displays itself in the discrete spectrum of the quasi-momentum $p_0$ with a step $\sim N^{- 1}$. Integrals over $\d p_0$ in the expressions for diagrams are replaced by discrete sums. Due to the factors $(p_0 \pm i \varepsilon)^{- k}$ (at small $p_0$), $k \geq 1$, under the integral sign, these sums, in turn, can be approximated by the integrals, including at $\varepsilon \to 0$, only if the step $\sim N^{- 1}$ can be neglected in comparison with $\varepsilon$,
\begin{equation}\label{eN>>1}                                              %93
N^{- 1} \ll | \varepsilon |, ~~~ | \varepsilon | N \to \infty .
\end{equation}

\noindent An upper bound on $ | \cS_{\rm ghost} |$ from ghost loop diagrams is $\propto N$, and $ \cS_{\rm ghost} $ equal to $O( \varepsilon )$ is, due to (\ref{eN>>1}), insufficient to guarantee that $ \cS_{\rm ghost} $ disappears as $\varepsilon$ tends to 0. If, however, $ \cS_{\rm ghost} = O( \varepsilon^2 )$, then the requirement
\begin{equation}\label{eeN<<1}                                             %94
\varepsilon^2 N \to 0 .
\end{equation}

\noindent provides $ \cS_{\rm ghost} \to 0$. Both (\ref{eN>>1}) and (\ref{eeN<<1}) are fulfilled, for example, at $N \sim |\varepsilon|^{- 3 / 2}$.

Here we have considered a sufficient condition for the integral quantity $\cS_{\rm ghost}$ to vanish. Usually, however, in diagrammatic technique we are interested in local values such as amplitudes. These quantities are given directly by the values of the Feynman diagrams. Then the vanishing of the values of these diagrams is sufficient for these amplitudes to vanish, regardless of whether they tend to zero as $O ( \varepsilon^2 )$, $O ( \varepsilon )$, or something else.

Thus, an $O( \varepsilon^2 )$ contribution to $ \cS_{\rm ghost} $ can be disregarded at $\varepsilon \to 0$. Returning to the integral over $\Xi$ (\ref{1/Phi=int-exp-F}), we see that the term with $\mm$ being $O ( \varepsilon^2 )$ in the exponent can be neglected. We can then continue dealing with the expansion of this integral in terms of the non-Gaussian correction $O(( \xi )^3)$ (in the exponent), suppressed by powers of $\varepsilon$, and obtain for $\cPhi$ an expression proportional to $\Det (\occO \cmM^{- 1} \ccO )^{1 / 2}$ in the leading approximation in $\varepsilon$. Or, instead of expanding, we can initially consider the theory with the gauge-fixing term at $\mm = 0$.

That is, we can take the gauge-violating term as $\ccF = - \frac{1}{4} \sum_{\rm sites} \cmf_\rho \cLambda^{\rho \kappa} \cmf_\kappa$. In this case, we can act similarly to the standard way and consider the family of gauges
\begin{equation}                                                           %95
\cmf_\lambda [ g ] = \rkappa_\lambda
\end{equation}

\noindent parameterized by a vector function on sites $\rkappa_\lambda$. The functional integral in such a gauge follows by introducing the delta-function factor $\cPhi_0 [ g ] \prod_{\rm sites , \lambda} \delta (\cmf_\lambda [ g ] - \rkappa_\lambda )$ under the integral sign. Here $\cPhi_0$ is the normalization factor. Then we can perform exponential averaging of the functional integral over $\rkappa_\lambda$ with the exponential weight $\exp \left ( - \frac{ i }{ 4 } \sum_{\rm sites} \rkappa_\lambda \cLambda^{\lambda \mu} \rkappa_\mu \right )$. If we can confine ourselves to the leading order over metric variations, then adding functional-integral contributions from other simplicial structures will restore symmetry and independence of the functional integral from the non-invariant factor parameterized here by $\rkappa_\lambda$ on sites. Then the exponential averaging of the functional integral leaves it the same up to an inessential constant. On the other hand, the exponential averaging under the functional integral sign reproduces, by integrating the delta-functions, the gauge-violating term $ \ccF $ in the action. And the factor $\cPhi$ turns out to be just $\cPhi_0$, which, in turn, follows according to the standard procedure, subjecting $\cmf_\lambda [ g ]$ to a gauge (diffeomorphism) transformation, now infinitesimal,
\begin{equation}                                                           %96
\delta \cmf_\rho = - \ccO_\rho{}^\nu \xi_\nu , ~ \cPhi = \Det \ccO .
\end{equation}

\subsection{Ghost contribution in the limiting case \texorpdfstring{$\cA \to 0$}{A → 0}}\label{A=0ghost}

In the limiting case $\cA \to 0$, we take $G^{\rm eff}$ instead of $\cG$, for which ${\mathrm O}^{\lambda \mu}_\rho$ is the same as for $G$ (\ref{OforA=0}). This gives for the corresponding $\cO$ (from (\ref{ccO=crOD}) with no check mark)
\begin{eqnarray}\label{cO}                                                 %97
{\cal O}_\rho{}^\nu = \left( \rnu \Delta^{(s )} - \frac{ \varepsilon^2 }{ \rnu \Delta^{(s )} } \right) \delta_\rho^\nu + \left[ 1 + \frac{ \varepsilon^2 }{ (\rnu \Delta^{(s )} )^2 } \right] \Delta^{(s )}_\rho \rnu^\nu - \frac{ 2 \varepsilon^2 }{ (\rnu \Delta^{(s )}) \Delta^{(s) 2}_\perp } \Delta^{(s )}_\rho \Delta^{(s ) \nu}_\perp \nonumber \\ - 2 \rnu^\mu \Gamma^\nu_{\rho \mu} + \frac{ 2 \varepsilon^2 }{ (\rnu \Delta^{(s )}) \Delta^{(s) 2}_\perp } \Delta^{(s ) \mu}_\perp \Gamma^\nu_{\rho \mu} + \frac{ 2 \varepsilon^2 }{ (\rnu \Delta^{(s )}) \Delta^{(s) 2}_\perp } \Delta^{(s )}_\rho \left( \frac{ \Delta^{(s ) \lambda}_\perp }{ \Delta^{(s) 2}_\perp } - \frac{ \rnu^\lambda }{ \rnu \Delta^{(s )} } \right) \Delta^{(s ) \mu}_\perp \Gamma^\nu_{\lambda \mu}.
\end{eqnarray}

Here, we can single out the free part, $\cO_{(0 )} \stackrel{\rm def}{=} \cO |_{ \Gamma^\nu_{\lambda \mu} = 0 }$, whose inverse plays the role of a ghost propagator,
\begin{eqnarray}\label{1/O(0)}                                             %98
& & \hspace{-10mm} {\cal O}_{ ( 0 ) \nu}^{ - 1 }{}^\rho = \frac{ \rnu \Delta^{(s )} }{(\rnu \Delta^{(s )} )^2 - \varepsilon^2} \delta_\rnu{}^\rho - \frac{1}{2} \frac{ (\rnu \Delta^{(s )})^2 + \varepsilon^2 }{[ (\rnu \Delta^{(s )})^2 - \varepsilon^2 ]^2} \Delta^{(s )}_\nu \rnu^\rho + \frac{ \varepsilon^2 \rnu \Delta^{(s )} }{[ (\rnu \Delta^{(s )})^2 - \varepsilon^2 ]^2} \frac{ \Delta^{(s )}_\nu \Delta^{(s ) \rho}_\perp }{ \Delta^{(s) 2}_\perp }.
\end{eqnarray}

\noindent It can be noted that it is similar to the regularized one using the principal value prescription in the sense considered here.

The $\Det \cO$ of interest, up to a normalization constant, is equal to $\Det ( {\cal O}_{ ( 0 ) }^{ - 1 } {\cal O} )$. In ${\cal O}_{ ( 0 ) }^{ - 1 } {\cal O}$, we can omit terms that contribute $O( \varepsilon^2 )$ to $ S_{\rm ghost} $, but we do not set $\varepsilon^2$ equal to zero everywhere, in particular, leaving those $\varepsilon^2$ that provide regularization,
\begin{eqnarray}\label{O/O(0)}                                             %99
& & ( {\cal O}_{ ( 0 ) }^{ - 1 } {\cal O} )_\rho{}^\nu = \delta_\rho{}^\nu - \frac{2 (\rnu \Delta^{(s )})}{(\rnu \Delta^{(s )})^2 - \varepsilon^2} \rnu^\mu \Gamma^\nu_{\rho \mu} + \frac{ (\rnu \Delta^{(s )})^2 + \varepsilon^2 }{[ (\rnu \Delta^{(s )})^2 - \varepsilon^2 ]^2} \Delta^{(s )}_\rho \rnu^\lambda \rnu^\mu \Gamma^\nu_{\lambda \mu} + \dots .
\end{eqnarray}

\noindent Some fields entering here are small for $\varepsilon \to 0$:
\begin{equation}                                                          %100
\rnu^\lambda \rnu^\mu \Gamma^\nu_{\lambda \mu} = \rnu^\lambda \rnu^\mu g^{\nu \sigma} \left( \Delta^{(s )}_\mu g_{\sigma \lambda} - \frac{1}{2} \Delta^{(s )}_\sigma g_{\lambda \mu} \right) = O ( \varepsilon ) .
\end{equation}

\noindent This smallness is indirectly ensured by the gauge-fixing factor $\exp\{ - \frac{i}{4} \sum_{\rm sites} \mf_\rho [ g ] \Lambda^{\rho \kappa} \mf_\kappa [ g ] \}$ in the functional integral with $ \Lambda = \varepsilon^{-2} \mM^{- 1} $ providing $\rnu^\mu w_{\lambda \mu} = O ( \sqrt{ \varepsilon^2 } ) = O ( | \varepsilon | ) = O ( \varepsilon )$.

More precisely, this can be expressed as a typical value of the correlator of $\rnu^\mu w_{\lambda \mu}$ with any other metric component:
\begin{eqnarray}\label{nGeff}                                             %101
& & \rnu^\mu G^{\rm eff}_{\lambda \mu \sigma \tau} = \frac{1}{2} \rnu^\mu \left[ G^{\rm eff}_{\lambda \mu \sigma \tau} ( n, n ) + G^{\rm eff}_{\lambda \mu \sigma \tau} ( \on, \on ) \right] \nonumber \\ & & = \frac{1}{2} \left[ \left( n^\mu + \varepsilon \frac{\Delta^{(s ) \mu }_\perp}{\Delta^{(s ) 2 }_\perp} \right) G^{\rm eff}_{\lambda \mu \sigma \tau} ( n, n ) + \left( \on^\mu - \varepsilon \frac{\Delta^{(s ) \mu }_\perp}{\Delta^{(s ) 2 }_\perp} \right) G^{\rm eff}_{\lambda \mu \sigma \tau} ( \on, \on ) \right] = O ( \varepsilon ),
\end{eqnarray}

\noindent since $n^\mu G^{\rm eff}_{\lambda \mu \sigma \tau} ( n, n ) = n^\mu G_{\lambda \mu \sigma \tau} ( n, n ) = O ( \ralpha ) = O ( \varepsilon^2 )$ and the same for $n \Rightarrow \on$.

The estimate $\rnu^\mu w_{\lambda \mu} = O ( \varepsilon )$ also leads to the following partition of the field $\rnu^\mu \Gamma^\nu_{\rho \mu}$ into $O ( 1 )$ and $O ( \varepsilon )$ parts:
\begin{eqnarray}\label{O(1)+O(e)}                                         %102
& & 2 \rnu^\mu \Gamma^\nu_{\rho \mu} = g^{\nu \lambda} ( \rnu \Delta^{(s )} ) g_{\lambda \rho } + g^{\nu \lambda} ( \Delta^{(s )}_\rho g_{\lambda \mu } - \Delta^{(s )}_\lambda g_{\rho \mu } ) \rnu^\mu, \nonumber \\ & & g^{\nu \lambda} ( \Delta^{(s )}_\rho g_{\lambda \mu } - \Delta^{(s )}_\lambda g_{\rho \mu } ) \rnu^\mu = O ( \varepsilon ), \nonumber \\ & & g^{\nu \lambda} ( \rnu \Delta^{(s )} ) g_{\lambda \rho } = O ( 1 ) .
\end{eqnarray}

When expanding the effective ghost action $S_{\rm ghost} = - i \ln \Det \{ 1 + [ \cO_{ ( 0 ) }^{ - 1 } \cO - 1] \}$ over $\cO_{ ( 0 ) }^{ - 1 } \cO - 1$, we consider the possible $O ( \varepsilon )$ and $O ( 1 )$ contribution (the $O ( \varepsilon^2 )$ contribution, as we consider in the paragraph with equations (\ref{eN>>1}), (\ref{eeN<<1}), can be omitted at $\varepsilon \to 0$). First consider the terms (that is, the diagrams) with one $O ( \varepsilon )$ field $\rnu^\lambda \rnu^\mu \Gamma^\nu_{\lambda \mu}$ and an arbitrary number $n - 1$ of $O ( 1 )$ fields $g^{\nu \lambda} ( \rnu \Delta^{(s )} ) g_{\lambda \rho }$. These fields enter such a term as
\begin{eqnarray}\label{nnG}                                               %103
& & [ ( \rnu \Delta^{(s )} ) g_{\nu_n \lambda_n } ] g^{ \lambda_n \nu_{n - 1}} \otimes \dots \otimes [ ( \rnu \Delta^{(s )} ) g_{\nu_j \lambda_j } ] g^{ \lambda_j \nu_{ j - 1}} \otimes \dots \otimes [ ( \rnu \Delta^{(s )} ) g_{\nu_3 \lambda_3 } ] g^{ \lambda_3 \nu_2} \nonumber \\ & & \otimes [ ( \rnu \Delta^{(s )} ) g_{\nu_2 \lambda_2 } ] g^{ \lambda_2 \nu_1} \otimes \Delta^{(s )}_{\nu_1} \Gamma^{\nu_n}_{\lambda_1 \mu} \rnu^{\lambda_1} \rnu^{\mu} , \quad \Delta^{(s )}_{\nu_1} \equiv \Delta^{(s )}_{\perp \nu_1} + ( \rnu^2 )^{- 1} ( \rnu \Delta^{(s )} ) \rnu_{\nu_1} .
\end{eqnarray}

\noindent The factors in this tensor product are generally taken at different sites (with different coordinates $x^0$). For the product of the first $n-1$ factors to be $O ( 1 )$, the indices $\rnu_j$ should be nonzero. In particular, $\rnu_1 \neq 0$, and therefore $\Delta^{(s )}_{\nu_1} = \Delta^{(s )}_{\perp \nu_1}$. This is the only dependence on $\Delta^{(s )}_{\perp }$ in this term, and integration over $\d^3 p_{\perp }$ gives zero for such a contribution to $S_{\rm ghost}$. Thus, the considered diagram is in fact $O( \varepsilon^2 )$.

Then consider the contribution of the terms with all possible numbers of the field $g^{\nu \lambda} ( \rnu \Delta^{(s )} ) g_{\lambda \rho }$ and at least one field $g^{\nu \lambda} ( \Delta^{(s )}_\rho g_{\lambda \mu } - \Delta^{(s )}_\lambda g_{\rho \mu } ) \rnu^\mu$. Leaving only these fields in (\ref{O/O(0)}), we write for $ ( \cO_{ ( 0 ) }^{ - 1 } \cO \| g \| )_{\rho \nu}$ ($\| g \|$ is the metric matrix):
\begin{eqnarray}                                                          %104
& & ( {\cal O}_{ ( 0 ) }^{ - 1 } {\cal O} \| g \| )_{\rho \nu} = s_{\rho \nu} + a_{\rho \nu} + \dots , \nonumber \\ & & s_{\rho \nu} = g_{\rho \nu} - \frac{ \rnu \Delta^{(s )} }{( \rnu \Delta^{(s )} )^2 - \varepsilon^2 } [ ( \rnu \Delta^{(s )} ) g_{\rho \nu } ], \nonumber \\ & & a_{\rho \nu} = \frac{ ( \rnu \Delta^{(s )} ) \rnu^\mu }{( \rnu \Delta^{(s )} )^2 - \varepsilon^2 } [ ( \Delta^{(s )}_\nu g_{\rho \mu } ) - ( \Delta^{(s )}_\rho g_{\nu \mu } ) ].
\end{eqnarray}

\noindent Then we find
\begin{eqnarray}                                                          %105
\ln \left[ \Det ( \cO_{ ( 0 ) }^{ - 1 } \cO ) \prod_{\rm sites} \det \| g \| \right] = \Tr \ln s + \Tr \left( s^{- 1} a \right) - \frac{1}{2} \Tr \left( s^{- 1} a s^{- 1} a \right) + \dots .
\end{eqnarray}

\noindent Since $\Tr \left( s^{- 1} a \right) = 0$, the contribution of the terms with instances of the field $g^{\nu \lambda} ( \Delta^{(s )}_\rho g_{\lambda \mu } - \Delta^{(s )}_\lambda g_{\rho \mu } ) \rnu^\mu$ under consideration is actually equal to $O ( \varepsilon^2 )$.

The only remaining terms a priori larger than $O ( \varepsilon^2 ) $ are those with solely instances of the field $g^{\nu \lambda} ( \rnu \Delta^{(s )} ) g_{\lambda \rho } = O ( 1 )$. Singling out the total contribution of these terms, we have:
\begin{eqnarray}\label{lnDetO/O(0)}                                       %106
& & \hspace{-10mm} \ln \Det ( {\cal O}_{ ( 0 ) }^{ - 1 } {\cal O} ) = \ln \Det ( s \| g \|^{- 1} ) + \dots \nonumber \\ & & \hspace{-10mm} = \ln \Det \left\{ (\rnu \Delta^{(s )}) \rnu^\mu \left[ \Delta^{(s )}_\mu g_{\rho \lambda} - ( \Delta^{(s )}_\mu g_{\rho \lambda} ) \right] g^{\lambda \nu} - \varepsilon^2 \delta^\nu_\rho \right\} - \ln \Det \left[ (\rnu \Delta^{(s )})^2 - \varepsilon^2 \right] + \dots .
\end{eqnarray}

\noindent Here $ \Delta^{(s )}_\mu g_{\rho \lambda} - ( \Delta^{(s )}_\mu g_{\rho \lambda} )$ would be $ g_{\rho \lambda} \Delta^{(s )}_\mu$ in the continuum limit, but since the Leibnitz rule for differentiating a product is violated for finite differences, the commutator of $\Delta^{(s )}_\mu$ and $g_{\rho \lambda}$ is $( \Delta^{(s )}_\mu g_{\rho \lambda} )$ only up to corrections of higher order in $\Delta$,
\begin{equation}                                                          %107
\left[\Delta^{(s )}_\mu, g_{\rho \lambda} \right] = ( \Delta^{(s )}_\mu g_{\rho \lambda} ) - \frac{1}{2} \oDelta_\mu ( \Delta_\mu g_{\rho \lambda} ) \Delta_\mu ,
\end{equation}

\noindent that is,
\begin{equation}                                                          %108
\Delta^{(s )}_\mu g_{\rho \lambda} - ( \Delta^{(s )}_\mu g_{\rho \lambda} ) = g_{\rho \lambda} \Delta^{(s )}_\mu - \frac{1}{2} \oDelta_\mu ( \Delta_\mu g_{\rho \lambda} ) \Delta_\mu .
\end{equation}

\noindent Thus, we obtain
\begin{eqnarray}\label{lnDetO/O(0)O(D4)}                                  %109
& & \ln \Det ( {\cal O}_{ ( 0 ) }^{ - 1 } {\cal O} ) = \ln \frac{ \Det \left[ ( \rnu \Delta^{(s )} ) \| g \| ( \rnu \Delta^{(s )} ) \| g \|^{- 1} + O ( ( \Delta )^4 ) - \varepsilon^2 \right] }{ \Det \left[ ( \rnu \Delta^{(s )} )^2 - \varepsilon^2 \right] } + \dots .
\end{eqnarray}

\noindent Here we can neglect the term $O ( ( \Delta )^4 )$ for small metric variations from site to site, but not in the general case. On the other hand, its preservation can be considered as an excess of calculation accuracy. Indeed, its form is defined by the (approximate) diffeomorphism variation $g^\Xi_{\lambda \mu} - g_{\lambda \mu}$ (\ref{gXi-g}), which is fixed only in the leading order over metric variations from site to site. Namely, $( \Delta^{(s )}_\mu g_{\rho \lambda} ) g^{\lambda \nu}$ in (\ref{lnDetO/O(0)}) is a part of $\Gamma^\nu_{\rho \mu}$ entering $g^\Xi_{\rho \mu} - g_{\rho \mu}$ as $\Gamma^\nu_{\rho \mu} \xi_\nu$. Therefore, we can improve the accuracy of the formula for $g^\Xi_{\lambda \mu} - g_{\lambda \mu}$ (\ref{gXi-g}) so as to eliminate the terms $O ( ( \Delta )^4 )$ from (\ref{lnDetO/O(0)O(D4)}). Such an improvement of the transformation formula and the related revision of the ghost contribution are discussed below in Subsection \ref{diff-improve}. Here we will consider the result of such a refinement - formula (\ref{lnDetO/O(0)O(D4)}) without terms of order $O ( ( \Delta )^4 )$.

Thus, we need to analyze $\Det \left[ ( \rnu \Delta^{(s )} ) \| g \| ( \rnu \Delta^{(s )} ) \| g \|^{- 1} - \varepsilon^2 \right]$. Remind that, as we discussed in Subsection \ref{ghost-general} (the paragraph with formulas (\ref{eN>>1}), (\ref{eeN<<1}) and the previous one), at an intermediate stage we place the system in a box. The size of this box along the direction of time $x^0$ should tend to infinity at $\varepsilon \to 0$, so that the discretization step of the quasi-momentum $p_0$ will be negligible compared to $\varepsilon$, and the discrete sums over $p_0$ in the expressions for diagrams can be approximated by the integrals. Another way to interpret this is the boundary effect, which in the definition of $(\rnu \Delta^{(s )} - \varepsilon)^{- 1}$, where $\rnu \Delta^{(s )} - \varepsilon$ is a $2N \times 2N$ matrix, is proportional to $\exp ( - \varepsilon N )$ and disappears at $\varepsilon N \to \infty$. At the same time, the expansion of $\Det \left[ ( \rnu \Delta^{(s )} ) \| g \| ( \rnu \Delta^{(s )} ) \| g \|^{- 1} - \varepsilon^2 \right]$ over $\varepsilon$ goes over the effective parameter $\varepsilon^2 N$. As $\varepsilon^2 N$ goes to zero, $\varepsilon$ should be omitted here, $\Det$ factorizes, and $\ln \Det ( {\cal O}_{ ( 0 ) }^{ - 1 } {\cal O} ) \to 0$.

\subsection{Ghost contribution at \texorpdfstring{$\cA \neq 0$}{A ̸= 0}}\label{Aneq0ghost}

For $\cA \neq 0$, the changes concern mainly the resolution of singularities in the expressions of interest to us, similar to the discussion of the finiteness of the gauge-fixing term in Subsection (\ref{finiteAnot0}). We consider the ghost propagator $\ccO_{(0 )}^{- 1}$.
\begin{eqnarray}\label{calO=structures}                                   %110
& & \ccO_{(0 ) \rho}{}^\nu \xi_\nu \stackrel{\rm def}{=} \crO^{\lambda \mu}_\rho \left( \Delta^{(s )}_\mu \xi_\lambda + \Delta^{(s )}_\lambda \xi_\mu \right); \nonumber \\ & & \ccO_{(0 ) \lambda \mu} \stackrel{\rm def}{=} \co_P P_{\lambda \mu} + \co_\rnu \frac{ \rnu_\lambda \rnu_\mu }{ \rnu^2 } + \co_\Delta \frac{ \Delta^{(s) }_{ \perp \lambda} \Delta^{(s) }_{ \perp \mu } }{ \Delta^{(s) 2 }_{ \perp } } + \co_{\rnu \Delta } \rnu_\lambda \Delta^{(s) }_{ \perp \mu } + \co_{ \Delta \rnu } \Delta^{(s) }_{ \perp \lambda } \rnu_\mu.
\end{eqnarray}

\noindent Using $\crO^{\lambda \mu}_\rho$ (\ref{OatAnot0}), we obtain the coefficients of the five structures in $\ccO_{(0 ) }$ (\ref{calO=structures}). We do not present the dependence on $\rbeta = O ( \varepsilon^2 )$ due to its cumbersomeness, assuming $\rbeta$ to be equal to zero, which means omitting the terms $O ( \varepsilon^2 \rbeta ) = O ( \varepsilon^4 )$ in the coefficients; only $\co_P$ will be presented exactly for clarity.
\begin{eqnarray}\label{checked_o}                                         %111
& & \hspace{-10mm} \co_P = ( \rnu \Delta^{(s )} ) \left[ 1 - \varepsilon^2 \frac{ 1 - \cA ( \rnu \Delta^{(s )} )^{- 2} \rbeta_P \oDelta \Delta }{ ( \rnu \Delta^{(s )} )^2 + \cA \rnu^2 } \right] , ~ \co_\rnu = 2 ( \rnu \Delta^{(s )} ), \nonumber \\ & & \hspace{-10mm} \co_\Delta = ( \rnu \Delta^{(s )} ) + \frac{ \varepsilon^2 ( \rnu \Delta^{(s )} ) }{ ( \rnu \Delta^{(s )} )^2 + \cA \rnu^2 } \left[ - 3 + \frac{1}{2} \frac{ \cA }{ \Delta^{(s ) 2}_\perp } - \frac{1}{2} \frac{ \cA \rnu^2 }{ ( \rnu \Delta^{(s )} )^2 + \cA \rnu^2 } \right] , \nonumber \\ & & \hspace{-10mm} \co_{\rnu \Delta } = - 2 \frac{ \varepsilon^2 }{ \rnu^2 \Delta^{(s ) 2}_\perp } , ~ \co_{ \Delta \rnu } = 1 + \frac{ \varepsilon^2 }{ ( \rnu \Delta^{(s )} )^2 + \cA \rnu^2 } \left[ 1 + \frac{1}{2} \frac{ \cA }{ \Delta^{(s ) 2}_\perp } - \frac{1}{2} \frac{ \cA \rnu^2 }{ ( \rnu \Delta^{(s )} )^2 + \cA \rnu^2 } \right] .
\end{eqnarray}

\noindent The reciprocal matrix reads
\begin{eqnarray}\label{1/checked_o}                                       %112
& & ( \ccO_{(0 )}^{- 1} )_{\lambda \mu} = \bco_P P_{\lambda \mu} + \bco_\rnu \frac{ \rnu_\lambda \rnu_\mu }{ \rnu^2 } + \bco_\Delta \frac{ \Delta^{(s) }_{ \perp \lambda} \Delta^{(s) }_{ \perp \mu } }{ \Delta^{(s) 2 }_{ \perp } } + \bco_{\rnu \Delta } \rnu_\lambda \Delta^{(s) }_{ \perp \mu } + \bco_{ \Delta \rnu } \Delta^{(s) }_{ \perp \lambda } \rnu_\mu , \nonumber \\ & & \bco_P = \frac{1}{ \co_P }, ~ \bco_\rnu = \frac{ \co_\Delta }{ \mathrm{det}_\ccO }, ~ \bco_\Delta = \frac{ \co_\rnu }{ \mathrm{det}_\ccO }, ~ \bco_\rnu = \frac{ \co_\Delta }{ \mathrm{det}_\ccO }, ~ \bco_{\rnu \Delta} = \frac{ - \co_{\rnu \Delta} }{ \mathrm{det}_\ccO }, ~ \bco_{\Delta \rnu} = \frac{ - \co_{\Delta \rnu} }{ \mathrm{det}_\ccO }, \nonumber \\ & & \mathrm{det}_\ccO = \co_\rnu \co_\Delta - \rnu^2 \Delta^{(s) 2 }_{ \perp } \co_{\rnu \Delta} \co_{\Delta \rnu } \nonumber \\ & & = 2 ( \rnu \Delta^{(s )} )^2 \left\{ 1 + \frac{\varepsilon^2 }{ ( \rnu \Delta^{(s )} )^2 + \cA \rnu^2 } \left[ - 3 + \frac{1}{2} \frac{ \cA }{ \Delta^{(s ) 2}_\perp } - \frac{1}{2} \frac{ \cA \rnu^2 }{ ( \rnu \Delta^{(s )} )^2 + \cA \rnu^2 } \right] \right\} \nonumber \\ & & + 2 \varepsilon^2 \left\{ 1 + \frac{\varepsilon^2 }{ ( \rnu \Delta^{(s )} )^2 + \cA \rnu^2 } \left[ 1 + \frac{1}{2} \frac{ \cA }{ \Delta^{(s ) 2}_\perp } - \frac{1}{2} \frac{ \cA \rnu^2 }{ ( \rnu \Delta^{(s )} )^2 + \cA \rnu^2 } \right] \right\} .
\end{eqnarray}

In $\ccO_{ ( 0 ) }^{ - 1 } \ccO$ we omit the terms that contribute $O( \varepsilon^2 )$ to $ S_{\rm ghost} $,
\begin{eqnarray}\label{O/O(0)Aneq00}                                      %113
& & ( \ccO_{ ( 0 ) }^{ - 1 } \ccO )_\rho{}^\nu = \delta_\rho{}^\nu - 2 (\ccO_{ ( 0 ) }^{ - 1 })_\rho{}^\lambda \rnu^\mu \Gamma^\nu_{\lambda \mu} = \delta_\rho{}^\nu - 2 \left\{ \frac{1}{ \co_P } P_\rho{}^\lambda \right. \nonumber \\ & & \left. + \frac{1}{ \mathrm{det}_\ccO } \left[ ( \rnu \Delta^{(s )} ) \frac{ \rnu_\rho \rnu^\lambda }{ \rnu^2 } + 2 ( \rnu \Delta^{(s )} ) \frac{ \Delta^{(s )}_{\perp \rho} \Delta^{(s ) \lambda }_\perp }{ \Delta^{(s ) 2 }_\perp } - \Delta^{(s )}_{\perp \rho} \rnu^\lambda \vphantom{ \frac{ \rnu_\rho \rnu^\lambda }{ \rnu^2 } } \right] \right\} \rnu^\mu \Gamma_{\lambda \mu}^\nu .
\end{eqnarray}

\noindent Now, when $\cA \neq 0$, $\co_P$ and $\mathrm{det}_\ccO$ pass through zero at some real $p_0$ ($\mathrm{det}_\ccO$ at $\sin p_0 \approx \pm \varepsilon$, and $\co_P$ at $p_0 = 0$), and here we can speak of the principal value integrability over $p_0$ not in the sense of the substitution $( \rnu \Delta^{(s )} )^{- j} \Rightarrow \frac{1}{2} \{ [ ( \rnu \Delta^{(s )} ) + \varepsilon ]^{- j} + [ ( \rnu \Delta^{(s )} ) - \varepsilon ]^{- j} \}$, as in the ghost diagrams in the limiting case $\cA \to 0$ or in the graviton diagrams, but just in the Cauchy sense, which means that the integrand vanishes in the interval $( p_{0 0} - \delta , p_{0 0} + \delta )$ around the difficult point $p_{0 0}$ in the limit $\delta \to 0$. In particular,
\begin{equation}                                                          %114
\mathrm{p.v.} \int^\pi_{- \pi} \frac{ \d p_0 }{ \sin p_0 - c } = 0, \quad -1 < c < 1 .
\end{equation}

\noindent The important thing is that such a definition gives finite values, which are then multiplied by coefficients that tend to zero as $\varepsilon \to 0$.

In $\ccO_{ ( 0 ) }^{ - 1 } \ccO$, keeping $\varepsilon$ in the denominators is required to avoid the occurrence of squares of singular factors. But related terms can be “reduced to a common denominator” with an accuracy of $O ( \varepsilon^2 )$ using simple relations:
\begin{eqnarray}                                                          %115
& & \frac{1}{ \mathrm{det}_\ccO } - \frac{1}{ 2 [ ( \rnu \Delta^{(s )} )^2 + \varepsilon^2 ] } = \frac{ 2 [ ( \rnu \Delta^{(s )} )^2 + \varepsilon^2 ] - \mathrm{det}_\ccO }{ 2 [ ( \rnu \Delta^{(s )} )^2 + \varepsilon^2 ] \mathrm{det}_\ccO } = O ( \varepsilon^2 ) , \nonumber \\ & & \frac{1}{ \co_P } - \frac{ \rnu \Delta^{(s )} }{ ( \rnu \Delta^{(s )} )^2 + \varepsilon^2 } = \frac{ ( \rnu \Delta^{(s )} )^2 + \varepsilon^2 - ( \rnu \Delta^{(s )} ) \co_P }{ [ ( \rnu \Delta^{(s )} )^2 + \varepsilon^2 ] \co_P } = O ( \varepsilon^2 ) , \nonumber \\ & & \frac{ ( \rnu \Delta^{(s )} )^2 }{ ( \rnu \Delta^{(s )} )^2 + \varepsilon^2 } = 1 + O ( \varepsilon^2 ) .
\end{eqnarray}

\noindent As a result, we obtain
\begin{eqnarray}\label{O/O(0)Aneq0}                                       %116
& & ( \ccO_{ ( 0 ) }^{ - 1 } \ccO )_\rho{}^\nu = \frac{1}{ ( \rnu \Delta^{(s )} )^2 + \varepsilon^2 } \left[ ( \rnu \Delta^{(s )} )^2 \delta_\rho{}^\nu - 2 ( \rnu \Delta^{(s )} ) \rnu^\mu \Gamma_{\rho \mu}^\nu + \Delta^{(s )}_\rho \rnu^\lambda \rnu^\mu \Gamma_{\lambda \mu}^\nu \right] .
\end{eqnarray}

The further consideration repeats the analysis of the contributions from $O ( 1 )$ and $O ( \varepsilon )$ terms in ${\cal O}_{ ( 0 ) }^{ - 1 } {\cal O}$ (\ref{O/O(0)}) with the similar formula (\ref{O/O(0)Aneq0}). In this case, the estimate of the order of magnitude of the field $\rnu^\mu w_{\lambda \mu}$ with respect to $\varepsilon$ is defined as typical value of its correlator with any other component of the metric, $\rnu^\mu \cG_{\lambda \mu \sigma \tau}$, where $\cG$ is the half-sum of $\cG (n, n)$ and $\cG (\on, \on)$, and it is equal to $O ( \varepsilon )$, as in the estimate of $\rnu^\mu G^{\rm eff}_{\lambda \mu \sigma \tau}$ (\ref{nGeff}), since $n^\mu \cG_{\lambda \mu \sigma \tau} ( n, n ) = O ( \ralpha ) = O ( \varepsilon^2 )$ and the same for $n \Rightarrow \on$, which can be obtained using the expansion of $\cG (n, n)$ ($\cG (\on, \on)$) in powers of $\cA$ with respect to $G (n, n)$ ($G (\on, \on)$), like the expansion of $\cG (\rnu, \rnu)$ in powers of $\cA$ with respect to $G (\rnu, \rnu)$ (\ref{ntGn=nGn+}). Up to dependence on $\varepsilon^2$ we obtain (\ref{lnDetO/O(0)O(D4)}), where the dependence on the metric appears in the form $( \rnu \Delta^{(s )} ) \| g \| ( \rnu \Delta^{(s )} ) \| g \|^{- 1} + O ( ( \Delta )^4 )$. As considered at the end of Subsection \ref{A=0ghost}, the term $O ( ( \Delta )^4 )$ can either be neglected for small metric variations from site to site or eliminated at all if we improve the accuracy of the formula for $g^\Xi_{\lambda \mu} - g_{\lambda \mu}$ (\ref{gXi-g}) and specify it in a certain way at non-leading orders, as we consider in Subsection \ref{diff-improve}. As a result, we obtain the disappearance of $\ln \Det ( \ccO_{ ( 0 ) }^{ - 1 } \ccO )$ in the limit $\varepsilon \to 0$.

\subsection{Refinement of the approximate diffeomorphism formula}\label{diff-improve}

To eliminate the terms $O ( ( \Delta )^4 )$ from (\ref{lnDetO/O(0)O(D4)}), it is sufficient to replace $( \Delta^{(s )}_\mu g_{\rho \lambda} )$ entering $\Gamma_{\lambda , \rho \mu }$ with the commutator $ \Delta^{(s )}_\mu g_{\rho \lambda} - g_{\rho \lambda} \Delta^{(s )}_\mu$. To preserve symmetry between the coordinates, this substitution must be performed for any finite difference $\Delta^{(s )}_\mu$, not just for $\rnu \Delta^{(s )}$. Then $\Gamma^\nu_{\rho \mu}$ becomes an operator depending on finite differences, not simply a function,
\begin{eqnarray}\label{GammaGamma}                                        %117
& &  g^\Xi_{\lambda \mu} - g_{\lambda \mu} = - \Delta^{(s) }_\mu \xi_\lambda - \Delta^{(s) }_\lambda \xi_\mu + 2 \Gamma \! \! \! \Gamma^\nu_{\lambda \mu} \xi_\nu , \nonumber \\ & & \Gamma \! \! \! \Gamma^\nu_{\rho \mu} \stackrel{\rm def}{=} \frac{1}{2} [\mathfrak{D}_\mu ( g_{\rho \lambda} ) + \mathfrak{D}_\rho ( g_{\lambda \mu} ) - \mathfrak{D}_\lambda ( g_{ \rho \mu} ) ] g^{\lambda \nu} , \quad \mathfrak{D}_\mu ( g_{\rho \lambda} ) \stackrel{\rm def}{=} \Delta^{(s )}_\mu g_{\rho \lambda} - g_{\rho \lambda} \Delta^{(s )}_\mu .
\end{eqnarray}

\noindent Here $\mathfrak{D}_\mu ( g_{\rho \lambda} )$ is not simply a function, but an operator depending on $\Delta_\mu$. This refined $\Gamma \! \! \! \Gamma$ should be substituted into $\cO$ (\ref{cO}), $\cO_{ ( 0 ) }^{ - 1 } \cO$ (\ref{O/O(0)}) (or (\ref{O/O(0)Aneq00}), (\ref{O/O(0)Aneq0}) for $\cA \neq 0$) instead of $\Gamma$. The above conclusions on the contribution to $ S_{\rm ghost} $ (or $ \cS_{\rm ghost} $) from the fields $\rnu^\mu \Gamma_{\rho \mu}^\nu$ and $\rnu^\lambda \rnu^\mu \Gamma_{\lambda \mu}^\nu$ can be repeated for the fields $\rnu^\mu \Gamma \! \! \! \Gamma_{\rho \mu}^\nu$ and $\rnu^\lambda \rnu^\mu \Gamma \! \! \! \Gamma_{\lambda \mu}^\nu$, only the estimate for the case of $\rnu^\lambda \rnu^\mu \Gamma \! \! \! \Gamma_{\lambda \mu}^\nu$ should be made more accurately. Namely, we have
\begin{equation}\label{nG!!!G}                                            %118
\rnu^\lambda \rnu^\mu \Gamma \! \! \! \Gamma_{\lambda \mu}^\nu = \left[ \rnu^\lambda \rnu^\mu \mD_\lambda ( g_{\mu \rho} ) - \frac{1}{2} \rnu^\lambda \rnu^\mu \mD_\rho ( g_{\lambda \mu} ) \right] g^{\rho \nu}.
\end{equation}

\noindent Here we must proceed more carefully, since the second term in square brackets depends on $\Delta_\rho$, including $\Delta_{\perp \rho}$. However, it can be neglected since $\rnu^\lambda \rnu^\mu w_{\lambda \mu} = O ( \varepsilon^2 )$ in the sense that its correlator with any other metric component is $O ( \varepsilon^2 )$:
\begin{eqnarray}\label{nnGeff}                                            %119
& & 2 \rnu^\lambda \rnu^\mu G^{\rm eff}_{\lambda \mu \sigma \tau} = \left( n^\lambda + \varepsilon \frac{\Delta^{(s ) \lambda }_\perp}{\Delta^{(s ) 2 }_\perp} \right) \left( n^\mu + \varepsilon \frac{\Delta^{(s ) \mu }_\perp}{\Delta^{(s ) 2 }_\perp} \right) G^{\rm eff}_{\lambda \mu \sigma \tau} ( n, n ) \nonumber \\ & & + \left( \on^\mu - \varepsilon \frac{\Delta^{(s ) \mu }_\perp}{\Delta^{(s ) 2 }_\perp} \right) \left( \on^\lambda - \varepsilon \frac{\Delta^{(s ) \lambda }_\perp}{\Delta^{(s ) 2 }_\perp} \right) G^{\rm eff}_{\lambda \mu \sigma \tau} ( \on, \on ) = O ( \varepsilon^2 ),
\end{eqnarray}

\noindent since $n^\mu G^{\rm eff}_{\lambda \mu \sigma \tau} ( n, n ) = n^\mu G_{\lambda \mu \sigma \tau} ( n, n ) = O ( \ralpha ) = O ( \varepsilon^2 )$ and the same for $n \Rightarrow \on$. For $\cA \neq 0$, we replace $G^{\rm eff}$ in (\ref{nnGeff}) by $\cG$ and take into account that $n^\mu \cG_{\lambda \mu \sigma \tau} ( n, n ) = O ( \ralpha ) = O ( \varepsilon^2 )$ and the same for $n \Rightarrow \on$, which can be obtained using the expansion of $\cG (n, n)$ ($\cG (\on, \on)$) in powers of $\cA$ with respect to $G (n, n)$ ($G (\on, \on)$), like the expansion of $\cG (\rnu, \rnu)$ in powers of $\cA$ with respect to $G (\rnu, \rnu)$ (\ref{ntGn=nGn+}).

The first term in square brackets in (\ref{nG!!!G}) is $O ( \varepsilon )$, but depends on $\rnu \Delta$, not on $\Delta_\perp$. Thus, as for $\rnu^\lambda \rnu^\mu \Gamma_{\lambda \mu}^\nu$ in (\ref{nnG}), the contribution of such a diagram with the field $\rnu^\lambda \rnu^\mu \Gamma \! \! \! \Gamma_{\lambda \mu}^\nu$ is $O ( \varepsilon^2 )$.

Thus, the absence of the ghost contribution in the limit $\varepsilon \to 0$ also takes place in this case.

\subsection{Generalization to the gauge-fixing term bilinear in \texorpdfstring{$( - g )^\alpha g_{\lambda \mu}$}{(−g)α gλμ}}\label{(-g)^alpha*g}

Let us consider the gauge-fixing term that follows from the principal value type gauge-fixing term (\ref{F[g]}) by replacing the variable $w_{\lambda \mu}$, in which it is bilinear, with $( - g )^\alpha w_{\lambda \mu}$, $\alpha = \mathrm{const}$. The bilinear part of this term and, therefore, of the total action remain the same, but new interaction vertices arise. The question is, firstly, how the ghost contribution changes and, secondly, what will be the contribution to the diagram technique due to the new vertices. The following related discussion holds for both the limiting case of $\cA$ tending to 0 and the case of $\cA$ not equal to 0; the minor differences lie in the effective mechanism for regularizing the singularities and are accounted for by references to different formulas for the required quantities associated with the propagators.

For the ghost contribution, we subject ${\mathrm O}^{\lambda \mu}_\rho ( - g )^\alpha w_{ \lambda \mu }$ to an infinitesimal (approximate) diffeomorphism variation which acts on $( - g )^\alpha w_{ \lambda \mu }$ according to
\begin{equation}\label{dXi-gw}                                            %120
( - g )^{- \alpha} \delta^\Xi [ ( - g )^\alpha w_{ \lambda \mu } ] = \delta^\Xi g_{ \lambda \mu } + \alpha w_{ \lambda \mu } g^{\nu \rho} \delta^\Xi g_{ \nu \rho } .
\end{equation}

\noindent We track the $O ( \varepsilon )$ and $O ( 1 )$ contributions to the effective ghost action. At this level of precision, ${\mathrm O}^{\lambda \mu}_\rho$ commutes with $( - g )^\alpha$. Therefore, it is convenient to define the operator $\cO$ in this case as
\begin{equation}\label{Oxi=gOdgw}                                         %121
\cO_\rho{}^\nu \xi_\nu \stackrel{\rm def }{ = } - ( - g )^{ - \alpha } \rO^{\lambda \mu}_\rho \delta^\Xi [ ( - g )^\alpha w_{ \lambda \mu } ]
\end{equation}

\noindent (and the same for $\cO  \Rightarrow \ccO$, $\rO  \Rightarrow \crO$ for the case $\cA \neq 0$). At the same time, terms of order $O ( \varepsilon^2 )$ in $\cO$ are taken into account in the (exact) ghost propagator $\cO_{ ( 0 ) }^{ - 1 }$ (when $\Gamma^\nu_{\lambda \mu} = 0$, $g =  \mathrm{const}$), where they play the regularization role. This propagator is the same, as for $\alpha = 0$, (\ref{1/O(0)}) (or (\ref{1/checked_o}) for $\cA \neq 0$). The field $\rnu^\mu \Gamma^\nu_{\rho \mu}$ in ${\cal O}_\rho{}^\nu$ (\ref{cO}) and then in $( {\cal O}_{ ( 0 ) }^{ - 1 } {\cal O} )_\rho{}^\nu$ (\ref{O/O(0)}) (or (\ref{O/O(0)Aneq00}), (\ref{O/O(0)Aneq0}) for $\cA \neq 0$) corresponds to varying $w_{\lambda \mu}$ in $( - g )^\alpha w_{\lambda \mu}$ by $\delta^\Xi$ (the first term on the right hand side of (\ref{dXi-gw})) and is modified by adding a term originating from the second term on the right hand side of (\ref{dXi-gw}) (corresponding to varying $( - g )^\alpha$):
\begin{equation}\label{nG+anwG}                                           %122
\rnu^\mu \Gamma^\nu_{\rho \mu} \Rightarrow \rnu^\mu \Gamma^\nu_{\rho \mu} + \alpha \rnu^\mu w_{\rho \mu} (g^{\lambda \sigma} \Gamma^\nu_{\lambda \sigma} - g^{\lambda \nu} \Delta^{(s )}_\lambda ) .
\end{equation}

\noindent It is tempting to use the refined version of the underlying approximate diffeomorphism formula, Subsection \ref{diff-improve}. This is equivalent to the Christoffel symbol $\Gamma^\nu_{\lambda \mu}$ becoming, at the level of non-leading orders in metric variations from site to site, the finite difference operator $\gGamma^\nu_{\lambda \mu}$ (\ref{GammaGamma}). The contribution $O( 1 )$ to the ghost action is potentially determined by the first term on the right-hand side of (\ref{nG+anwG}); the new second term on the right-hand side of (\ref{nG+anwG}) is of order $O( \varepsilon )$, since $\rnu^\mu w_{\rho \mu} = O ( \varepsilon )$ ((\ref{nGeff}) or the discussion at the end of Subsection \ref{Aneq0ghost} for $\cA \neq 0$) and one can single out the part $O( 1 )$ in the factor:
\begin{eqnarray}                                                          %123
& & 2 \rnu^\mu \gGamma^\nu_{\rho \mu} = [ \rnu \Delta^{(s )}, g_{\lambda \rho} ] g^{\lambda \nu} + O ( \varepsilon ), \nonumber \\ & & 2 \alpha \rnu^\mu w_{\rho \mu} (g^{\lambda \sigma} \gGamma^\nu_{\lambda \sigma} - g^{\lambda \nu} \Delta^{(s )}_\lambda ) = \alpha \rnu^\mu w_{\rho \mu} ( 2 \Delta^{(s )}_\lambda - g^{\sigma \tau} \Delta^{(s )}_\lambda g_{\sigma \tau} ) g^{\lambda \nu} + O ( \varepsilon^2 ) .
\end{eqnarray}

\noindent The possible $O ( \varepsilon )$ contribution to the effective ghost action can be due to the diagrams with one $O ( \varepsilon )$ field $\alpha \rnu^\mu w_{\rho \mu} ( 2 \Delta^{(s )}_\lambda - g^{\sigma \tau} \Delta^{(s )}_\lambda g_{\sigma \tau} ) g^{\lambda \nu}$ (in fact, operator) and an arbitrary number $n - 1$ of $O ( 1 )$ fields/operators $[ \rnu \Delta^{(s )}, g_{\lambda \rho} ] g^{\lambda \nu}$. These fields enter such a diagram as
\begin{eqnarray}                                                          %124
& & [ \rnu \Delta^{(s )} , g_{\nu_n \lambda_n } ] g^{ \lambda_n \nu_{n - 1}} \otimes \dots \otimes [ \rnu \Delta^{(s )} , g_{\nu_j \lambda_j } ] g^{ \lambda_j \nu_{ j - 1}} \otimes \dots \otimes [ \rnu \Delta^{(s )} , g_{\nu_3 \lambda_3 } ] g^{ \lambda_3 \nu_2} \nonumber \\ & & \otimes [ \rnu \Delta^{(s )} , g_{\nu_2 \lambda_2 } ] g^{ \lambda_2 \nu_1} \otimes \alpha \rnu^{\lambda_1} w_{\lambda_1 \nu_1} ( 2 \Delta^{(s )}_\mu - g^{\rho \sigma} \Delta^{(s )}_\mu g_{\rho \sigma} ) g^{\mu \nu_n}.
\end{eqnarray}

\noindent To ensure the product of the first $n-1$ factors to be $O ( 1 )$, the indices $\rnu_j$ should be nonzero. In particular, $\rnu_n \neq 0$; then, if $\mu = 0$, then $g^{\mu \nu_n} = O ( \varepsilon )$; if $\mu \neq 0$, then $\Delta^{(s )}_\mu = \Delta^{(s )}_{\perp \mu}$ and integration over $\d^3 p_\perp$ gives zero due to the antisymmetry of the whole expression with respect to $p_{\perp \mu}$. Here the symmetry of the ghost propagator ${\cal O}_{ ( 0 ) }^{ - 1 }$ (\ref{1/O(0)}) or the combined expression ${\cal O}_{ ( 0 ) }^{ - 1 } {\cal O}$ (\ref{O/O(0)}) (or (\ref{1/checked_o}), (\ref{O/O(0)Aneq00}), (\ref{O/O(0)Aneq0}) for $\cA \neq 0$) plays a role. Thus, this diagram is, in fact, of order $O ( \varepsilon^2 )$. Together with the estimates of the contribution to $\ln \Det ( {\cal O}_{ ( 0 ) }^{ - 1 } {\cal O} )$ of other $O ( 1 )$ and $O ( \varepsilon )$ fields in this Subsection, \ref{A=0ghost}, above to be of order $O ( \varepsilon^2 )$ (including the case of the refined diffeomorphism formula, as noted in Subsection \ref{diff-improve}) this means that $\ln \Det ( {\cal O}_{ ( 0 ) }^{ - 1 } {\cal O} ) \to 0$ at $\varepsilon \to 0$.

According to the definition of $\cO$, (\ref{Oxi=gOdgw}), this provides the normalization factor under the functional integral sign to be
\begin{equation}\label{g^(4a)}                                            %125
\Phi = \Det [ ( - g)^\alpha \cO ] = \Det \cO \prod_{\rm sites} ( - g)^{4 \alpha} \to \prod_\mathrm{sites} ( - g)^{4 \alpha}
\end{equation}

\noindent for $\varepsilon \to 0$.

For the diagrammatic contribution of the new interaction vertices, it is tempting to perform the following change of variables in the functional integral,
\begin{equation}\label{w=w'/g^a}                                          %126
w_{0 \lambda} = \rgamma^{- \alpha} w^\prime_{0 \lambda} , \quad w_{j k} = w^\prime_{j k} , \quad \rgamma = \det \| g_{j k} \| ,
\end{equation}

\noindent for this simultaneously leads to two improvements, first, making the gauge-fixing term close to a bilinear one (thus simplifying the structure of vertices caused by this term), second, almost cancelling the found measure factor $( - g)^{4 \alpha}$ (\ref{g^(4a)}), which would complicate the diagram technique. Here $\rgamma$ is the determinant of the spatial block of the metric.

The matrices of the bilinear forms of the action plus the gauge-fixing term are the same for $w_{\lambda \mu}$ and for $w^\prime_{\lambda \mu}$. Correspondingly, the propagators are the same. Then a diagram with external lines $\rnu^\mu w_{\lambda \mu}$ and any of $w_{\sigma \tau}$, possibly with internal interaction vertices, already contains a factor $\varepsilon$ and is zero in the limit $\varepsilon \to 0$, and the same applies to a diagram in which these external fields are replaced by primed fields. It is convenient to omit the prime. If $w^\prime_{0 \lambda}$ are dummy variables (which seems to be the usual case, since one usually concentrates on the correlators of only the physical (spatial) components $w_{j k}$), it is even more appropriate to omit the prime.

With taking into account the Jacobian of (\ref{w=w'/g^a}), the measure factor becomes
\begin{eqnarray}\label{(g/g)^(4a)}                                        %127
& & \prod_\mathrm{sites} \left( \frac{ - g }{ \rgamma } \right)^{4 \alpha} = \prod_\mathrm{sites} \left( 1 + w_{0 j} \rgamma^{j k} w_{k 0} - w_{0 0} \right)^{4 \alpha} ,
\end{eqnarray}

\noindent where $\rgamma^{j k}$ is the reciprocal to the spatial block of the metric. Since $w_{0 j} = O ( \varepsilon )$, $w_{0 0} = O ( \varepsilon^2 )$ in the sense of their correlators with any components of the metric ((\ref{nGeff}), (\ref{nnGeff}) or the discussion at the end of Subsection \ref{Aneq0ghost} for $\cA \neq 0$), the equivalent contribution to the effective action $-i \sum_\mathrm{sites} \ln \Det ( - g / \rgamma )^{4 \alpha} = \sum_\mathrm{sites} O ( \varepsilon^2 )$ can be considered equal to zero at $\varepsilon \to 0$, and the coefficient (\ref{(g/g)^(4a)}) to one.

After this operation - substituting $\rnu^\mu w_{\lambda \mu} = \rgamma^{- \alpha} \rnu^\mu w^\prime_{\lambda \mu}$ and omitting the prime - the expression $\rO^{\lambda \mu}_\rho ( - g )^\alpha w_{ \lambda \mu } $ entering the gauge-fixing term takes the form
\begin{eqnarray}\label{Vw}                                                %128
& & \rO^{\lambda \mu}_\rho ( - g )^\alpha w_{ \lambda \mu } \Rightarrow \rO^{\lambda \mu}_\rho w_{ \lambda \mu } + \mathrm{V}^{\lambda \mu}_\rho w_{ \lambda \mu } , \nonumber \\ & & \mathrm{V}^{\lambda \mu}_\rho w_{ \lambda \mu } = \rnu^\mu w_{\rho \mu} \left[ \left( \frac{ - g }{ \rgamma } \right)^\alpha - 1 \right] + \frac{ \varepsilon^2 }{ 2 } \frac{ \Delta_{ \perp}^{(s) \nu } }{ \Delta_{ \perp}^{(s) 2 } } G^{(0) }_{\rho \nu \pi \zeta} \rnu^\pi \rlambda^{ \zeta 0} \frac{ \Delta_{ \perp}^{(s) j } }{ \Delta_{ \perp}^{(s) 2 } } w_{0 j} \left[ \left( \frac{ - g }{ \rgamma } \right)^\alpha - 1 \right] \nonumber \\ & & + \frac{ \varepsilon^2 }{ 2 } \frac{ \Delta_{ \perp}^{(s) \nu } }{ \Delta_{ \perp}^{(s) 2 } } G^{(0) }_{\rho \nu \pi \zeta} \rnu^\pi \rlambda^{ \zeta j} \frac{ \Delta_{ \perp}^{(s) k } }{ \Delta_{ \perp}^{(s) 2 } } w_{j k} \left[ \left( - g \right)^\alpha - 1 \right]
\end{eqnarray}

\noindent (and the same for $\rO  \Rightarrow \crO$, $G  \Rightarrow \cG$ in the case $\cA \neq 0$). Here $\mathrm{V}^{\lambda \mu}_\rho = \mathrm{V}^{\lambda \mu}_\rho ( g, \rgamma)$, a function of $g, \rgamma$ starting from order 1 over $w$ if expanded; effectively, $\mathrm{V}^{\lambda \mu}_\rho w_{ \lambda \mu } = O ( \varepsilon^2 )$ (setting $\rnu^\mu w_{\lambda \mu} = O ( \varepsilon )$, $\rnu^\lambda \rnu^\mu w_{\lambda \mu} = O ( \varepsilon^2 )$ in it). For the gauge-fixing term, we have
\begin{eqnarray}\label{wg^aOg^aw/e^2}                                     %129
& & \hspace{-5mm} \frac{ 1 }{ \varepsilon^2 } w_{\lambda \mu} ( - g )^\alpha \orO^{\lambda \mu}_\rho \left( \mM^{ - 1 } \right)^{\rho \kappa } \rO^{\sigma \tau}_\kappa ( - g )^\alpha w_{\sigma \tau} \Rightarrow \frac{ 1 }{ \varepsilon^2 } w_{\lambda \mu} \orO^{\lambda \mu}_\rho \left( \mM^{ - 1 } \right)^{\rho \kappa } \rO^{\sigma \tau}_\kappa w_{\sigma \tau} + \frac{ 1 }{ \varepsilon^2 } w_{\lambda \mu} \orV^{\lambda \mu}_\rho \nonumber \\ & & \hspace{-5mm} \cdot \left( \mM^{ - 1 } \right)^{\rho \kappa } \rO^{\sigma \tau}_\kappa w_{\sigma \tau} + \frac{ 1 }{ \varepsilon^2 } w_{\lambda \mu} \orO^{\lambda \mu}_\rho \left( \mM^{ - 1 } \right)^{\rho \kappa } \rV^{\sigma \tau}_\kappa w_{\sigma \tau} + \frac{ 1 }{ \varepsilon^2 } w_{\lambda \mu} \orV^{\lambda \mu}_\rho \left( \mM^{ - 1 } \right)^{\rho \kappa } \rV^{\sigma \tau}_\kappa w_{\sigma \tau} .
\end{eqnarray}

\noindent Here the first term for the general $w$ is $O ( \varepsilon^{- 2} )$ and is bilinear; upon using the propagator provided by the first term together with the action, the second and third terms are of order $O ( \varepsilon )$: the lowest order in $\varepsilon$ in $\rO w$ is $O ( \varepsilon )$ (the $\rnu w$ term), and in $\rV w$ it is $O ( \varepsilon^2 )$ (the third term with $w_{j k}$ in (\ref{Vw})), and there is the overall factor $\varepsilon^{-2}$. Then the appearance in a diagram of interaction vertices from the second and third terms leads to the smallness of at least $O ( \varepsilon )$ of this diagram and its disappearance at $\varepsilon \to 0$. Finally, the fourth term in (\ref{wg^aOg^aw/e^2}) is of order $O ( \varepsilon^2 )$, so using interaction vertices from it in any diagram also leads to the disappearance of this diagram in the limit $\varepsilon \to 0$.

In the action $\cS_{\rm g}$, replacing $\rnu w$ with $\rnu w + ( \rgamma^{- \alpha} - 1 ) \rnu w$ leads to the appearance of new interaction vertices proportional to $\alpha$ and its higher powers, but containing $\rnu w = O ( \varepsilon )$; therefore, the dependence on $\alpha$ disappears at $\varepsilon \to 0$.

Thus, the diagrammatic effect in the calculation of any specific amplitudes, caused by the appearance of new interaction vertices due to the replacement of $w_{\lambda \mu}$ by $( - g )^\alpha w_{\lambda \mu}$ in the gauge-fixing term, disappears as $\varepsilon$ tends to 0. Also the effective ghost contribution vanishes in this limit (more exactly, it reduces to a simple power volume factor in the functional measure, which is then cancelled by scaling the gauge variables $\rnu w$ by a power volume factor).

\subsection{Electromagnetic (Yang-Mills) analogy}\label{YM-ghost}

A priori non-trivial ghost contribution may arise if the electromagnetic field is generalized to the Yang-Mills field. The bilinear form of the action plus the gauge-fixing term $(- 1 / 2) \sum_{\rm sites} ( n^\lambda A_{\lambda a} ) \rlambda ( n^\mu A_{\mu b} ) \delta^{a b}$ is the sum of such forms for independent copies of the electromagnetic field numbered by the color index, $A_{ \lambda a }$. Correspondingly, the propagators, up to the trivial color factor $\delta^{a b}$, are the electromagnetic field propagators. The non-triviality of the ghost contribution is determined by the (approximate) infinitesimal symmetry transformation
\begin{equation}                                                          %130
\delta A^a_\lambda = \Delta^{(s )}_\lambda u^a - g t^{ a b c } A_{\lambda b} u_c ,
\end{equation}

\noindent so that the normalization factor inserted into the functional integral is
\begin{eqnarray}                                                          %131
& & \hspace{-10mm} \Det \ccO , \mbox{ where } \ccO^{a b} u_b \stackrel{\rm def}{=} \crO^\lambda \delta A^a_\lambda , \quad \ccO^{a b} = \rnu \Delta^{(s )} \left[ 1 - \frac{ \varepsilon^2 }{ ( \rnu \Delta^{(s )} )^2 + \cA ( \rnu^2 + \ralpha \oDelta \Delta ) } \right] \delta^{a b} \nonumber \\ & & + g t^{ a c b } \rnu^\lambda A_{\lambda c} - \varepsilon^2 g \frac{ \rnu \Delta^{(s )} }{ ( \rnu \Delta^{(s )} )^2 + \cA ( \rnu^2 + \ralpha \oDelta \Delta ) } \frac{ \Delta^{(s ) \lambda }_\perp }{ \Delta^{(s ) 2 }_\perp } t^{ a c b } A_{\lambda c} .
\end{eqnarray}

\noindent Here we can single out the free part $ \ccO_{ ( 0 ) } \stackrel{\rm def }{=} \ccO |_{ A_{ \lambda a } = 0 } $, the inverse
\begin{equation}                                                          %132
( \ccO_{ ( 0 ) }^{ - 1 } )_{ a b } = \frac{ [( \rnu \Delta^{(s )} )^2 + \cA ( \rnu^2 + \ralpha \oDelta \Delta )] \delta_{a b} }{ \rnu \Delta^{(s )} [( \rnu \Delta^{(s )} )^2 - \varepsilon^2 + \cA ( \rnu^2 + \ralpha \oDelta \Delta )] }
\end{equation}

\noindent of which plays the role of a ghost propagator. Note that these free part and propagator resemble the maximal spin (=2) parts of the free part $\ccO_{(0 ) \lambda \mu}$ of $\ccO_{\lambda \mu}$ ($\co_P P_{\lambda \mu}$ (\ref{checked_o})) and of the propagator $( \ccO_{(0 )}^{- 1} )_{\lambda \mu}$ ($ \bco_P P_{\lambda \mu} = \co_P^{- 1} P_{\lambda \mu}$ (\ref{1/checked_o})) for gravity.

For $\cA = 0$, this propagator turns out to exactly match the principal value prescription $( \rnu \Delta^{(s )} )^{- 1} \Rightarrow \frac{1}{2} \{ [ ( \rnu \Delta^{(s )} ) + \varepsilon ]^{- 1} + [ ( \rnu \Delta^{(s )} ) - \varepsilon ]^{- 1} \}$ for $( \rnu \Delta^{(s )} )^{- 1} \delta_{a b}$ (not in the Cauchy sense). If $\cA \neq 0$, then it is natural to prescribe the principal value precisely in the true Cauchy sense.

It is convenient to analyze $\Det ( {\cal O}_{ ( 0 ) }^{ - 1 } {\cal O} )$, where
\begin{eqnarray}\label{O/O(0)_em}                                         %133
& & ( {\cal O}_{ ( 0 ) }^{ - 1 } {\cal O} )^{a b} = \delta_{a b} + \frac{ ( \rnu \Delta^{(s )} )^2 + \cA ( \rnu^2 + \ralpha \oDelta \Delta ) }{ \rnu \Delta^{(s )} [( \rnu \Delta^{(s )} )^2 - \varepsilon^2 + \cA ( \rnu^2 + \ralpha \oDelta \Delta )] } g t^{a c b} \rnu^\lambda A_{\lambda c} \nonumber \\ & & -  \frac{ \varepsilon^2 g }{ ( \rnu \Delta^{(s )} )^2 - \varepsilon^2 + \cA ( \rnu^2 + \ralpha \oDelta \Delta ) } \frac{ \Delta^{(s ) \lambda }_\perp }{ \Delta^{(s ) 2 }_\perp } t^{ a c b } A_{\lambda c} .
\end{eqnarray}

\noindent The typical value of $\rnu^\lambda A_{\lambda a}$ can be estimated as the typical value of its correlator with any other field component, $\rnu^\lambda D^{\rm eff}_{\lambda \mu}$ in the limiting case $\cA \to 0$ or $\rnu^\lambda \cD_{\lambda \mu}$ for $\cA \neq 0$. Here $D^{\rm eff}$ or $\cD$ are $[D^{\rm eff} (n, n) + D^{\rm eff} (\on, \on)] / 2$ or $[\cD (n, n) + \cD (\on, \on)] / 2$, respectively; $D^{\rm eff} (n, n)$ differs from $\cD (n, n)$, naively taken at $\cA = 0$, by scaling its $O ( \ralpha^0 ) = O ( 1 )$ part by a factor of $\Delta^{(s) 2} ( - \oDelta \Delta)^{- 1} $, as discussed in Subsection \ref{em_finite} above equation (\ref{1/M_em}). The estimate is of the type of that one in equation (\ref{nGeff}) for gravity. We use $n^\lambda D^{\rm eff}_{\lambda \mu } ( n, n ) = n^\lambda D_{\lambda \mu } ( n, n ) = O ( \alpha ) = O ( \varepsilon^2 )$ and the same for $n \Rightarrow \on$ or $n^\lambda \cD_{\lambda \mu } ( n, n ) = O ( \alpha ) = O ( \varepsilon^2 )$ and the same for $n \Rightarrow \on$.
For $\cD ( n, n )$ ($\cD ( \on, \on )$) this also follows from the expansion of $\cD (n, n)$ ($\cD (\on, \on)$) in powers of $\cA$ with respect to $D (n, n)$ ($D (\on, \on)$), similar to the expansion of $\cG (\rnu, \rnu)$ in powers of $\cA$ with respect to $G (\rnu, \rnu)$ (\ref{ntGn=nGn+}) in gravity. Thus we get that $\rnu^\lambda A_{\lambda a} = O ( \varepsilon )$.

The contribution to $\ln \Det ( \ccO_{ ( 0 ) }^{ - 1 } \ccO )$, linear in the field $\rnu^\lambda A_{\lambda a}$ (from the second term in (\ref{O/O(0)_em})) is zero for either of two reasons: due to the antisymmetry of $t^{a c b} A_{\lambda c}$ in $a$ and $b$ or due to the antisymmetry of this term in $ \rnu \Delta^{(s )} $. The nonzero contribution to $\ln \Det ( \ccO_{ ( 0 ) }^{ - 1 } \ccO )$ starts with a contribution bilinear in the field $\rnu^\lambda A_{\lambda a}$, or with the third term in (\ref{O/O(0)_em})), both of order $O ( \varepsilon^2 )$. Thus, $\ln \Det ( \ccO_{ ( 0 ) }^{ - 1 } \ccO )$ vanishes as $\varepsilon$ tends to 0.

Thus, in this more simple system we are faced with the same mechanisms of providing well-definiteness and eventually vanishing the ghost contribution for the principal value type gauge-fixing term as in gravity, and these mechanisms are different for $\cA \neq 0$ and for the limiting $\cA \to 0$ case: in the $\cA \to 0$ case, the ghost propagator is regularized at $\rnu \Delta^{(s )} \to 0$ by effectively using the principal value prescription $( \rnu \Delta^{(s )} )^{- j} \Rightarrow \frac{1}{2} \{ [ ( \rnu \Delta^{(s )} ) + \varepsilon ]^{- j} + [ ( \rnu \Delta^{(s )} ) - \varepsilon ]^{- j} \}$ (not in the Cauchy sense), and in the $\cA \neq 0$ case, the singularities are like those at $ ( \rnu \Delta^{(s )} )^2 + \cA ( \rnu^2 + \ralpha \oDelta \Delta ) \to 0 $, and it is natural to prescribe the principal value precisely in the true Cauchy sense. Since the result eventually scales to zero, only its finiteness matters in the intermediate step.

\section{Conclusion}

How can one constructively obtain the required true zero-approximation finite-difference form of the action taking into account non-leading orders in metric variations? The point is that the action is a function of solely $g_{\lambda \mu}$ at the sites only in the leading order over metric variations; at non-leading orders the action also depends on lattice-specific simplicial metric components additional to $g_{\lambda \mu}$. In the dimension $d=4$ under consideration, we have $2^d - 1 = 15$ edge lengths per site in the aforementioned simplest periodic simplicial structure with a hypercubic cell (consisting of 4!=24 4-simplices), of which only $d ( d + 1 ) / 2 = 10$ are the number of $g_{\lambda \mu}$s, and 5 are the number of additional components per site. In principle, varying the additional metric components is sufficient to fine-tune (at the non-leading order level) one action function. Thus, assuming that the additional metric components are chosen correctly, the action can be considered to have the required form simply as a mini-superspace action (more precisely, as a "mini-superspace-in-mini-superspace" simplicial action).

Above, we arrived at the synchronous gauge as one that allows us to construct a consistent discrete perturbative expansion (and also allows us to integrate functionally over the connection and find the functional measure in terms of edge lengths in closed form, if we take the Regge calculus approach as a basis). We also noted the possibility of summing/averaging over all gauges to obtain the value of the original functional integral and the exact physical value in question. However, with such summation/averaging we can again restrict ourselves to the class of synchronous gauges if we generalize $g_{0 \lambda }^{(0 ) }$ not only as a constant but also as a function of the lattice site. Obviously, by summing the functional integral over all possible $g_{0 \lambda }^{(0 ) }$s, we go through all possible configurations. As $\varepsilon$ tends to 0 and $\rlambda^{\lambda \mu}$ was previously taken to be $O ( \varepsilon^{- 2} )$, the most singular part of the gauge-fixing term $\ccF$ leads to a delta-function factor in $\exp ( i \ccF )$ according to the limit exponential representation of the delta-function $\delta ( x ) = \lim_{\varepsilon \to 0} \pi^{- 1 / 2} \varepsilon^{- 1} \exp [ i ( \varepsilon^{- 2} x^2 - \pi / 4 ) ]$:
\begin{equation}                                                          %134
\prod_\mathrm{sites} ( 4 \pi )^{- 2} \exp \left( - \frac{ i }{ 4 } w_{0 \lambda} \rlambda^{\lambda \mu} w_{0 \mu} \right) \Rightarrow ( \Det \rlambda )^{ - 1 / 2 } \prod_\mathrm{sites} \delta^4 ( g_{0 \lambda} - g_{0 \lambda}^{(0 )} ) ,
\end{equation}

\noindent and integrating over $\prod_\mathrm{sites} \d^4 g_{0 \lambda}^{(0 )}$ returns us to the original functional integral symbol without any gauge fixing. It can be assumed that the methodology for obtaining the average value for all gauges is based on the methodology for working with the synchronous gauge.

Thus, the discrete perturbative expansion for gravity can be correctly formulated to correspond to the continuum expansion, in particular to reproduce those Feynman diagrams or their structures that are finite, at distances significantly larger than the elementary length scale. This is achieved, firstly, by the correct choice of the zero-approximation gravity action, taking into account the non-leading terms in the finite differences; secondly, by the correct choice of the gauge, which should fix the symmetry, namely the diffeomorphism symmetry in the leading order over metric variations from site to site.

This correct action contains both symmetrized $\Delta^{(s ) }_\lambda$ and standard advanced $\Delta_\lambda$ finite-difference derivatives (and we found the optimal distribution of these derivative forms over the terms); the proper gauge is the soft synchronous gauge in the principal value type prescription.

We have analyzed the principal value type graviton propagator and the corresponding term that needs to be added to the action. It is important that this term can be written (may be, up to $O ( \varepsilon^2 )$ terms) as a non-simple bilinear form of actually four gauge conditions.

This gauge-fixing term is a function of the "hard" synchronous gauge propagator, and a priori it is not clear whether it is non-singular or not; meanwhile, this is important if we use a nonlinear parametrization of the metric and this term becomes a source of interaction vertices (as we have in our recent paper \cite{khat2}). We find that this term can be defined in a finite way. Also the ghost contribution is found to vanish in the limit $\varepsilon \to 0$.

More generally, the principal value gauge-fixing term may differ from the bilinear term, roughly speaking, by a factor that is a power of $( - g )$ (again, as in our paper \cite{khat2}) and itself serve as a source of interaction vertices already without parameterizing the metric. We have analyzed such a term and have found that these interaction vertices make a vanishing contribution to the diagram technique in the limit $\varepsilon \to 0$; the effective ghost contribution vanishes in this limit up to a factor in the functional measure that is a power of $( - g )$.

The mechanism for ensuring the finiteness of the gauge-fixing term for the principal value type prescription operates somewhat differently in the limiting $\cA \to 0$ case (i. e., when non-leading orders over metric/field variations from site to site are neglected) and for $\cA \neq 0$ (i. e., when non-leading orders over finite differences are taken into account "as is"). In the limiting $\cA \to 0$ case, the possible singularities at $\Delta^{(s ) }_0 \to 0$ are explicitly cancelled; in the $\cA \neq 0$ case, some singularities appear, but allow a finite definition in the sense of the Cauchy principal value. What is important to us is the possibility of a non-infinite definition of some expressions independently of $\varepsilon$, where the Cauchy principal value is simply the most symmetric of these definitions; these expressions are then multiplied by arbitrarily small values at $\varepsilon \to 0$, so that the differences in the specific definitions become unimportant.

The gravitational propagator (\ref{vpG=}) for the improved finite-difference form of the action $\cS_{\rm g}$ (\ref{cSg}) is rather bulky, but in most cases we can use in (\ref{vpG=}) its effective form $G^{\rm eff}$ (\ref{Geff},\ref{G}) for small quasi-momenta or the one restricted to the spatial-spatial metric components $\cG_{\alpha \beta \gamma \delta}$ (\ref{Gabgd}) neglecting terms of normal order $O ( \varepsilon^2 )$. The interaction vertices are read from $\cS_{\rm g}$.

The existence of the consistent perturbative expansion means that it approximately reproduces the finite continuum corrections to the Newtonian potential as arising from the small momentum region; in the RC approach (Subsection \ref{Regge}), additional vertices and contributions also arise.

\section*{Acknowledgments}

The present work was supported by the Ministry of Education and Science of the Russian Federation.

\end{document}